\documentclass{article}[12pt]

\usepackage[english]{babel}

\usepackage[letterpaper,top=2cm,bottom=2cm,left=3cm,right=3cm,marginparwidth=1.75cm]{geometry}

\usepackage{amsmath,amsfonts,amssymb,bm}
\usepackage{graphicx}
\usepackage[colorlinks=true, allcolors=blue]{hyperref}
\usepackage{hyzsyntax}
\usepackage{setspace}
\usepackage{epigraph}
\usepackage{longtable}
\usepackage{cite}
\usepackage{setspace}
\usepackage{float}
\LTcapwidth=\textwidth

\counterwithin{equation}{section}
\counterwithin{figure}{section}
\counterwithin{table}{section}

\newcommand{\U}{\textup{U}}
\newcommand{\SU}{\textup{SU}}
\newcommand{\SO}{\textup{SO}}
\newcommand{\Sp}{\textup{Sp}}

\title{\Huge Atomic Higgsings of 6D SCFTs \vspace{1cm}}

\author{\Large Jiakang Bao and Hao Y. Zhang \vspace{0.5cm} \\
\href{jiakang.bao@ipmu.jp}{jiakang.bao@ipmu.jp}, \href{hao.zhang@ipmu.jp}{hao.zhang@ipmu.jp}\vspace{1cm}}

\date{}

\begin{document}
\maketitle

\onehalfspacing

\begin{tabular}{ll}
    & Kavli Institute for the Physics and Mathematics of the Universe (WPI), \\
    & University of Tokyo, Kashiwa, Chiba 277-8583, Japan \\
\end{tabular}

\vspace{2cm}

\begin{abstract}
In this paper, we study the full Higgs branch Hasse diagrams for any given 6d $\mathcal{N}=(1,0)$ SCFTs constructed via F-theory. This can be done by a procedure of determining all the minimal Higgsings on the generalized quivers of the 6d SCFTs. We call this procedure the atomic Higgsing, which can be implemented iteratively. We present our general algorithm with many concrete examples of Hasse diagrams. We also compare our algorithm with the Higgsings determined by the 3d $\calN = 4$ magnetic quivers. For the cases where the magnetic quivers are unitary, we can reproduce the full Hasse diagrams. We also construct the orthosymplectic magnetic quivers from the Type IIA brane systems for some new examples. Our approach, based on F-theory, applies to the known and new orthosymplectic cases, as well as theories that do not have known descriptions in terms of magnetic quivers. We expect our geometry-based approach to help extend the horizon of the RG flows of the 6d SCFTs.
\end{abstract}

\newpage

\tableofcontents

\epigraph{I say, type $A$ and type $D$ singularities are so exceptional
that they don't show the generic behavior. Type $E$ is the generic case.}{Yuji Tachikawa \cite{Yujislides}}

\section{Introduction and Summary}\label{intro}
The renormalization group (RG) flow is a powerful perspective for understanding the connections among quantum field theories with different energy scales. Supersymmetry provides more structure to study RG flows. For example, a sufficiently high amount of supersymmetry gives rise to the moduli spaces of vacua, which are parametrized by the vacuum expectation values (VEVs) along which the potential energy vanishes. At an RG fixed point, the symmetry structure might be enhanced to the superconformal symmetry, which is mathematically described by superconformal algebras. In such cases, we get SuperConforml Field Theories (SCFTs).

\subsection{Recap of 6D SCFTs via F-Theory}\label{recapFthy}
Six is the largest spacetime dimensionality in which superconformal symmetry could possibly occur \cite{Nahm:1977tg}. Therefore, the 6d SCFTs serve as a family of master theories that facilitate the study of their lower-dimensional descendants.

Moreover, the interest in studying the 6d SCFTs also comes from their strongly-coupled nature which is intrinsically tied to that of string theory. The 6d $\mathcal{N}=(2,0)$ SCFTs may be constructed as worldvolume theories of NS5-branes in Type IIA or IIB string theory on the Kleinian singularities \cite{Seiberg:1996qx,Witten:1996hc}. Their lower supersymmetry cousins, the 6d $\mathcal{N}=(1,0)$ theories, also admit string constructions, such as M5-branes probing the transverse Kleinian singularity or the M9 end-of-the-world brane (or both), or suspended D6-D8-NS5 brane configurations in Type IIA. On the one hand, such string constructions can provide us with a large family of 6d $(1,0)$ SCFTs. On the other hand, the non-perturbativeness of the M5-brane worldvolume is an intrinsic property of M-theory.

The most general string-theoretic construction of 6d SCFTs \cite{Heckman:2013pva,Heckman:2015bfa} makes use of F-theory \cite{Vafa:1996xn,Morrison:1996na}. The F-theory is the non-perturbative version of Type IIB that allows us to geometrize the axio-dilaton profile at strong coupling. It has been by far the most general framework for geometrically engineering quantum field theories. There, the basic objects are the 7-branes given by Weierstrass fibre singularities of complex codimension 1. The bifundamental matters are then determined at complex codimension 2 where further fibre enhancements occur, using the Katz-Vafa branching approach \cite{Katz:1996xe}. The Type IIB string theory on $\bbC^2/\Gamma_{\SU(2)}$ ($\Gamma_{\SU(2)}<\SU(2)$ a finite subgroup) can be viewed as F-theory on the same base space but with no interesting fibre profile. To go down to $\mathcal{N}=(1,0)$ supersymmetry following \cite{Heckman:2013pva}, we can replace $\bbC^2/\Gamma_{\SU(2)}$ by $\bbC^2/\Gamma_{\U(2)}$ to temporarily break supersymmetry, but then we need to include certain non-trivial fibre profile to restore supersymmetry. In particular, 6d anomaly cancellation imposes stringent conditions that are always capable of uniquely fixing the gauge symmetry and the matter content. This was given in detail in \cite{Heckman:2013pva}. Moreover, the full list of possible adjacency patterns of complex curves on the base was worked out in \cite{Heckman:2015bfa}.

With the most general known tool to construct 6d SCFTs, it is also natural to study the rich structure of possible RG flows among them, as done in \cite{Heckman:2015bfa,Heckman:2015ola,Heckman:2016ssk,Heckman:2018pqx,Mekareeya:2016yal,Frey:2018vpw,Hassler:2019eso,Baume:2021qho,Fazzi:2022hal,Fazzi:2022yca,Baume:2023onr,DelZotto:2023myd,Fazzi:2023ulb}. In particular, in \cite{Heckman:2016ssk}, a subset of 6d RG flows has been explicitly matched to a hierarchy of nilpotent VEVs inside the flavour symmetry of a particular UV theory.

However, there are clearly more RG flows than those that are known to admit elegant algebraic descriptions. For example, one could consider semi-simple RG flows by changing an infinitely long chain of curves, or RG flows that split a single 6d SCFT into multiple irreducible ones. All of these are not yet understood in terms of elegant algebraic structures as in the nilpotent hierarchy case. Therefore, on the 6d side, we could get inspiration from various approaches in order to expand our power in analyzing larger families of RG flows.

One very specific question goes as follows. Given any specific 6d SCFT, are we able to find all its descendants via \textit{minimal} Higgsings? A \textbf{minimal Higgsing} is a Higgsing $\calT_A \rightarrow \calT_B$ such that it is impossible to find a different $\calT_C$ with $\calT_A \rightarrow \calT_C \rightarrow \calT_B$. In other words, we need an algorithm to generate the full Higgs branch of this theory. This would be the main focus of the paper. To encode the structure of the Higgs branch, one would often use the Hasse diagram (aka phase diagram) that describes the partial ordering of the theories along the RG flows.

\subsection{Recap of Magnetic Quivers}\label{recapMQs}
It is expected that the Higgs branch of the 6d theory in question is a symplectic singularity. Therefore, the problem of finding the Higgs branch structure can be translated into understanding the stratification of the singularity. In particular, the partial ordering in the Hasse diagram would coincide with the one for the symplectic leaves. For any pair of leaves $(\mathcal{L}_1,\mathcal{L}_2)$, we have $\mathcal{L}_1<\mathcal{L}_2$ if $\mathcal{L}_1\subset\overline{\mathcal{L}_2}$. Then identifying the minimal flows is equivalent to finding the elementary transverse slice between the two leaves.

This brings us another important source of inspiration, namely the 3d $\mathcal{N}=4$ magnetic quivers \cite{Cabrera:2019izd,Cabrera:2019dob}. One may think of the basic idea as some sort of electromagnetic duality. For two 3d theories, when their Higgs and Coulomb branches are exchanged, we have the 3d mirror symmetry, which is also known as the symplectic duality. Now, this perspective may be extended to a $p$-dimensional QFT with 8 supercharges for $p=3,4,5,6$, in the sense that its \emph{Higgs} branch could be studied via the \emph{Coulomb} branch (aka dressed space of monopole operators) of a 3d $\mathcal{N}=4$ magnetic quiver. More specifically, we have the following identity as an equality of moduli spaces:
\begin{equation}
    \mathcal{H}^{p\text{d}}(\text{electric theory})=\mathcal{C}^{3\text{d}}(\text{magnetic quiver}).\label{H=C}
\end{equation}
With this at hand, we may invoke many tools that have been vigorously developed for the 3d $\mathcal{N}=4$ Coulomb branches/moduli spaces of dressed monopole operators \cite{Cremonesi:2013lqa,Cremonesi:2014kwa,Cremonesi:2014xha,Nakajima:2015txa,nakajima2015questions,Hanany:2016gbz,Braverman:2016wma,Braverman:2016pwk,Nakajima:2016owa,Cabrera:2017ucb,Hanany:2017ooe,Ferlito:2017xdq,Braverman:2017ofm,Nakajima:2017bdt,Hanany:2018uzt,Cabrera:2018ldc,Hanany:2018xth,Cabrera:2018jxt,Braverman:2018gvt,Hanany:2019tji,Nakajima:2019olw,Bourget:2020gzi,Bourget:2020xdz,Bourget:2021siw,Bourget:2022ehw,Bourget:2022tmw,Akhond:2020vhc,Akhond:2021knl,Akhond:2022jts,Akhond:2021ffo}. In particular, we have the quiver subtractions \cite{Cabrera:2018ann} and the quiver decays and fissions \cite{Bourget:2023dkj,Bourget:2024mgn} to help us obtain the Hasse diagrams for the Higgsings \cite{Bourget:2019aer}.

For $p=6$, when one moves along the tensor branch, there would be massless degrees of freedom arising from the tensionless BPS strings at a singularity. As a result, the Higgs branch would emanate which would change discontinuously when at least one scalar field in the tensor multiplet/inverse gauge coupling becomes zero. Examples include discrete gaugings \cite{Hanany:2018vph,Hanany:2018dvd,Hanany:2018cgo,Bourget:2020bxh,Hanany:2023uzn}, small $E_8$ instanton transitions \cite{Ganor:1996mu,Mekareeya:2017sqh,Hanany:2018uhm}, etc. In our story, we would move to the origin of the tensor branch, which is the CFT point, where all the gauge couplings are infinite.

Now, to make use of \eqref{H=C}, the first task would be to find the corresponding magnetic quivers. In this paper, we shall mainly use the D6-D8-NS5 brane setup in Type IIA to obtain the magnetic quivers. Given a theory with its tensor branch description encoded by the generalized/electric quiver, it could be possible to construct it via the brane system. Then after Hanany-Witten (HW) brane transitions \cite{Hanany:1996ie,Cabrera:2016vvv,Cabrera:2017njm,Cabrera:2019izd,Cabrera:2019dob,Hanany:2022itc}, we can move to the magnetic phase and read off the magnetic quiver.

When there is a unitary magnetic quiver description, the quiver decays and fissions \cite{Bourget:2023dkj,Bourget:2024mgn} would give all the descendant theories of the UV theory. The idea is to find all the ``smaller'' quivers (to be explained more precisely later) of the parent quiver, and their partial ordering would naturally be the one for the symplectic leaves. However, for many 6d theories, we need to introduce orientifolds to construct the Type IIA brane system (possibly with negatively charged branes allowed \cite{Mekareeya:2016yal,Hanany:2022itc}). This would then yield orthosymplectic magnetic quivers. For those orthosymplectic cases, the quiver substractions and the quiver decays and fissions are still under development. For instance, one cannot tell whether a theory is bad only from the appearance of the underbalanced nodes. For theories associated with non-special nilpotent orbits, the existence of the magnetic quivers is not clear, and simply applying the same logic as for the unitary cases does not seem to work. As we will see, there could even be cases where we have both unitary and orthosymplectic magnetic quivers in the same Hasse diagrams.

Moreover, in general, there is no uniform algorithm to identify the magnetic quivers, and one has to look for it in a case-by-case approach. In fact, we are not aware of any physical principle supporting the existence of the magnetic quiver for an arbitrary electric theory of interest. With this situation in mind, it would be very fruitful to look for alternative methods that could study the Higgsings of the theories whose magnetic quivers are not yet known. This provides another motivation for our work.

\subsection{Conventions}\label{conventions}
Before summarizing our results, let us first state the notations and conventions in this paper. In the literature, there are two ways of drawing the Hasse diagrams, where the process of Higgsings either goes from the top to the bottom or goes from the bottom to the top. Here, we shall adopt the former convention (except for one example in Appendix \ref{A3rank3sympsing}).

Since the Hasse diagrams often involve many flows in our examples, we shall omit all the multiplicities to avoid clutter. For instance, in the conformal matter theory
\begin{equation}
    [\SO(8)] \ \ 1 \ \ \overset{\kso(8)}{4} \ \ \dots \ \ \overset{\kso(8)}{4} \ \ 1 \ \ [\SO(8)],
\end{equation}
there are obviously two $[\SO(8)]$ flavours one can start to Higgs. As the two routes are identical, we shall only draw ``half'' of the Hasse diagram. Where there would be multiple identical routes and where these routes would merge into one child theory should be clear from the generalized quivers of the theories.

Notice that when the slice of a flow is the union of $\mu$ copies of an elementary slice $\mathcal{S}$, we shall denote it as $\mu\cdot\mathcal{S}$. Sometimes, there is an ``outer action''/monodromy on the slice. When this is known, we shall follow the notations in \cite{fu2017generic,juteau2023minimal} to label them (see \S\ref{moreslices} for the notations).

We shall also use the following shorthand notations for the generalized quivers. For $n$ copies of the same piece, we shall use $\{\dots\}^{\otimes n}$. For instance,
\begin{equation}
    \{[E_8] \ \ 1\}^{\otimes2}=[E_8] \ \ 1 \ \ \sqcup \ \ 1 \ \ [E_8].
\end{equation}
When there are $n$ same legs attached to the same curve of self-intersection $-m$, we shall write $m(\dots)^{\otimes n}$. For instance,
\begin{align}
    [\SO(8)] \ \ 1 \ \ &\overset{\kso(8)}{4} \ \ 1 \ \ [\SO(8)]\nonumber\\
    \overset{\kso(8)}{4} \ \ (1 \ \ [\SO(8)])^{\otimes3}\quad=\quad\quad\quad\quad\quad\quad&\ \ 1\label{D4hat1notation}\\ 
    [&\SO(8)]\nonumber.
\end{align}

In this paper, we will consider various examples, and we shall adopt the following labeling system. The starting UV theory $\mathcal{T}$ would be denoted as $G^r$, where $G$ is the flavour group on at least one end in the generalized quiver. For instance, it would be the group $G$ in the $(G,G)$ conformal matter theory. As many cases have families of the generalized quivers with arbitrary lengths, the superscript $r$ would denote the rank. For example, $D_4^3$ refers to the $(D_4,D_4)$ conformal matter theory of rank 3:
\begin{equation}
    [\SO(8)] \ \ 1 \ \ \overset{\kso(8)}{4} \ \ 1 \ \ \overset{\kso(8)}{4} \ \ 1 \ \ \overset{\kso(8)}{4} \ \ 1 \ \ [\SO(8)].
\end{equation}
In particular, $G_k^{e_8}$ would be the $G_k$-type orbi-instanton theory, where one end of the generalized quiver has flavour $G_k$ and the other end always has $E_8$. Moreover, for the example in \eqref{D4hat1notation}, we shall denote it as $\widehat{D}_4^1$ as this is obtained by attaching an extra leg (which is identical to the other two) to the $D_4^1$ conformal matter theory. Now, for all the descendants, that is, all the nodes in the Hasse diagram below the starting UV theory $\mathcal{T}$, we shall write them as $(\mathcal{T},n)$. Here, the number $n$ is just a label and does not have any special meaning (although theories with larger $n$ would often live deeper in the IR on the whole). The starting UV theory $\mathcal{T}$ is then denoted as $(\mathcal{T},0)$.

As a given type theory could possibly have an infinite family of generalized quivers with different lengths, there would be differences in long and short quivers \cite{Hassler:2019eso}. For a long (resp.~short) quiver, the Higgsings of different flavour symmetries would be uncorrelated (resp.~correlated). When we say that we are considering the Hasse diagram of long quivers, we would often mean that all the descendant theories are also long quivers. In other words, there could be cases where the UV theory is a long quiver while there are some short quivers in the IR.

We also need to specify the convention for the symplectic groups. Here, we choose the quaternionic notation $\Sp(n)$, that is, $\Sp(n)=\Sp(2n,\mathbb{C})$.

\subsection{Summary}\label{summary}
In this paper, we will try to look for a more systematic approach to identify all possible \textit{minimal} Higgsings for a given 6d SCFT via F-theory geometry. There would be four types of minimal Higgsings:
\begin{itemize}
    \item First, we have the minimal nilpotent orbit Higgsings which, as the name suggests, are associated with the minimal nilpotent orbits of the flavour symmetries.
    \item We also have the minimal plateau Higgsings which are triggered by semi-simple parts of the flavour symmetries (and hence are often called semi-simple Higgsings in the literature). Here, we call them the plateau Higgsings due to the curve configurations of the theories in such cases (as will be made precise below).
    \item Besides, we have the endpoint-changing Higgsings. They would change the curve configurations of the generalized quivers.
    \item Moreover, we need to introduce the notion of the combo Higgsings. A combo Higgsing is a combination of individually forbidden steps that are collectively allowed. Some combo Higgsings can be thought of as DE-type analogues of the plateau/semi-simple Higgsings, but other ``sporadic'' combo Higgsings do not come in the infinite families.
\end{itemize}

The first two types of Higgsings are well-known and have been extensively studied such as in \cite{Heckman:2016ssk,Hassler:2019eso,Baume:2021qho}. Here, we have two more types. With the algorithm to perform these ``atomic Higgsings'', we can in principle obtain the complete Hasse diagram for any given 6d SCFT.

Besides the story from the F-theory, the magnetic quivers provide another inspiration for the paper. In particular, the quiver decays and fissions are also designed to find the full Hasse diagrams. However, when there are orthosymplectic magnetic quivers, the complete rule of quiver decays and fissions is still under development. There could even be cases whose existence of the magnetic quivers is not known. Here, we would only need the tensor branch descriptions to recover all the nodes in the Hasse diagrams.

Nevertheless, we shall still discuss the magnetic quivers in this paper for various purposes:
\begin{itemize}
    \item We can use the unitary magnetic quivers as a cross-check of our algorithm.
    \item For some minimal Higgsings, the transverse slices are not obvious from the generalized quivers. These slices might be identified with the help of magnetic quivers (including some orthosymplectic ones).
    \item We can find the orthosymplectic quiver decays and fissions for some cases in light of our algorithm. There could also be cases whose Hasse diagrams contain both the unitary and the orthosymplectic magnetic quivers. Our algorithm could indicate how they could be related under the RG flows and identify the transverse slices in some cases. Of course, here, we are not aiming for complete rules for the orthosymplectic quiver subtractions or the decays and fissions.
\end{itemize}
Overall, all the tools and approaches that appear in the literature are very crucial and useful to obtain the Higgs branch structures of 6d SCFTs, and we hope that our algorithm could also shed light on this.

Regarding the magnetic quivers, we shall also have some discussions on the validity of the Type IIA constructions, such as the long/short quivers, as well as the non-special nilpotent orbits. Moreover, we would construct some new magnetic quivers for some of the theories.

\paragraph{Outlook} There are still many aspects of understanding the RG flows of 6d SCFTs that are worth exploring in the future. For some minimal Higgsings, such as some endpoint-changing Higgsings, our algorithm is based on a brute-force search of the possible IR theories. It would be desirable to find a more practical and efficient way for such flows, especially for certain $\mathcal{N}=(1,0)$ theories.

For each minimal flow, there is an elementary slice. In our algorithm, it would be non-trivial to write down the specific slices except for some special cases, and we have not provided a complete rule to identify all the slices. In general, it is not clear what VEVs would trigger various flows. When other methods are available, we can use these methods to find the elementary slices. However, there are still many cases where the slices are unknown. A more systematic way to identify the transverse slices is required.

When a theory has an associated pair of nilpotent orbits, it could be possible that an orbit is very even. From the generalized quivers (and also the magnetic quivers), the two very even orbits corresponding to the same partition do not show any differences. However, as argued in \cite{Distler:2022yse}, the theories associated to the very even pairs $(\mathcal{O}^\text{I},\mathcal{O}^\text{II})$ and $(\mathcal{O}^\text{I},\mathcal{O}^\text{II})$ would have different Higgs branches (the one with $(\mathcal{O}^\text{II},\mathcal{O}^\text{II})$ is the same as the one with $(\mathcal{O}^\text{I},\mathcal{O}^\text{I})$). In \cite{Distler:2022yse}, they have extra labels indicating the positive and negative chiralities of the spinor representations of the D-type flavour symmetries.

In this paper, we shall not distinguish the two cases and temporarily treat them as the same. Nevertheless, we still need to find a way to tell the different Higgsings from one another and incorporate this into our algorithm. At the current stage, little is known about the differences in the Higgsings of these very even cases. Let us just make a brief comment here. Denote the two theories as $\mathcal{T}^\text{I,I}$ and $\mathcal{T}^\text{I,II}$. Since they would still eventually flow to the same set of IR theories (such as those associated with other nilpotent orbit pairs), this indicates that (at least) one of the two theories should have some extra descendant theories. In other words, our current algorithm ignoring the differences between $\mathcal{T}^\text{I,I}$ and $\mathcal{T}^\text{I,II}$ would give a subdiagram without these extra descendant IR theories in the actual full Hasse diagram. However, it is even harder to tell which one would have the extra nodes and flows.

From the perspective of magnetic quivers, we can also see that the orthosymplectic quivers are more complicated than the unitary ones. In particular, more insights on the quiver subtractions and quiver decays and fissions are needed for the orthosymplectic cases. For instance, for theories involving non-special nilpotent orbits, it is not even clear whether one could construct the corresponding magnetic quivers. In some cases as discussed in \S\ref{sec:magneticquivers}, the ``multiplicities'' are also not clear from the orthosymplectic magnetic quivers. For example, the $(E_6,E_6)$ conformal matter theory has two $E_6$ flavours that can be Higgsed, which is evident in the tensor branch description. However, they seem to lie in the same position in the magnetic quiver in the sense of quiver decays and fissions. Moreover, there seems to be an endpoint-changing flow for the $\widehat{D}_4^1$ theory, and if so, it should correspond to a quiver fission. This is not obvious from the magnetic quiver if we simply extrapolate the knowledge of the unitary magnetic quivers to the orthosymplectic ones.

\paragraph{Outline} The rest of the paper is organized as follows. In \S\ref{sec:review}, we give a brief review of 6d SCFTs and their F-theory engineerings, as well as some basic aspects of the magnetic quivers. In \S\ref{sec:algorithm}, we give the general procedure of identifying the minimal Higgsings for a given 6d SCFT. We discuss the elementary slices in \S\ref{sec:elementaryslices}, especially for all the minimal RG flows we have in our examples. In \S\ref{sec:examples}, we consider various examples of the complete Hasse diagrams. We consider the magnetic quivers in \S\ref{sec:magneticquivers}, where known results can be reproduced and there are also some new identifications of the magnetic quivers in several 6d SCFTs. In Appendix \ref{table6d}, we collect the results of the matter contents and the flavour symmetries in 6d SCFTs for convenience. In Appendix \ref{TypeIIAbranes}, we present the Type IIA brane constructions, from which the magnetic quiver descriptions can be deduced. In Appendix \ref{Sp3conformalmatters}, we have considered the magnetic quivers for the C-type conformal matter theories as another example. In Appendix \ref{quiverdecayfission}, we review the quiver decay and fission algorithm for the unitary magnetic quivers. In Appendix \ref{A3rank3sympsing}, we list the magnetic quivers in the Hasse diagram of a 6d $(1,0)$ theory with an A-type base and $\ksu$-type fibre decorations, which also shows up as part of some D-type Hasse diagrams, as supplementary material.

\section{A Lightening Review of 6D SCFTs}\label{sec:review}
Building upon our recap of the study of 6d SCFTs, we now give the necessary technical background which will be assumed throughout our paper (see \cite{Heckman:2018jxk} for a more extensive review). The first two subsections are devoted to the construction of the 6d SCFTs via F-theory. Then we shall recall the study of their Higgs branches via the Coulomb branches/moduli spaces of dressed monopole operators of their dual 3d $\mathcal{N}=4$ magnetic quivers.

\subsection{Classification of Bases and Fibres}\label{basesfibres}
The F-theory construction of 6d SCFTs involves the internal space of a non-compact elliptically fibred Calabi-Yau threefold \cite{Morrison:1996na,Morrison:1996pp}, with a base configuration given by the orbifold
\be
    \bbC^2/\Gamma_{\U(2)}.
\ee
Here, $\Gamma_{\U(2)}$ is a finite subgroup of the isomorphism $\U(2)$ acting on the covering space $\bbC^2$ in the following way:
\be
    (z_1, z_2) \rightarrow (\omega_p z_1, \omega_p^q z_2), \ \ \text{s.t.} \ \ \omega_p^p = 1.
\ee
If we were to compactify the weakly-coupled string theory (i.e., Type IIB) on a generic orbifold of the above type, we would break supersymmetry whenever $\Gamma_{\U(2)}$ is not a subgroup of $\SU(2) \subset \U(2)$. However, the non-trivial fibre profile in the elliptic fibration that is used as the F-theory internal manifold makes the background strongly coupled, so supersymmetry is restored.

Nevertheless, it has been a long-standing open problem to obtain an intrinsic description of such strongly coupled QFTs. Therefore, it is instructive to try to describe them on the tensor branches, i.e., by giving VEVs to the tensor multiplets and considering the resulting supersymmetric QFTs that are not conformal.

For a 6d $(2,0)$ theory, going onto the tensor branch can be understood from a top-down approach via the resolution of the Kleinian singularity into a collection of complex projective spaces $\bbC\bbP^1$ (we will call them complex curves for brevity), whose self- and mutual-intersection patterns are encoded in the ADE Dynkin diagram via the McKay correspondence \cite{mckay1983graphs}. Specifically, each curve would have self-intersection $-2$, whereas some pairs of the curves would have mutual intersections 1 according to the Dynkin diagram.

Following \cite{Heckman:2013pva,Heckman:2015bfa}, the tensor branch configuration of a $(1,0)$ theory can be treated similarly as its $(2,0)$ counterpart again by resolving the singularity. However, this time, we need to carefully keep track of the gauge symmetries that are ``paired'' to all the curves. The first difference is that the intersection pairing matrix of the resolved curves no longer has to be one of the ADE Dynkin diagrams, but they are instead enumerated in \cite{Heckman:2013pva}. For each sought-after curve configuration, in case there are any $-1$ curves (namely, curves with self-intersection $-1$), they can always be iteratively blown down until there are no more $-1$ curves. One then reaches the so-called endpoint configuration.

Now, each curve $C_i$ could be paired to a gauge group factor that is a single Lie group $\kg_i$, meaning that the gauge coupling of $\kg_i$ is inversely proportional to the volume of $C_i$. However, the allowed options of $\kg_i$ depend on the self-intersection of $C_i$ via a set of stringent anomaly cancellation conditions \cite{Heckman:2013pva,Heckman:2015bfa}. Specifically, for a $-n$ curve with $n \geq 3$, its paired $\kg_i$ must be non-trivial. Thus, one gets instances of non-Higgsable clusters (NHCs), i.e., minimally paired gauge group(s) associated with a given curve configuration:
\be
    \overset{\ksu(3)}{3}, \quad \overset{\kso(8)}{4}, \quad \overset{\kf_4}{5}, \quad \overset{\ke_6}{6}, \quad \overset{\ke_7}{7} + \frac{1}{2} \mathbf{56}, \quad \overset{\ke_7}{8}, \quad \overset{\ke_8}{12}.
\ee
In addition, there are three more NHC configurations with more than one curve (with some extra matter contents left implicit):
\be
    \overset{\ksu(2)}{2} \ \ \overset{\kg_2}{3}, \quad\quad 2 \ \ \overset{\ksu(2)}{2} \ \ \overset{\kso(7)}{3}, \quad\quad \overset{\ksu(2)}{2} \ \ \overset{\kso(7)}{3} \ \ \overset{\ksu(2)}{2}.
\ee
In general, anomaly cancellation conditions are strong enough to determine (1) what the allowed gauge configuration is given a curve configuration, and (2) what the matter content is given both the curve and the gauge configurations. We reproduce the relevant table in the literature in Appendix \ref{table6d}. Via a complete set of rules of building blocks and their gluings, an atomic classification of 6D SCFTs based on their tensor branch descriptions is made available in \cite{Heckman:2015bfa}.

\subsection{Previous Knowledge of RG Flows among 6D SCFTs}\label{RGflows}
With a classification of 6d $(1,0)$ SCFTs from F-theory, it is now tempting to ask if their RG flows can be studied within the same framework. Before answering this question, let us recall some simpler RG flows of 6d SCFTs using Type II string theories. In Type IIB, one can consider a stack of $k$ NS5-branes realizing the $A_{k-1}$ type $(2,0)$ theory. Then the operation of splitting it into two stacks of NS5-branes is an example of RG flows. In the dual Type IIA description involving a $\bbC^2/\bbZ_k$ geometric singularity, such a flow should be described as a complex structure deformation, resulting in two separate singularities with lower orders.

However, with the F-theory construction, one can get larger families of RG flows, which involve some flows that cannot be accessed by simple operations as above. We now introduce two aspects of them.

\subsubsection{T-Brane Deformations}\label{Tbranes}
In \cite{Cecotti:2010bp}, T-branes (where ``T'' comes from the word triangular) were introduced to describe a special type of brane configuration that is a non-Abelian analogue of intersecting branes. One can obtain them by taking the UV brane configuration and turning on a VEV corresponding to a nilpotent element in the flavour symmetry (hence the name nilpotent VEV). For unitary gauge groups and flavour symmetries, the resulting brane configuration can honestly be described as a suspended brane configuration only involving D-branes and NS5-branes. Here, we reserve the terminology for more general cases and use the term \textit{T-brane VEV} interchangeably with the term \textit{nilpotent VEV}. More interpretations on T-brane VEVs in terms of moduli space geometry can be found in \cite{Anderson:2013rka}.

Nilpotent VEVs take value in the nilpotent orbits of the UV flavour symmetry group. They admit a partial ordering defined as follows (distinct orbits themselves are disjoint by definition, so we need to instead take the closure):
\be
    \calO_1 \leq \calO_2 \quad \text{if} \quad \calO_1 \subset \overline{\calO}_2.
\ee
The collection of all nilpotent orbits forms a Hasse diagram under the above partial ordering relation. In \cite{Heckman:2016ssk}, the Hasse diagram of nilpotent orbits is found to be precisely embedded into the Hasse diagram of 6d SCFTs. Physically, the authors discovered that for some specific UV theories ($k$ M5-branes probing $\bbC^2/\Gamma_{\kg}$ transverse singularity), all nilpotent VEVs would land on CFTs. For two IR theories triggered by two nilpotent orbits $\calO_1, \calO_2$ from the same UV theory $\calT_\text{UV}$, the existence of the RG flow between them is implied whenever there is a closure inclusion between $\calO_1, \calO_2$. In other words, we can have a 6d SCFT $\calT[\calO_i]$ for each $\calO_i$. However, one can ask if all possible 6d SCFTs between $\calT[\calO_1]$ and $\calT[\calO_2]$ would take the form $\calT[\calO_3]$, i.e., whether it must come from a third nilpotent orbit $\calO_3$. For the case where only Higgsing on a single flavour symmetry $G$ is involved, the one-to-one correspondence holds. The resulting Hasse diagram is thus called the \textit{nilpotent hierarchy} of the long quiver type. Here, ``long'' indicates that we consider an infinite repeating pattern of the generalized quiver, and we assume the other end not to be Higgsed.

For theories with multiple flavour symmetries (such as $G_1, G_2$) and the nilpotent Higgsings for all of these flavour symmetries (such as $\calO_{G_1}, \calO'_{G_2}$), one could also get a more complicated double Hasse diagram of the \textit{short quiver type} \cite{Hassler:2019eso}. In such cases, however, we will see later that between the UV theory and the IR theory of a specific short quiver nilpotent hierarchy, not all 6d SCFTs can be described by a pair of nilpotent orbits.

\subsubsection{Flat Connections via Homomorphisms into \texorpdfstring{$E_8$}{e8}}\label{homE8}
In the M-theory configuration of $N$ M5-branes simultaneously probing a transverse $\bbC^2/\Gamma_{\kg}$ singularity and an M9-brane (where the singularity sits in the directions of the M9-brane that are transverse to the M5-branes), the flavour symmetry on the tensor branch would be $G \times E_8$. To understand the Higgs branch of such theory, we can not only consider the T-brane Higgsings with respect to this $G$ factor but also perform another type of Higgsings associated with $E_8$, related to the flat $E_8$ gauge configurations on the asymptotic boundary $S^3/\Gamma_{\kg}$. Such flat connections are known to be topologically classified by their holonomies, i.e., discrete homomorphisms of the finite group $\Gamma_{\kg}$ into $E_8$. Turning on such a flat gauge configuration on the asymptotic boundary also amounts to performing a Higgsing on the 6d SCFT, and we shall call such Higgsings discrete homomorphism Higgsings for brevity.

Results matching the discrete homomorphism Higgsings to the 6d SCFTs can be found in \cite{Heckman:2015bfa,Frey:2018vpw}. However, in the limit that the number of M5-branes goes to infinity, we remark that the Higgsing Hasse diagram for the long quiver SCFT also goes to infinity. It would be ideal to have a first-principle understanding to determine the IR SCFT directly using the data of such discrete homomorphisms. This is particularly necessary since a partial ordering for the collection of discrete homomorphisms of finite subgroups of $\SU(2)$ into $E_8$ does not seem to be mathematically known as of now.

\subsection{Anomaly Polynomials and Higgs Branch Dimensions}\label{IanddH}
A robust observable of a 6d SCFT is its anomaly polynomial. It is a degree 8 homogeneous polynomial of the characteristic classes of the 6d tangent bundle, the R-symmetry bundle, and the flavour symmetry bundle.

The anomaly polynomial of a 6d $(2,0)$ theory of A-type was understood back in \cite{Harvey:1998bx} by taking the 6d $(2,0)$ theory to live on a stack of $k$ overlapping M5-branes and considering the anomaly inflow of the M-theory bulk topological term onto those M5 branes. A similar idea applies to 6d E-string theories that have $\calN = (1,0)$ supersymmetry by adding an extra ingredient of M9-brane in the same picture \cite{Ohmori:2014pca}.

Later, the anomaly polynomial of a general 6d $(1,0)$ theory was determined in \cite{Ohmori:2014kda} by using the 't Hooft anomaly matching and analyzing the anomaly on the tensor branch. Using the idea of the 't Hooft anomaly matching, the total anomaly on the tensor branch (where the supersymmetry and the R-symmetry are preserved) should match with that at the superconformal fixed point. On the tensor branch, one can have fermions in the hyper, vector, and tensor multiplets running in the loop, producing an 8-form anomaly polynomial $I^8_{\text{1-loop}}$. In addition, since the gauge anomaly (coming from the fermions in the vector multiplet running in the loop) has to be canceled, one always needs to use the Green-Schwarz-Wess-Sagnotti mechanism to determine the Green-Schwarz term $I^8_{\text{GS}}$. Then the total anomaly only involves non-dynamical symmetries (R-symmetry,  flavour symmetries, and background curvature):
\be
    I_{8} = I_{\text{1-loop}} + I_{\text{GSWS}} =  \alpha c_2(R)^2 + \beta c_2(R) p_1(T) + \gamma p_1(T)^2 + \delta p_2(T).
\ee

Rich information is contained in the above anomaly polynomial. For example, the Weyl $a$-anomaly in 6d can be expressed as a linear combination of the above quantities \cite{Cordova:2015fha}:
\be
    a = \frac{16}{7}(\alpha - \beta + \gamma) + \frac{6}{7}\delta,
\ee
which is expected to monotonically decrease along the RG flows\footnote{In \cite{Heckman:2015axa}, a large family of candidate invariants that change monotonically under RG flows is identified. See also \cite{Elvang:2012st,Cordova:2015vwa,Heckman:2021nwg,Fazzi:2023ulb,Baume:2023onr} for more studies on $a$-theorem in 6d.}.

One such quantity that is particularly useful for us would be the dimension of the Higgs branch. At least for RG flows that do not change the endpoint configuration of the 6d SCFT, the change in the Higgs branch (quaternionic) dimension is directly proportional to the change in the $\delta$ coefficient \cite{Mekareeya:2016yal}: 
\be
    \Delta d_\bbH =  -1440 \Delta (\delta).
\ee

Technically, computing $\delta$ has the following advantage as compared to computing other quantities. The GSWS term takes the form of $\frac{1}{2} \Omega^{ij} I^4_i I^4_j$, so it could not possibly produce terms proportional to $p_2(T)$. Therefore, to determine $\delta$, it is sufficient to perform the 1-loop computation and skip the rest.

\subsection{Magnetic Quivers}\label{magnetic}
As mentioned in \S\ref{intro}, magnetic quivers\footnote{In this paper, as we shall focus on the infinite coupling phases (for all gauge couplings), the magnetic quivers are always unframed.} provide a powerful tool to study the Higgs branches of SCFTs in various dimensions, following the identity \eqref{H=C}. There are two useful methods to analyze the stratifications of the 3d $\mathcal{N}=4$ Coulomb branches/moduli spaces of dressed monopole operators. One is the quiver subtraction \cite{Cabrera:2018ann,Bourget:2019aer}, and the other is the quiver decay and fission \cite{Bourget:2023dkj,Bourget:2024mgn}. Starting from the same 6d theory $\mathcal{T}$, the resulting Hasse/phase diagrams should have the same structure using the two methods. The difference lies in the different parts of the Hasse diagram they describe. Suppose that there is a Higgsed theory $\mathcal{T}'$. The magnetic quiver $\mathtt{Q}_1$ obtained from the quiver subtraction gives the closure of the symplectic leaf transverse to $\mathcal{H}(\mathcal{T}')$ in $\mathcal{H}(\mathcal{T})$. On the other hand, the magnetic quiver $\mathtt{Q}_2$ from the quiver decay/fission concerns the Higgs branch $\mathcal{H}(\mathcal{T}')=\mathcal{C}(\mathtt{Q}_2)$. An illustration of this can be found in \cite[Figure 21]{Bourget:2024mgn}. As our algorithm in \S\ref{sec:algorithm} computes the Higgsed theories using the tensor branch descriptions, each of them should correspond to a magnetic quiver under the quiver decay and fission (if it admits a magnetic quiver description). Therefore, we shall mainly focus on the quiver decay and fission algorithm here.

Let us first recall some preliminaries of the 3d $\mathcal{N}=4$ quivers. The balance of a gauge node is given by
\begin{equation}
    b=\begin{cases}
        \#(\text{hypers})-k,&\U(k)\text{ node},\\
        \frac{1}{2}\#(\text{half-hypers})-k+1,&\text{(S)O}(k)\text{ node},\\
        \frac{1}{2}\#(\text{half-hypers})-2k-1,&\Sp(k)\text{ node}.
    \end{cases}
\end{equation}
We say a node is overbalanced (balanced, resp.~underbalanced) if $b>0$ ($b=0$, resp.~$b<0$), and we shall colour it black (white, resp.~grey).

The (minimal amount of the) global symmetry is encoded by the balanced nodes in the quiver (although there could be further enhancements in some cases). In the UV, the Coulomb branch has the topological symmetry $\U(1)^r$ where the rank $r$ is equal to the number of $\U(1)$ factors in the gauge group. The global symmetry gets enhanced in the IR such that the UV symmetry is the maximal torus of the IR symmetry. One can read off the global symmetry as follows. For a unitary quiver, the balanced nodes form the Dynkin diagram of the non-abelian part of the IR global symmetry, multiplied by the abelian part $\U(1)^{\#(\text{unbalanced nodes})-1}$. The dimension of the Coulomb branch is the sum of the ranks of all the gauge nodes minus one. For an orthosymplectic quiver, a chain of $p$ balanced nodes gives an enhancement of $\SO(p+1+m)$ symmetry if there is an (S)O(2) gauge node at neither (one, resp.~each) end of the chain for $m=0$ ($m=1$, resp.~$m=2$). The dimension of the Coulomb branch is the sum of the ranks of all the gauge nodes.

In this paper, the main strategy for obtaining the magnetic quiver is to construct the Type IIA brane system and then move to the magnetic phase. See Appendix \ref{TypeIIAbranes}. One may think of the quiver decay and fission algorithm as moving some D6- and/or NS5-branes to infinity such that the remaining branes give a ``smaller'' child quiver of the parent quiver. For unitary quivers, the quiver decay and fission process has been worked out in \cite{Bourget:2023dkj,Bourget:2024mgn}. For orthosymplectic quivers, the full story is yet to be completed due to various complications such as determining when the theory becomes bad, the existence of non-special nilpotent orbits etc\footnote{There are also some discussions on the cases with mixed unitary and orthosymplectic nodes in \cite{Lawrie:2023uiu}.}. The precise statement of the quiver decay and fission algorithm is reviewed in Appendix \ref{quiverdecayfission}.

\section{Minimal Higgsings}\label{sec:algorithm}
In this section, we shall give the general algorithm that determines the full Higgs branch Hasse diagram\footnote{For the elementary slices, we shall discuss them in \S\ref{sec:elementaryslices}.} of a 6d SCFT. The idea is simply as follows:
\begin{itemize}
    \item Take \textit{any} 6d SCFT (with an F-theory tensor branch description) as the UV theory and perform all the possible minimal Higgsings. By definition, a minimal Higgsing is a Higgsing that cannot be further decomposed into multiple steps of Higgsings.
    \item For each possible IR theory of any such minimal Higgsing, iterate the above step until we reach the trivial theory.
\end{itemize}
In practice, such a seemingly straightforward algorithm would hardly simplify the question at all unless we give an explicit procedure of determining all the possible minimal Higgsings for any given 6d SCFT.

Therefore, giving such a procedure would be the main task in this section. Among these Higgsings, we will first cover the minimal nilpotent VEV Higgsings associated with the flavour symmetries in \S\ref{subsec:nilporb}. We shall then discuss the minimal plateau Higgsings in \S\ref{subsec:plateau}, which would happen on a chain of curves with identical fibre decorations. They can be described by the VEVs of semi-simple elements in the flavour symmetries as opposed to the nilpotent elements. In \S\ref{endptchanging}, we shall introduce the endpoint-changing Higgsings, which, as the name suggests, would change the curve configurations. As the last ingredient, we will identify an infinite number of new (minimal) RG flows which we call combo Higgsings in \S\ref{subsec:combo}. They are combinations of several \textit{individually forbidden} steps of \textit{naive Higgsings}.

It will be useful to combine our notion of the minimal Higgsings with the well-known family of the nilpotent Higgsings \cite{Heckman:2016ssk} by turning on a VEV given by the minimal nilpotent orbit inside the flavour symmetry. These Hasse diagrams are sometimes called the nilpotent hierarchies. We remark that only the \textit{minimal} nilpotent VEVs are guaranteed to be the minimal Higgsings, whereas any larger VEVs necessarily correspond to multiple/reducible Higgsings. Moreover, as suggested by the above outline of this section, the nilpotent Higgsings would generically only cover part of the complete Higgs branch Hasse diagrams.

\subsection{Minimal Nilpotent Orbit Higgsings}\label{subsec:nilporb}
For any non-Abelian flavour symmetry in a 6d SCFT, we can always turn on its minimal nilpotent orbit. The dimension of the minimal nilpotent orbits of all semi-simple Lie algebras can be found in \cite{MR1251060}\footnote{In \cite{MR1251060}, all the dimensions are complex dimensions while we always use the quaternionic dimensions here.}. They are always equal to the dual Coxeter numbers of the corresponding Lie algebras minus 1:
\begin{equation}
    \mathrm{dim}_{\mathbb{H}}(\calO_{\min}(\mathfrak{g})) = h^\vee_G - 1.
\end{equation}
In the F-theory tensor branch description, the minimal nilpotent orbit Higgsings split into two distinct possibilities, depending on whether the curve with the attached flavour symmetry admits any fibre decoration or not:
\begin{itemize}
\item{\textbf{Breaking of Gauge Algebras:}} When the flavour symmetry is carried by an ordinary hypermultiplet that is charged under a gauge symmetry, the minimal nilpotent orbit of this flavour symmetry also amounts to Higgsing the gauge symmetry.
\item{\textbf{Blow-down of $-1$ Curves}:} When the flavour symmetry is carried by a $-1$ curve with no paired gauge algebra, the minimal nilpotent orbit of this flavour symmetry always amounts to blowing down this $-1$ curve. In this case, and the minimal nilpotent orbit gives a Higgsing at the origin of the tensor branch. One could also think of the Higgsing as triggered by a VEV of the $-1$ curve as a conformal matter.
\end{itemize}

\paragraph{Minimality of such flows} Such flows are mostly minimal. Nevertheless, for the cases where the flavour symmetry is attached to an E-string that is connected to more than one other curve, the flow can be non-minimal.

\subsection{Minimal Plateau Higgsings}\label{subsec:plateau}
This type of minimal RG flow occurs when there is a chain of $-n$ curves all with non-trivial fibre decorations. Specifically, there are the following cases when this can happen:
\begin{itemize}
\item A chain of $-n$ curves can be Higgsed to a chain of $-(n-1)$ curves. For instance, we have $444\dots44\rightarrow333\dots33$, triggered by $\U(1) \subset \text{S}[\U(4) \times \U(4) \times \U(1)]$.
\item For a chain of $-2$ curves with fibre decoration, it can be Higgsed from $I_1 - \ksu(2) - \ksu(3) \dots \ksu(k+1) - \cdots - \ksu(k+1) \dots \ksu(3) - \ksu(2) - I_1$ to $I_1 - \ksu(2) - \ksu(3) \dots \ksu(k) - \cdots - \ksu(k) \dots \ksu(3) - \ksu(2) - I_1$, where the new $\ksu(k)$ fibre comes form either an originally $\ksu(k)$ or an $\ksu(k+1)$ fibre\footnote{Often, we would omit the $I_1$ fibres when writing the tensor branch descriptions.}. This can be understood as triggered by the VEV of a delocalized $\U(1)$ flavour symmetry.
\item On a chain of $141414\dots14$ curve, we have the Higgsing from alternating $\ksp(k+1) - \kso(2k+10) - \ksp(k+1) - \kso(2k+10) - \cdots$ to $\ksp(k) - \kso(2k+8) - \ksp(k) - \kso(2k+8) - \cdots$. This is triggered by a VEV of $\U(1) \unlhd \text{O}(2) \subset \text{O}(10)$.
\end{itemize}

One could either have a chain of $-2$ curves, each with an $\ksu(k)$ gauge symmetry, or one could have a chain of alternating $-4$ and $-1$ curves, with an $\kso(2k+8)$ flavour symmetry on each $-4$ curve and an $\ksp(k)$ flavour symmetry on each $-1$ curve. In fact, such a flow is also called a semi-simple flow in \cite{Heckman:2018pqx} as it is triggered by a VEV in the semi-simple part of the flavour symmetry.

Despite the fact that they are indeed triggered by a semi-simple part of the flavour symmetry, we choose to call them the minimal plateau Higgsings. This emphasizes the consequence that such a Higgsing reduces the height of the ``plateau'' by a minimal possible unit.

At this point, it is natural to wonder why we do not have any exceptional analogue of the plateau Higgsings. The answer is that they indeed exist, but in a different form, as (special cases of) the combo Higgsings. This will be discussed in \S\ref{subsec:combo}.

\subsection{Endpoint-Changing Higgsings}\label{endptchanging}
In this part, we discuss a type of the Higgs branch RG flows that would change the Dirac pairings (i.e., the curve configurations in the F-theory description). Such flows already exist in $(2,0)$ theories, whose stringy origins depend on various string constructions. In fact, they also exist in $(1,0)$ theories, which is a more difficult subject that evades any systematic study.

Here, we shall explain our approach to understanding these flows. At the current stage, our algorithm requires a brute-force search of all the possible descendant theories. This enumerative approach crucially relies on the monotonicity of certain quantities along RG flows. More concretely, for a given theory, one can in principle list all candidate IR theories (possibly reducible) with smaller endpoint configurations than the theory in question. Then the IR theories should have smaller Higgs branch dimensions $d_{\mathbb{H}}$. Moreover, we assume the ``$a$-theorem'' in 6d \cite{Mekareeya:2016yal,Fazzi:2023ulb} which states that the $a$ central charge
\begin{equation}
    a=\frac{16}{7}(\alpha-\beta+\gamma)+\frac{6}{7}\delta
\end{equation}
should decrease under the RG flows.

With these descendant theories, one can then find all the possible Higgsings based on the compatibility of the 6d tensor branch description. For such flows, we shall only treat the $(2,0)$ cases and some $(1,0)$ theories. A more systematic treatment of the conformal matter theories (i.e., those Higgsable to A-type $(2,0)$ theories) is be presented in our follow-up work \cite{Bao:2025pxe}\footnote{For theories whose endpoint configurations consist of not only $-2$ curves but also curves of different self-intersections, such endpoint-changing flows are still poorly understood. As a first challenge, it is unclear how one should attempt to construct a systematic approach to compute the Higgs branch dimensions of such theories.}.

For some families of theories, we can give a more algorithmic manipulation on the curve configurations with our current knowledge at hand:
\begin{itemize}
    \item For theories Higgsable to non-trivial $(2,0)$ ones, such flows can always be understood in relation to the tensor-changing flows in the $(2,0)$ theory. They are triggered by turning on the VEVs of the scalars in the $(2,0)$ tensor multiplets which belong to the hypers after decompositions into the $(1,0)$ language. To realize this, one deletes a node in the $(2,0)$ theory and takes the resulting curve configuration, either reducible or non-reducible, to be the $(2,0)$ descendant of the IR theory. For these flows among such theories, we need some suitable uplifts by adding gauge symmetries/conformal matter theories, which can be at least studied enumeratively for a given 6d SCFT.
    \item For theories that are Higgsable to the trivial theory, one can always delete a $-1$ curve and turn it into a theory Higgsable to a non-trivial $(2,0)$ theory (unless it is a rank-1 E-string theory). Then we perform Step 1 and attach back the affine $-1$ curve for each reducible component.
    \item For a theory $\calT$ not Higgsable to $(2,0)$, we ``affinize'' the theory by attaching $k$ $(-1)$ curves so that it becomes a theory $\calT_k$ Higgsable to a $(2,0)$ one. Then we shall still perform the flows of the theories with $-1$ endpoints and remove all the added $k$ $(-1)$ curves to detect possible flows for the original theory $\calT$ in question.
\end{itemize}

\subsubsection{String Engineering without T-brane VEVs}\label{generalstory}
The 6d $\mathcal{N}=(2,0)$ theories have $(2,0)$ tensor multiplets, and each $(2,0)$ tensor multiplet contains a $(1,0)$ tensor multiplet and a $(1,0)$ hypermultiplet. It is the scalar in this hypermultiplet that can acquire a VEV, so all $(2,0)$ theories would have a non-trivial Higgs branch, whose dimension is given by the dimension of its tensor multiplets. One version of such a minimal RG flow could cause fission of the tensor branch curve configuration, which would turn one irreducible (component of a) 6d SCFT into a disjoint union of two or three irreducible components. In the case of $\mathcal{N}=(2,0)$ theories, explicit examples of such flows can be found, for example, in \cite{Lawrie:2024zon}.

In such context of the $(2,0)$ theories, such flows can be detected in the string constructions. In the setup of the Type IIB string theory on a Kleinian singularity $\bbC^2/\Gamma_{\SU(2)}$, such flows manifest themselves as complex structure deformations of the singularity, resulting in a collection of ``smaller'' singularities. This was mathematically worked out in \cite{katz1992gorenstein}. For instance, we have
\begin{equation}
    xy = z^{n_1 + n_2} \ \ \longrightarrow \ \ xy = (z - z_1)^{n_1}(z - z_2)^{n_2}
\end{equation}
so that an $A_{n_1 + n_2 - 1}$-type $(2,0)$ theory flows to the disjoint union of an $A_{n_1-1}$-type $(2,0)$ theory and an $A_{n_2 - 1}$-type $(2,0)$ theory. In general, the geometric analysis tells us that we should delete a single node in the \textit{non-affine} ADE Dynkin diagram, and the remaining collection of Dynkin diagrams describes the collection of the IR theories.

From the dual Type IIA string theory perspective, we can also understand the same A-type theories by a mere separation of the stack of the $(n_1+n_2)$ NS5-branes into two stacks, each with $n_1$ and $n_2$ branes. The analysis can be generalized to the D-types (but not the E-types), for which one needs to introduce an ON$^-$-plane that remains in one of the smaller stacks.

The general story of such flows for $\mathcal{N}=(1,0)$ theories has not been very well understood in the literature, with the difficulty coming from a precise understanding of how the fibre enhancements would change along such flows. For $(1,0)$ theories with $\ksu$-type gauge enhancements that are Higgsable to A-type $(2,0)$ theories, the complete Hasse diagrams can be algorithmically obtained by unitary magnetic quivers \cite{Bourget:2023dkj,Bourget:2024mgn}.

The simplest family of such $(1,0)$ RG flows can be understood by a generalization of the above construction. Take the Type IIB construction of $(k_1+k_2)$ NS5-branes probing $\bbC^2/\bbZ_{n_1 + n_2}$. We can perform a complex structure deformation of the geometric singularity into a $\bbC^2/\bbZ_{n_1}$ singularity and a $\bbC^2/\bbZ_{n_2}$ singularity. Then one generically has the choice to split the NS5-branes into a stack of $k_1$ NS5-branes probing one singularity and another stack of $k_2$ NS5-branes probing the other singularity. This would always give an endpoint-changing flow. In the special case of $k_1 = 0$ (or $k_2 = 0$), one of the irreducible components of the resulting theory would be a $(2,0)$ theory. The Type IIA description of such flows is completely analogous, with the roles of the NS5-branes and the Kleinian singularities exchanged.

To clarify the difference between such endpoint-changing Higgs branch flows and the tensor branch flows that give infinitely large VEVs to the scalars in the $(1,0)$ tensor multiplets, it is particularly helpful to examine a family of M-theory constructions. Consider $N$ M5-branes probing $\bbR \times \bbC^2 / \Gamma_{\SU(2)}$. If we separate the M5-branes along the $\bbR$ direction, then we have the tensor branch flow which is not our focus here. Therefore, the desirable Higgs branch flow corresponds to moving the relative position of the M5-branes \textit{perpendicular to} $\bbR$, namely within $\bbC^2/\Gamma_{\SU(2)}$.

Contrary to what it seems, such endpoint-changing flows may not be minimal flows. We conjecture that a minimal endpoint-changing flow would always change the Higgs branch dimension by 1, but the above flows generically have dimension changes greater than 1. In practice, such a flow can often be decomposed into a (sequence of) minimal T-brane flow(s) and a minimal endpoint-changing flow.

\subsubsection{String Engineering with T-brane VEVs}\label{indorb}
When a theory is associated with a pair of nilpotent orbits, there could be a more systematic way to obtain its descendant theories under the endpoint-changing flows. In such cases, the generalized quivers can be split into multiple pieces under the Higgsings.

There are two situations. First, for a plateau of curves of rank $r_1$, such as the A-type theories with $222\dots2$ and the D-type conformal matter theories with $1414\dots141$, let us denote the theory as $\mathcal{T}_{r_1}$. Then it can have the following minimal endpoint-changing flow:
\begin{equation}
    \mathcal{T}_{r_1}\rightarrow\mathcal{T}_{r_2} \ \sqcup \ \underbrace{2 \ 2 \ \dots \ 2}_{r_1-r_2-1}\ ,\label{fission1}
\end{equation}
with a plateau of rank $r_2$ and a chain of $-2$ curves of length $r_1-r_2-1$. When there is further a trivalent- or quadrivalent-leg configuration on the plateau, besides the ones in \eqref{fission1}, we can also get a collection of generalized quivers simply by covering the central curve where the legs intersect. For instance, we have
\begin{align}
    &2 \nonumber\\
    2 \ & 2 \ \underbrace{2 \ \dots \ 2}_{r} \ \rightarrow \ 2 \ \sqcup \ 2 \ \sqcup \ \underbrace{2 \ \dots \ 2}_{r}.
\end{align}
Likewise,
\begin{equation}
    \overset{\kso(8)}{4} \ (1 \ [\SO(8)])^{\otimes4} \ \rightarrow \ \{1 \ [E_8]\}^{\otimes4},
\end{equation}
where we have compensated each $-1$ curve for an extra $\SO(8)$ flavour symmetry, and this further causes an enhancement to $E_8$.

There is another situation with more non-trivial splits for the long quivers. Given a theory associated with a pair of nilpotent orbits in $\mathfrak{g}$, it could be possible that both orbits can be written as some induced orbits from $\mathfrak{l}$, where $\mathfrak{l}$ is a Levi subalgebra of $\mathfrak{g}$. Suppose that the two orbits $\mathcal{O}$ and $\mathcal{O}'$ in $\mathfrak{g}$ are induced by the orbits $\mathcal{O}_{\mathfrak{l}}$ and $\mathcal{O}'_{\mathfrak{l}}$ respectively:
\begin{equation}
    \mathcal{O}=\text{Ind}^{\mathfrak{g}}_{\mathfrak{p}}(\mathcal{O}_l),\quad\mathcal{O}'=\text{Ind}^{\mathfrak{g}}_{\mathfrak{p}}(\mathcal{O}'_l).
\end{equation}
Since the orbit induced from an orbit in $\mathfrak{l}$ only depends on the Levi subalgebra $\mathfrak{l}$, but not on the choice of the parabolic subalgebra $\mathfrak{p}$ containing it, we shall use the notations $\text{Ind}^{\mathfrak{g}}_{\mathfrak{p}}$, $\text{Ind}^{\mathfrak{g}}_{\mathfrak{l}}$ and $\text{Ind}^{\mathfrak{g}}$ interchangeably. If the orbits $\mathcal{O}_{\mathfrak{l}}$ and $\mathcal{O}'_{\mathfrak{l}}$ have the decompositions
\begin{equation}
    \mathcal{O}_{\mathfrak{l}}=\mathcal{O}_1\oplus\dots\oplus\mathcal{O}_r,\quad\mathcal{O}'_{\mathfrak{l}}=\mathcal{O}'_1\oplus\dots\oplus\mathcal{O}'_r,
\end{equation}
then there is an endpoint-changing flow sending the parent theory associated to $\mathcal{O}_{\mathfrak{g}}\text{-}\mathcal{O}'_{\mathfrak{g}}$ to a collection of generalized quivers. Each would be associated to a pair $\mathcal{O}_i\text{-}\mathcal{O}'_j$. Every possible combination of the pairs from the decompositions of $\mathcal{O}_{\mathfrak{l}}$ and $\mathcal{O}'_{\mathfrak{l}}$ would correspond to such a flow.

The above discussions are expected to work for any Lie algebras. Here, let us illustrate this for the classical algebras whose orbits have corresponding partitions. Given an A-type theory $\mathcal{T}\left(\mathcal{O}_{\mathfrak{sl}_n},\mathcal{O}'_{\mathfrak{sl}_n}\right)$, it can flow to $\bigotimes\mathcal{T}\left(\mathcal{O}_i,\mathcal{O}'_j\right)$, where $\mathcal{O}_i$ would only appear once (and likewise for $\mathcal{O}'_j$). These $\mathcal{O}_i$ (as well as $\mathcal{O}'_j$) are given as follows \cite{kempken1983induced,MR1251060}. If the orbit $\mathcal{O}_{\mathfrak{sl}_n}$ can be written as $\mathcal{O}_{\sum\bm{p}}$, where $\sum\bm{p}$ denotes the partition with the $k^\text{th}$ part being a sum $p^1_k+p^2_k+\dots+p^r_k$, then we have
\begin{equation}
    \mathcal{O}_{\sum\bm{p}}=\text{Ind}^{\mathfrak{sl}_n}(\mathcal{O}),\quad\mathcal{O}=\mathcal{O}_1\oplus\dots\oplus\mathcal{O}_r,
\end{equation}
where each $\mathcal{O}_i$ is a nilpotent orbit in $\mathfrak{sl}_{d_i}$ with the partition $\bm{p}(i)=\left[p^i_1,\dots,p^i_n\right]$. There would be at most $r!$ distinct such flows. However, we conjecture that only when $r=2$, the flow would be minimal. In other words, a minimal flow should decompose the curve configuration into two pieces.

Therefore, given a long quiver specified by a plateau of $-k$ curves decorated by $\ksu(n)$ gauge symmetries and a pair of T-brane deformations corresponding to two partitions $\bm{p}$, $\bm{p}'$ of $n$, a minimal flow would lead to two pieces specified by the same data. The curves have the self-intersections satisfying $k_1+k_2+1=k$ with the gauge algebras satisfying $n_1+n_2=n$. The partitions should be determined by
\begin{equation}
    \bm{p}^\text{T}=\bm{p}_a^\text{T}\sqcup\bm{p}_b^\text{T},\quad (\bm{p}')^\text{T}=(\bm{p}'_c)^\text{T}\sqcup(\bm{p}'_d)^\text{T}.
\end{equation}
Here, the Latin subscripts take values in $\{1,2\}$ such that they can be concatenated into partitions. Then there are at most two distinct such minimal flows given by different combinations of $\bm{p}_{a,b,c,d}$. When a unitary magnetic quiver description is known, this corresponds to the quiver fission in \cite{Bourget:2023dkj,Bourget:2024mgn}.

For the BCD-type theories, given an orbit $\text{Ind}^{\mathfrak{g}}(\mathcal{O}_\mathfrak{l})$ with partition $\bm{p}$, we have the decomposition
\begin{equation}
    \mathcal{O}_{\mathfrak{l}}=\mathcal{O}_{\bm{d}}\oplus\mathcal{O}_{\bm{f}},\label{partitionsplitting}
\end{equation}
where $\bm{d}$ is a partition specifying an orbit in $\mathfrak{sl}_l$ and $\bm{f}$ is a partition specifying an orbit in $\mathfrak{g}'$ (determined by the Levi subalgebra $\mathfrak{l}$ and is of the same type as $\mathfrak{g}$). In particular, $\text{rank}(\mathfrak{g})=\text{rank}(\mathfrak{g}')+2l$. Define a new partition $\bm{q}$ such that $q_i=f_i+2d_i$. Then we have $\bm{p}$ being the B-, C- or D-collapse of $\bm{q}$. Such flows should always be minimal.

If there is an orthosymplectic magnetic quiver description, the quiver fission is not fully known in the literature. Nevertheless, the process of finding the descendant theory is similar to \eqref{partitionsplitting}, but with the Lusztig-Spaltenstein (LS) dual (recall that the LS dual in the A-type case is the same as the transpose):
\begin{equation}
    \mathtt{d}(\bm{p})=\mathtt{d}(\bm{p}_a)\sqcup\mathtt{d}(\bm{p}_b).
\end{equation}
Now, only $\bm{p}_a$ for $\mathfrak{g}'$ would be obvious in this decomposition while $\bm{p}_b$ is not an orbit in $\mathfrak{sl}_l$. Nevertheless, $\bm{p}_a$ would determine the orthosymplectic quiver in the descendant theory. Then the other piece, which is a unitary magnetic quiver, would simply be determined by the difference between the parent quiver and this descendant orthosymplectic quiver.

\subsection{Combo Higgsings}\label{subsec:combo}
We now introduce a new type of Higgsings, which would change both the curve types and the gauge algebras while preserving the endpoint configurations. We shall call such Higgsings the combo Higgsings. As will be illustrated below, we encourage the readers to think of the combo Higgsings as some sort of generalization of the plateau Higgsings for $(G, G)$-type conformal matters with $G \in \{D, E_{6, 7, 8}\}$.

Such Higgsings can happen in a connected segment of the generalized quiver when neither a T-brane Higgsing nor a plateau Higgsing is possible. In fact, their existence as a type of the minimal Higgsings is guaranteed by the principle that all conformal matter theories of positive ranks can be Higgsed via a chain of minimal Higgsings down to $(2,0)$ theories.

Concretely, we find an infinite number of minimal combo Higgsings, both in infinite families and in sporadic cases. They involve simultaneous reductions of gauge algebras and blow-downs of $-1$ curves. In these cases, each of these constituent operations \emph{cannot} be performed individually, but the combination of these operations is allowed as a \emph{single} Higgsing step.

Let us state the basic procedure of performing the combo Higgsings:
\begin{enumerate}
    \item To tell whether a combo flow is possible, we first perform a \textit{forced} blow-down of the $-1$ curve. Here, ``forced'' stands for the fact that the would-be IR theory is inconsistent because the adjacent curves have incompatible gauge symmetries. For instance, when only two tensors exist, the compatibility condition requires that $\kh_1 \not \subset \kf_2$ or $\kh_2 \not \subset \kf_1$ when the gauge algebras $\kf_i$ become $\kh_i$ after the Higgsing.
    \item To proceed, we need to check if we can perform a Higgsing within the gauge algebra paired with the tensor multiplet to turn the inconsistent theory into a consistent one.
    \begin{itemize}
        \item If yes, then we have obtained a combo flow. We can determine the change in the quaternionic dimension of this combo flow by comparing the UV theory with the IR theory. Such examples can be found in \eqref{eq_D4_combo_UV}$\sim$\eqref{eq_D4_combo_IR}.
        \item If no consistent IR theory can be reached by a mere gauge algebra Higgsing, but there are still possibilities to further perform Step (1), then we should keep performing Step (1) and Step (2) in an iterative manner. If we reach a valid IR theory, then we find a combo flow. As an example, this would happen for the infinite family of combo flows from the $(E_8, E_8)$ conformal matter down to the $(E_7, E_7)$ conformal matter (with decorated tails), as in \eqref{eq_E8_combo_UV}$\sim$\eqref{eq_E8_combo_IR}.
        \item If no consistent IR theory can be reached, nor are there any $-1$ curves that can be blown down (even in the ``forced'' sense), then this theory does not admit any combo flow, and should be viewed as the IR theory in the branch. We shall give such an example and remark on its generalization in \S\ref{forbiddencomboflow}.
    \end{itemize}
\end{enumerate}

\paragraph{Validity of such flows} We need to check whether such a flow is allowed. In other words, such Higgsings should satisfy the basic criteria for RG flows. Therefore, if the dimension of the Higgs branch increases or the $a$ central charge increases, this process would not be allowed.

Interestingly, in the Hasse diagram of the $E_7$ nilpotent orbits, we have $D_5\leq D_6(a_1)$ for these two neighbouring nodes. Therefore, it would be natural to expect a flow from $D_5$ to $D_6(a_1)$ in the $E_7$ nilpotent hierarchy:
\begin{equation}
    [\SU(2)] \ \ \overset{\kso(7)}{3} \ \ \overset{\ksu(2)}{2} \ \ 1 \ \ \underset{\underset{[\SU(2)]}{1}}{\overset{\ke_7}{8}} \ \ 1 \ \ \dots \ \ \rightarrow \ \ \overset{\kg_2}{3} \ \ \overset{\ksu(2)}{2} \ \ 2 \ \ \underset{[\SU(2)]}{1} \ \ \overset{\ke_7}{8} \ \ 1 \ \ \dots.
\end{equation}
However, this cannot be obtained from the combo Higgsings (or any other minimal Higgsings in our algorithm). In fact, one may check that $a$ increases from $D_5$ to $D_6(a_1)$. Following our algorithm and the current criterion for 6d RG flows, this seems to indicate that there should not be such a Higgsing although there is an ordering between $D_5$ and $D_6(a_1)$.

\paragraph{Minimality of such flows} We also need to check if such a flow is minimal. First of all, we should see if it can be written as a composite flow involving any known minimal nilpotent or plateau Higgsings. When this happens, the combo flow in question is not a minimal one.

Even if the above does not happen, there is a more subtle possibility that there is a ``smaller'' combo flow that can be obtained by a re-blow-up of a $-1$ curve. In other words, the net effect of this combo flow can be viewed as a combination of both the Higgsing of the gauge algebra and the small instanton transition. In the absence of non-Abelian flavour symmetries, we conjecture that a combo flow is minimal if and only if the (quaternionic) dimension of the Higgs branch is changed by 1.

\subsubsection{Infinite Families}\label{inffam}
Some infinite families admit the minimal combo flows, and they have the following features. The part of the 6d theory on the generalized quiver that undergoes the combo flow always has the structure of a rank $N$, $(G, G)$ conformal matter in the middle, and fixed tails on both ends. We can enumerate them according to different choices of $G$.

\paragraph{The D-types} Let us illustrate this with the bottom theory in the nilpotent hierarchy of the $(D_4,D_4)$ conformal matter theory:
\begin{equation}
    2 \ \ \overset{\mathfrak{su}(2)}{2} \ \ \overset{\mathfrak{g}_2}{3} \ \ 1 \ \ \overset{\mathfrak{so}(8)}{4} \ \  \dots \ \ \overset{\mathfrak{so}(8)}{4} \ \ 1
  \ \ \overset{\mathfrak{g}_2}{3} \ \ \overset{\mathfrak{su}(2)}{2} \ \  2.
  \label{eq_D4_combo_UV}
\end{equation}
Following the first step above, we write down the intermediate configuration obtained by forced blow-downs of the $-1$ curves\footnote{Notice that each $-3$ curve becomes a $-2$ curve as it is adjacent to one $-1$ curve. For each $-4$ curve, it is adjacent to two $-1$ curves. They would hence become $-2$ curves after blowing down all the $-1$ curves in the configuration.}:
\begin{equation}
    2 \ \ {\color{red} \overset{\mathfrak{su}(2)}{2} \ \ \overset{\mathfrak{g}_2}{2} \ \ \overset{\mathfrak{so}(8)}{2} \ \  \dots \ \ \overset{\mathfrak{so}(8)}{2} 
  \ \ \overset{\mathfrak{g}_2}{2} \ \ \overset{\mathfrak{su}(2)}{2}} \ \ 2,
\end{equation}
where the red part emphasizes the incompatible pairs of adjacent gauge nodes. To fix them, we do not blow the $-1$ curves back up, but instead, we perform the Higgsings on the $\mathfrak{g}_2$ and $\mathfrak{so}(8)$ curves to get the following IR theory:
\begin{equation}
    2 \ \ \overset{\mathfrak{su}(2)}{2}  \ \ \overset{\mathfrak{su}(3)}{2}\ \ \overset{\mathfrak{su}(4)}{2} \ \  \dots \ \ \overset{\mathfrak{su}(4)}{2} \ \ \overset{\mathfrak{su}(3)}{2} \ \ \overset{\mathfrak{su}(2)}{2} \ \  2.
    \label{eq_D4_combo_IR}
\end{equation}

\paragraph{The $E_6$ type} The bottom theory in the nilpotent hierarchy of the $(E_6, E_6)$ conformal matter theory of rank $r \geq 8$ is
\be
    2 \ \ \overset{\mathfrak{su}(2)}{2} \ \ \overset{\mathfrak{g}_2}{3} \ \ 1 \ \ \underbrace{\overset{\mathfrak{f}_4}{5} \ \ 1 \ \ \overset{\mathfrak{su}(3)}{3} \ \ 1 \ \ \overset{\mathfrak{e}_6}{6} \ \  \dots \ \  \overset{\mathfrak{e}_6}{6} \ \ 1 \ \ \overset{\mathfrak{su}(3)}{3} \ \ 1 \ \ \ \ \overset{\mathfrak{f}_4}{5}}_{\text{combo flow}} \ \ 1
  \ \ \overset{\mathfrak{g}_2}{3} \ \ \overset{\mathfrak{su}(2)}{2} \ \  2, \label{eqn:E6combo_UV}
\ee
and we have indicated where the combo flow takes place. A minimal combo flow takes it to
\be
    2 \ \ \overset{\mathfrak{su}(2)}{2} \ \ \overset{\mathfrak{g}_2}{3} \ \ 1 \ \ \overset{\mathfrak{so}(9)}{4} \ \ \overset{\mathfrak{sp}(1)}{1} \ \ \overset{\mathfrak{so}(10)}{4} \dots \ \ \overset{\mathfrak{so}(10)}{4} \ \ \overset{\mathfrak{sp}(1)}{1} \ \ \overset{\mathfrak{so}(9)}{4} \ \ 1
  \ \ \overset{\mathfrak{g}_2}{3} \ \ \overset{\mathfrak{su}(2)}{2} \ \  2
\ee
with $\Delta d_{\mathbb{H}}=1$. This flow is expected since for $\mathfrak{so}(9) \subset \mathfrak{f}_4$ and $\mathfrak{so}(10) \subset \mathfrak{e}_6$, the former algebras in both cases are maximal classical-type subalgebras of the latter.

Then another Higgsing brings us down to the minimal theory in the $(D_4, D_4)$ nilpotent hierarchy:
\be
    2 \ \ \overset{\mathfrak{su}(2)}{2} \ \ \overset{\mathfrak{g}_2}{3} \ \ 1 \ \ \overset{\mathfrak{so}(8)}{4} \ \ 1 \ \ \overset{\mathfrak{so}(8)}{4} \dots \ \ \overset{\mathfrak{so}(8)}{4} \ \ 1 \ \ \overset{\mathfrak{so}(8)}{4} \ \ 1
  \ \ \overset{\mathfrak{g}_2}{3} \ \ \overset{\mathfrak{su}(2)}{2} \ \  2
\ee
with $\Delta d_{\mathbb{H}}=1$.

\paragraph{The $E_7$ type} For the $(E_7, E_7)$ conformal matter theory with rank $r \geq 9$, the bottom theory in the nilpotent hierarchy is
\be
2 \ \ \overset{\mathfrak{su}(2)}{2} \ \ \overset{\mathfrak{g}_2}{3} \ \ 1 \ \ \overset{\mathfrak{f}_4}{5} \ \ 1 \ \
\underbrace{\overset{\mathfrak{g}_2}{3} \ \ \overset{\mathfrak{su}(2)}{2} \ \ 1 \ \ \overset{\mathfrak{e}_7}{8} \ \ \dots \ \ \overset{\mathfrak{e}_7}{8} \ \ 1 \ \ \overset{\mathfrak{su}(2)}{2} \ \ \overset{\mathfrak{g}_2}{3}}_{\text{combo flow}} \ \
 1 \ \ \ \ \overset{\mathfrak{f}_4}{5} \ \ 1
  \ \ \overset{\mathfrak{g}_2}{3} \ \ \overset{\mathfrak{su}(2)}{2} \ \  2.
\ee
The combo flow brings us to
\be
    2 \ \ \overset{\mathfrak{su}(2)}{2} \ \ \overset{\mathfrak{g}_2}{3} \ \ 1 \ \ \overset{\mathfrak{f}_4}{5} \ \ 1 \ \ \overset{\mathfrak{su}(3)}{3} \ \ 1 \ \ \overset{\mathfrak{e}_6}{6} \ \  \dots \ \  \overset{\mathfrak{e}_6}{6} \ \ 1 \ \ \overset{\mathfrak{su}(3)}{3} \ \ 1 \ \ \ \ \overset{\mathfrak{f}_4}{5} \ \ 1
  \ \ \overset{\mathfrak{g}_2}{3} \ \ \overset{\mathfrak{su}(2)}{2} \ \  2 \quad (\Delta d_{\bbH} = 1),
\ee
which then goes over the above combo flows for the $E_6$ and lower theories.

\paragraph{The $E_8$ type} For $E_8$, a combo flow can get more complicated, in that it combines multiple iterations of Higgsings and blow-downs of $-1$ curves. Thus it would pass through multiple invalid intermediate configurations before finally landing on the IR theory.

The shortest case in the infinite family is
\be
    2 \ \overset{\ksu(2)}{2} \ \overset{\kg_2}{3} \ 1 \ \overset{\kf_4}{5} \ 1 \ \overset{\kg_2}{3} \ \overset{\ksu(2)}{2} \ 2 \ 1 \ \underset{[N_e = 1]}{\overset{\ke_8}{11}} \ 1 \ 2 \ \overset{\ksu(2)}{2} \ \overset{\kg_2}{3} \ 1 \ \overset{\kf_4}{5} \ 1 \ \overset{\kg_2}{3} \ \overset{\ksu(2)}{2} \ 2 \ 1 \ \underset{[N_e = 1]}{\overset{\ke_8}{11}} \ 1 \ 2 \ \overset{\ksu(2)}{2} \ \overset{\kg_2}{3} \ 1 \ \overset{\kf_4}{5} \ 1 \ \overset{\kg_2}{3} \ \overset{\ksu(2)}{2} \ 2.
    \label{eq_E8_combo_UV}
\ee
A more general case would insert $k$ $(-12)$ curves with $\ke_8$ curves connected by $k$ segments of the $(E_8, E_8)$ conformal matter. Here, due to the limitation of the space, we are only explicitly writing down the case for $k = 0$.

We first perform a blow-down of all curves next to the $-11$ curves, leading to the forbidden configuration:
\be
   2 \ \overset{\ksu(2)}{2} \ \overset{\kg_2}{3} \ 1 \ \overset{\kf_4}{5} \ 1 \ \overset{\kg_2}{3} \ {\color{red} \overset{\ksu(2)}{2} \ 1 \ \underset{[N_e = 3]}{\overset{\ke_8}{9}} \ 1 \ \overset{\ksu(2)}{2}} \ \overset{\kg_2}{3} \ 1 \ \overset{\kf_4}{5} \ 1 \ \overset{\kg_2}{3} \ {\color{red} \overset{\ksu(2)}{2} \ 1 \ \underset{[N_e = 3]}{\overset{\ke_8}{9}} \ 1 \ \overset{\ksu(2)}{2}} \ \overset{\kg_2}{3} \ 1 \ \overset{\kf_4}{5} \ 1 \ \overset{\kg_2}{3} \ \overset{\ksu(2)}{2} \ 2.
\ee
One could then blow down all the $-1$ curves between the pair of $-9$ curves and simultaneously Higgs the $\ke_8$ down to $\ke_7$, $\kf_4$ down to $\kso (7)$, $\kg_2$ down to $\ksu(2)$, and $\ksu (2)$ (over $-2$ before the Higgsing) to empty. After all these steps, one eventually reaches an allowed 6d SCFT:
\be
    2 \ \overset{\ksu(2)}{2} \ \overset{\kg_2}{3} \ 1 \ \overset{\kf_4}{5} \ 1 \ \overset{\kg_2}{3} \ \overset{\ksu(2)}{2} \ 1 \ \overset{\ke_7}{8} \  1 \ \overset{\ksu(2)}{2} \overset{\kso(7)}{3} \ \overset{\ksu(2)}{2} \  1 \ \overset{\ke_7}{8} \ 1 \ \overset{\ksu(2)}{2} \ \overset{\kg_2}{3} \ 1 \ \overset{\kf_4}{5} \ 1 \ \overset{\kg_2}{3} \ \overset{\ksu(2)}{2} \ 2 \quad (\Delta d_{\bbH} = 1).
    \label{eq_E8_combo_IR}
\ee
It is interesting to notice that this multi-step procedure again only changes the quaternionic dimension of the 6d Higgs branch by 1.

This theory could eventually be Higgsed to a $(2,0)$ theory of A-type. To look for such a chain of minimal Higgsings, one would need to make use of the combo Higgsings of $D_4$-, $E_6$-, $E_7$-types that we have just discussed.

\subsubsection{Demonstration of a Forbidden Combo Flow}\label{forbiddencomboflow}
We start from the truncated D-type conformal matter as the UV theory:
\be
    \overset{\kso(8)}{4} \ \ 1 \ \ \overset{\kso(8)}{4}.\label{eqn:no_combo_theory}
\ee
Now, Step 1 tells us to blow-down the $-1$ curve, and we get
\be
    {\color{red} \overset{\kso(8)}{3} \ \ \overset{\kso(8)}{3}}.
\ee
We then perform Step 2. However, it cannot make this possible since a pair of $-3$ curves can never be adjacent to each other \cite{Heckman:2015bfa,Heckman:2018jxk}. More concretely, one can see this by noticing that non-trivial flavour symmetries on the $-3$ curves have to be of $\Sp$-type, into which the $\kso(n\geq7)$, $\kg_2$ and $\ksu(3)$ gauge symmetries do not embed. In addition, there are no $-1$ curves to blow down.

\subsubsection{Sporadic Cases}\label{sporadic}
Now, let us give a list of all the known combo flows that are \textit{sporadic} as in Table \ref{sporadiccases}. In other words, they do not extend to infinitely long quivers.
\begin{longtable}{|c|c|c|} \hline
$\calT_\text{UV}$ & $\calT_\text{IR}$ & $\Delta d_{\bbH}$ \\ \hline \hline
$ \overset{\kso(8)}{4} \ \ 1 \ \ \overset{\ksu(3)}{3}$ &  $[\Sp(1)]\ \  \overset{\kso(7)}{3} \ \ \overset{\ksu(2)}{2}$ & 1 \\ \hline
$\overset{\ksu(3)}{3} \ \ 1 \ \ \overset{\kf_4}{5} \ \ 1 \ \ \overset{\ksu(3)}{3}$ & $\overset{\ksu(2)}{2} \ \ \overset{\kso(7)}{3} \ \ \overset{\ksu(2)}{2}$ & 1 \\ \hline
$\overset{\kf_4}{5} \ \ 1 \ \ \overset{\ksu(3)}{3} \ \ 1 \ \ \overset{\kf_4}{5}$ & $\overset{\kso(9)}{4} \ \ \overset{\ksp(1)}{1} \ \ \overset{\kso(9)}{4}$ & 1 \\ \hline
$ \overset{\kso(8)}{4} \ \ 1 \ \ \overset{\ksu(3)}{3} \ \ 1 \ \ \overset{\ke_6}{6}$ & $\overset{\ksu(3)}{3} \ \ 1 \ \ \overset{\ke_6}{6} \ \ 1$ & 1 \\ \hline
$\overset{\kg_2}{3} \ \ \overset{\ksu(2)}{2} \ \ 1 \ \ \overset{\ke_6}{6}$ & $\overset{\ksu(3)}{3} \ \ 1 \ \ \overset{\ke_6}{5}$ & 1 \\ \hline
$\overset{\kg_2}{3} \ \ \overset{\ksu(2)}{2} \ \ 1 \ \ \overset{\ke_6}{6}$ & $\overset{\kg_2}{3} \ \ 1 \ \ \overset{\kf_4}{5}$ & 1 \\ \hline
$\overset{\kg_2}{3} \ \ \overset{\ksu(2)}{2} \ \ 1 \ \ \overset{\ke_7}{7}$ & $\overset{\ksu(3)}{3} \ \  \ \ 1 \ \ \overset{\ke_6}{6}$ & 1 \\ \hline
$\overset{\ksu(2)}{2} \ \ \overset{\kso(7)}{3} \ \ \overset{\ksu(2)}{2} \ \ 1 \ \ \overset{\ke_6}{6}$ & $\overset{\ksu(2)}{2} \ \ \overset{\kg_2}{3} \ \ 1 \ \ \overset{\kf_4}{5}$ & 1 \\ \hline
$([\SU(2)] \ \ 1)^{\otimes 3} \ \ \overset{\ke_7}{8} \ \ 1 \ \ \overset{\ksu(2)}{2}$ & $\overset{\kso(12)}{4} \ \ \overset{\ksp(1)}{1}$ & 1 \\ \hline
$\overset{\ke_6}{5} \ \ \underset{[\SU(2)]}{1} \ \ \overset{\ksu(2)}{2}$ & $\overset{\kso(10)}{4} \ \ \overset{\ksp(1)}{1}$ & 1  \\ \hline
$\overset{\kf_4}{5} \ \ \underset{[\SU(2)]}{1} \ \ \overset{\ksu(2)}{2}$ & $\overset{\kso(9)}{4} \ \ \overset{\ksp(1)}{1}$ & 1  \\ \hline
$\overset{\kf_4}{5} \ \ \underset{[\SU(2)]}{1} \ \ \overset{\ksu(2)}{2} \ \ \overset{\kso(7)}{3} \ \ \overset{\ksu(2)}{2}$ & $\overset{\kf_4}{4} \ \ 1 \ \ \ \ \overset{\kg_2}{3}  \ \ \overset{\ksu(2)}{2}$ & 1 \\ \hline
$\overset{\kf_4}{5} \ \ \underset{[\SU(2)]}{1} \ \ \overset{\ksu(2)}{2} \ \ \overset{\kso(7)}{3} \ \ \overset{\ksu(2)}{2}$ & $\overset{\kso(9)}{4} \ \ \underset{[\SO(3)]}{\overset{\ksp(1)}{1}} \ \ \ \ \overset{\kso(7)}{3}  \ \ \overset{\ksu(2)}{2}$ & 1 \\ \hline
$\overset{\kf_4}{5} \ \ 1 \ \ \overset{\ksu(2)}{2} \ \ \overset{\kso(7)}{3} \ \ \overset{\ksu(2)}{2}$ & $\overset{\kso(9)}{4} \ \ 1 \ \ \ \ \overset{\kso(7)}{3}  \ \ \overset{\ksu(2)}{2}$ & 1 \\ \hline
$\overset{\kso(8)}{4} \ \ 1 \ \ \overset{\kso(7)}{3} \ \ \overset{\ksu(2)}{2} \ \ 1 \ \ \overset{\ke_7}{8}$ & $\overset{\kso(7)}{3} \ \ \overset{\ksu(2)}{2} \ \ 1 \ \ \underset{1}{\overset{\ke_7}{8}}$ & 1 \\ \hline
$\overset{\kso(7)}{3} \ \ \overset{\ksu(2)}{2} \ \ 1 \ \ \overset{\ke_7}{7} \ \ 1 \ \ \overset{\ksu(2)}{2} \ \ \overset{\kso(7)}{3} \ \ \overset{\ksu(2)}{2}$ & $[\SU(2)] \ \ \overset{\kg_2}{3} \ \ 1 \ \ \overset{\kf_4}{5} \ \ 1 \ \ \overset{\kg_2}{3}$ & 1 \\ \hline
$\overset{\kg_2}{3} \ \ 1 \ \ \overset{\kf_4}{5} \ \ 1 \ \ \overset{\kg_2}{3} \ \ \overset{\ksu(2)}{2} \ \ 1 \ \ {\overset{\ke_7}{8}} \dots$ & $\overset{\ksu(2)}{2} \ \ \overset{\kso(7)}{3} \ \ \overset{\ksu(2)}{2} \ \ 1 \ \ \underset{1}{\overset{\ke_7}{8}} \ \ \dots$ & 1 \\ \hline
$\overset{\ksu(3)}{3} \ \ 1 \ \ \overset{\kf_4}{5} \ \ 1 \ \ \overset{\kg_2}{3} \ \ \overset{\ksu(2)}{2} \ \ 1 \ \ {\overset{\ke_7}{8}} \dots$ & $\overset{\ksu(2)}{2} \ \ \overset{\kso(7)}{3} \ \ \overset{\ksu(2)}{2} \ \ 1 \ \ {\overset{\ke_7}{7}} \ \ \dots$ & 1 \\ \hline
$\overset{\kso(7)}{3} \ \ 1 \ \ \overset{\kg_2}{3} \ \ \overset{\ksu(2)}{2}$ & $\overset{\ksu(4)}{2} \ \ \overset{\ksu(3)}{2} \ \ \overset{\ksu(2)}{2}$ & 1 \\ \hline
$\overset{\ksu(2)}{2} \ \ \overset{\kg_2}{3} \ \ 1 \ \ \overset{\kf_4}{5} \ \ 1 \ \ \overset{\ksu(3)}{3} \ \ 1 \ \ \overset{\ke_6}{6} \ \  ...$ & $2 \ \ \overset{\ksu(2)}{2} \ \ \overset{\kg_2}{3} \ \ 1 \ \ \overset{\kf_4}{5} \ \ ...$ & 1 \\ \hline
$\overset{\ksu(2)}{2} \ \ \overset{\kg_2}{3} \ \ 1 \ \ \overset{\kf_4}{5} \ \ 1 \ \ \overset{\kg_2}{3} \ \ \overset{\ksu(2)}{2} \ \ 1 \ \ \overset{\ke_7}{8} \ \ 1 \ \ \overset{\ksu(2)}{2} \ \ \overset{\kso(7)}{3} \ \ \overset{\ksu(2)}{2} \ \ 1$ & $2 \ \ \overset{\ksu(2)}{2} \ \ \overset{\kg_2}{3} \ \ 1 \ \ \overset{\kf_4}{5} \ \ 1 \ \ \overset{\kg_2}{3} \ \ \overset{\ksu(2)}{2} \ \  1 \ \ ...$ & 1 \\ \hline
$\mathcal{T}_{E_7}(\calO = E_6(a_1))$ & $\mathcal{T}_{E_7}(\calO = E_6)$ & 1 \\ \hline
$\mathcal{T}_{E_8}(\calO = E_6(a_1))$ & $\mathcal{T}_{E_8}(\calO = E_6)$ & 1 \\ \hline
$\mathcal{T}_{E_8}(\calO = E_7(a_1))$ & $\mathcal{T}_{E_8}(\calO = E_7)$ & 1 \\ \hline
$\mathcal{T}_{E_8}(\calO = E_8(a_1))$ & $\mathcal{T}_{E_8}(\calO = E_8)$ & 1 \\ \hline
\caption{The list of all known sporadic combo flows. It turns out that they are all incorporated in the $D_4, E_{6, 7, 8}$ nilpotent hierarchies. For the generalized quivers that are too long to fit in the table, we use the labels $\mathcal{T}_G(\mathcal{O})$. The explicit curve configurations can be found, for example, in \cite{Heckman:2016ssk,Chacaltana:2012zy,Mekareeya:2016yal}.}\label{sporadiccases}
\end{longtable}

For the flows collected in Table \ref{sporadiccases}, they can all be found in the $D_4, E_{6, 7, 8}$ nilpotent hierarchies in \cite{Heckman:2016ssk}. By the definition that they do not extend to infinitely long quivers, we know that such sporadicity is tied to the presence of the 322, 232, and 23 NHCs in the IR.

We comment that the nilpotent Higgsing from the sub-regular orbit to the regular orbit in each nilpotent hierarchy is usually given by a sophisticated combo Higgsing. If we take the Higgsing from the $E_6(a_1)$ orbit to the $E_6$ orbit in the $E_6$ nilpotent hierarchy, then we get the following combo Higgsing:
\be
\overset{\ksu(2)}{2} \ \ \overset{\kg_2}{3} \ \ 1 \ \ \overset{\kf_4}{5} \ \ 1 \ \ \overset{\ksu(3)}{3} \ \ 1 \ \ \overset{\ke_6}{6} \ \ 1 \ \ \dots \quad \rightarrow \quad 2 \ \ \overset{\ksu(2)}{2} \ \ \overset{\kg_2}{3} \ \ 1 \ \ \overset{\kf_4}{5} \ \ \dots.
\ee
The precise procedure is highly similar to the combo Higgsing of type $E_6$, except that one only performs it on all the curves up to the left of the first $-6$ curve (compared to \eqref{eqn:E6combo_UV}). For the Higgsing from $E_8(a_1)$ to $E_8$ which is left implicit in the table, the fact that we eventually lose one ``copy'' of the $(E_8, E_8)$ conformal matter is elaborated in \cite[Page 18]{Mekareeya:2016yal}. The specific procedure is again to perform the $E_8$ combo Higgsing on the part up to the left of the first $-12$ curve.

\section{Elementary Slices}\label{sec:elementaryslices}
When performing minimal Higgsings, besides the parent and child theories, there is another piece of information of great importance, that is, the types of flows. In terms of symplectic singularities, we need to determine the elementary slices transverse to the symplectic leaves in the larger leaves.

Given a minimal nilpotent orbit Higgsing, this can be directly extracted from the change of the flavour symmetry. The elementary slice is then the closure of the corresponding minimal nilpotent orbit. For a minimal plateau Higgsing, the flow is triggered by a semi-simple part of the flavour symmetry as aforementioned. In the endpoint-changing Higgsings, we find that all the slices are 1-dimensional, and they arise from quiver fissions on the magnetic quiver side. The combo flows also have 1-dimensional slices. Determining the slices for the minimal plateau and combo Higgsings is quite intricate and would require various knowledge from the generalized quivers and/or the magnetic quivers.

\subsection{Miminal Nilpotent Orbits in 6D Flavour Symmetries}\label{nilporbslices}
A well-known family of the elementary slices is the closures of minimal nilpotent orbits $\overline{\mathcal{O}}_{\min}$. This is often denoted as $g$ for the corresponding Lie algebra $\mathfrak{g}$. The slice is also the moduli space of one $\mathfrak{g}$ instanton. In Table \ref{slicesMQs}, we only list the corresponding magnetic quivers that we would encounter in this paper. The most up-to-date lists of all the magnetic quivers for the known elementary slices (not just restricted to $\overline{\mathcal{O}}_{\min}$) can be found for example in \cite{Bourget:2021siw,Bourget:2024mgn,Sperling:2021fcf}.
\begin{longtable}{|c|c|c|}
\hline
Slice & Magnetic quiver & $d_{\mathbb{H}}$ \\ \hline
$a_n$ & \includegraphics[width=2.5cm]{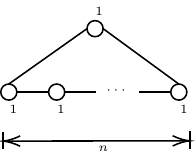} & $n$ \\
\hline
$b_n$ & \includegraphics[width=3.5cm]{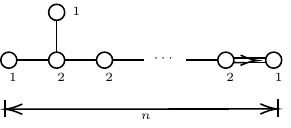} & $2n-2$ \\
\hline
$c_n$ & \includegraphics[width=3.5cm]{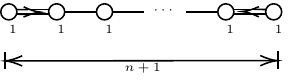} & $n$ \\
\hline
$d_n$ & \includegraphics[width=3.5cm]{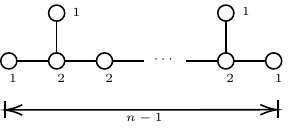} & $2n-3$ \\
\hline
$e_6$ & \includegraphics[width=2.5cm]{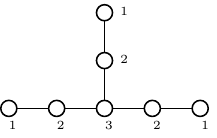} & 11 \\
\hline
$e_7$ & \includegraphics[width=3.5cm]{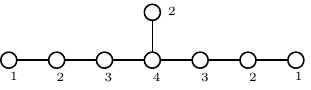} & 17 \\
\hline
$e_8$ & \includegraphics[width=4cm]{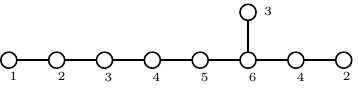} & 29 \\
\hline
$e_8$ & \includegraphics[width=7cm]{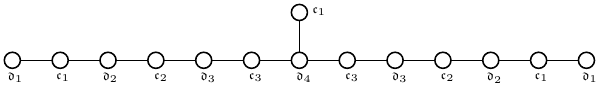} & 29 \\
\hline
$f_4$ & \includegraphics[width=2.5cm]{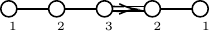} & 8 \\
\hline
$g_2$ & \includegraphics[width=1.3cm]{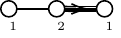} & 3 \\
\hline
\caption{Some magnetic quivers of the closures of minimal nilpotent orbits that we might encounter in this paper.}\label{slicesMQs}
\end{longtable}

Later, when we study the examples with orthosymplectic magnetic quivers, we would like to think of the Higgsings as quiver decays and fissions just like the unitary ones. If we know what the slices are, then we can match them with the differences of the parent and child magnetic quivers. In this paper, we mainly have the following cases:
\begin{itemize}
    \item the $d_n$ slice:
    \begin{equation}
        \includegraphics[width=7cm]{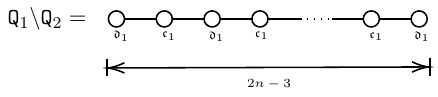},
    \end{equation}
    \item the $d_n$ slice:
    \begin{equation}
        \includegraphics[width=7cm]{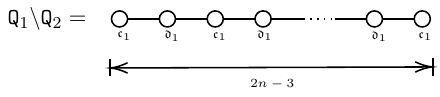}.
    \end{equation}
\end{itemize}
There is also an extraordinary case:
\begin{equation}
    \includegraphics[width=3cm]{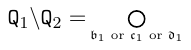}.
\end{equation}
Generically, we shall identify this with the $a_1=c_1=A_1$ slice. However, it turns out that there are differences $\mathtt{Q}_1\backslash\mathtt{Q}_2$ being such quivers but giving some different slices. They will be explicitly listed when we encounter them in our examples.

Sometimes, only the orthosymplectic magnetic quiver is known for the parent theory while only the unitary one is known for the child theory, or vice versa. In such cases, it is hard to obtain the slices from the difference of the quivers. They mostly come from the discussions in \S\ref{unknownslices} below. There is one case in this paper with a closure of minimal nilpotent orbit slice. This is an $e_6$ slice appearing in the rank 0 $(E_6,E_6)$ conformal matter theory in \S\ref{rank0E6ospMQ}. This would be expressed schematically in this paper. We shall draw the difference of the shapes of the quivers, and each node is simply labelled by $\mathfrak{g}\backslash\mathfrak{h}$ with the corresponding gauge algebras $\mathfrak{g}$ and $\mathfrak{h}$ in the parent and child quivers respectively. The $e_6$ slice can then be plotted as in \eqref{e6fromE6rank0}.

\subsection{Some Other Slices}\label{moreslices}
Besides the closures of minimal nilpotent orbits, there could be other elementary slices, such as the Kleinian singularities ($A_n$, $D_n$, $E_n$), the quasi-minimal singularities \cite{malkin2005minimal,Bourget:2021siw}\footnote{The term ``quasi-minimal'' for the singularities, introduced in \cite{malkin2005minimal}, should not be confused with the ``minimal'' in the Higgsings.}, the $\mathcal{J}_{k_1,k_2}$ slices \cite{Bourget:2022tmw} etc., in the process of Higgsings. Here, let us just mention the known slices that appear in our examples.

When there are minimal plateau and combo Higgsings, the slices would often be difficult to identify. If the magnetic quivers are known (especially the unitary ones), we could compare the differences between the parent and child quivers and recognize the transverse slice after rebalancing as recalled in \S\ref{magnetic}. In particular, we will have the following families in the quiver decays:
\begin{itemize}
    \item The A-type Kleinian singularity $A_n$ has the unitary magnetic quiver as
    \begin{equation}
        \includegraphics[width=1.5cm]{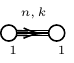}
    \end{equation}
    with $n$ edges, each $k$-laced. In particular, this 1-dimensional slice would appear when a $\U(k)$ node of balance $b>0$ becomes a $\U(k-1)$ node under the quiver decay. The transition is $A_{b+1}$.
    \item The $\mathcal{Y}(k)$ slice is given by the unitary magnetic quiver
    \begin{equation}
        \includegraphics[width=1.5cm]{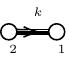}
    \end{equation}
    with a $k$-laced edge for $k\geq4$. In particular, $\mathcal{Y}(4)=a_2$. More properties of this 2-dimensional slice can be found in \cite{bellamy2023new,Bourget:2022tmw}.
\end{itemize}

In the quiver fissions which are always 1-dimensional transitions, there are two possible slices for the unitary cases \cite{Bourget:2023dkj,Bourget:2024mgn}. When the fission part is split into two identical sub-quivers, the slice is $A_1$. Otherwise, the slice is $m$ which is non-normal. See \cite{fu2017generic,vinberg1972class} for more details.

For theories associated with nilpotent orbits in our examples, we can also find some more different slices following \cite{slodowy1980simple,kraft1982geometry,fu2017generic,Cabrera:2017njm,juteau2023minimal,Bennett:2024loi}. Let us state the notation here. Sometimes, there is an ``outer'' action/monodromy $\mathfrak{S}_{n+1}$ on the slice, and we shall put $n$ plus signs as superscripts. For instance, $a_3^+$ indicates that there is an $\mathfrak{S}_2$ action on the $a_3$ slice. We shall also adopt the following notations:
\begin{equation}
    B_n=A_{2n-1}^+,\quad C_n=D_{n+1}^+,\quad F_4=E_6^+,\quad G_2=D_4^{++}.
\end{equation}
See \cite{slodowy1980simple,kraft1982geometry,fu2017generic,Cabrera:2017njm,juteau2023minimal,Bennett:2024loi} for more details on these slices.

\subsection{Unknown Cases}\label{unknownslices}
Despite the tools we have from the generalized quivers and the magnetic quivers, there are still some slices we could not identify. For our examples in this paper, most of them are 1-dimensional, so let us mention these 1-dimensional cases first. For the minimal plateau and combo Higgsings, one cannot read off the slice directly from the change of the flavour symmetry as in the minimal nilpotent orbit Higgsings. From the perspective of magnetic quivers, either the magnetic quivers are not known or the known magnetic quivers before and after the transition are of different types (unitary or orthosymplectic). In the latter case, we shall only denote the slice schematically using the difference of the quivers such as in \eqref{D4rank3ospuex} (which is similar to the $e_6$ slice in \eqref{e6fromE6rank0}).

For the endpoint-changing Higgsings, many of such flows correspond to quiver fissions on the magnetic side. When this happens to orthosymplectic quivers, we always get an orthosymplectic sub-quiver and a unitary sub-quiver. Although the exact geometry is still unclear, we expect the slice to be similar to $m$.

For the other 1-dimensional unknown slices, we shall use the question marks to indicate them in the Hasse diagrams. In Table \ref{D4rank2_table}, the flow from $(D_4^2,14)$ to $(D_4^2,21)$ is labelled by a double question mark. This is a slice of dimension 2. More explicitly, $(D_4^2,14)$ and $(D_4^2,21)$ are associated to the nilpotent orbit pairs $[3^2,1^2]\text{-}[3,1^5]$ and $[5,1^3]\text{-}[3,1^5]$ respectively. As we will see below, for long quivers, the transition between these two theories is $c_1$. However, this would become different when we have short quivers (such as the rank 2 case here). Following the discussions in Appendix \ref{TypeIIAbranes}, the magnetic quivers and their difference are given by
\begin{equation}
    \includegraphics[width=15cm]{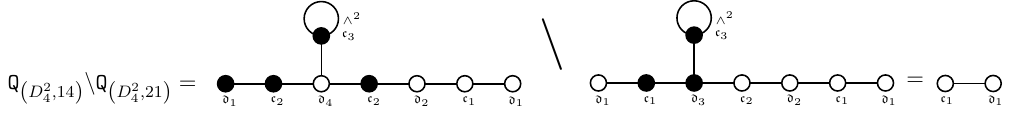}.
\end{equation}
From the generalized quivers, it is also not clear what exactly the slice is. As an odd number of alternating $\mathfrak{d}_1$ and $\mathfrak{c}_1$ nodes correspond to the $d_n$ slice, one might wonder if an even number of such alternating nodes would give rise to the $b_n$ slice, and this 2-dimensional unknown slice would then be $b_2=c_2$. However, to the best of our knowledge, for all the examples whose (orthosymplectic) magnetic quivers are known, the $b_n$ slices only appear when there are non-special orbits involved. Both finding the magnetic quiver for the non-special cases and identifying the above 2-dimensional slice still require further exploration.

There is another family of unknown slices when Higgsing the rank 1 $(E_6,E_6)$ conformal matter as in \S\ref{Etypes}. In some descendant theories, there are $\U(n)$ flavour symmetries that can be Higgsed. By inspecting the curve configurations before and after such transitions, one can see that they should indeed be minimal Higgsings that cannot be further decomposed. Since the slice for Higgsing an $\SU(n+1)$ flavour is $a_n$, we shall denote the slice for Higgsing a $\U(n)$ flavour as $a_n^*$, where $n$ still indicates the dimension of the slice\footnote{Some of them may be identified as some known slices due to the nilpotent orbits as will be mentioned in \S\ref{Etypes}.}.

\section{Examples of Complete Hasse Diagrams}\label{sec:examples}
In this section, we shall present the complete Hasse diagrams for a few UV theories. All of these examples follow our algorithm discussed above. Readers can also identify the type of Higgsing for any flow by inspecting the UV and the IR theories according to our classification. Hence, we shall not mention explicitly which minimal Higgsings the flows are in the examples below (although we shall still list the elementary slices). Before delving into any examples, it is worth mentioning that a subset of the Higgsed theories in the Hasse diagrams belongs to the nilpotent hierarchies \cite{Heckman:2016ssk}. The nilpotency follows from the nilpotent orbits in the Lie algebras. Given a UV theory with some flavour symmetry $G$, the Hasse diagram under RG flows can be determined by the Hasse diagram of the nilpotent orbits under the closure inclusion relation as each theory is in one-to-one correspondence with a pair of nilpotent orbits. Therefore, when there is an inclusion between a pair of orbits, a flow between the corresponding descendant theories is predicted.

In this paper, instead of focusing on the nilpotent hierarchy, we would like to consider all the possible minimal Higgsings. Therefore, our algorithm has the following features:
\begin{itemize}
    \item Generically, our approach generates a larger Hasse diagram that includes not only more nodes than the ones associated with the nilpotent orbits (and thus more flows involving new nodes) but also more flows among the existing nilpotent hierarchy.
    \item Our approach is completely iterative. If we take any descendant theory to be the new UV theory and re-run our algorithm, we will reproduce the exact Hasse sub-diagram of the original Hasse diagram formed by all the descendant nodes of the designated UV theory with the same set of RG flows.
    \item However, there is no clear generalization of the algebraic description as in the nilpotent hierarchy story. It would be very interesting if any similar structure can be identified.
\end{itemize}

\subsection{SU-Type Theories on \texorpdfstring{$-2$}{-2} Curves}\label{SUtype}
\begin{figure}[h]
    \centering
    \includegraphics[width=0.8\linewidth]{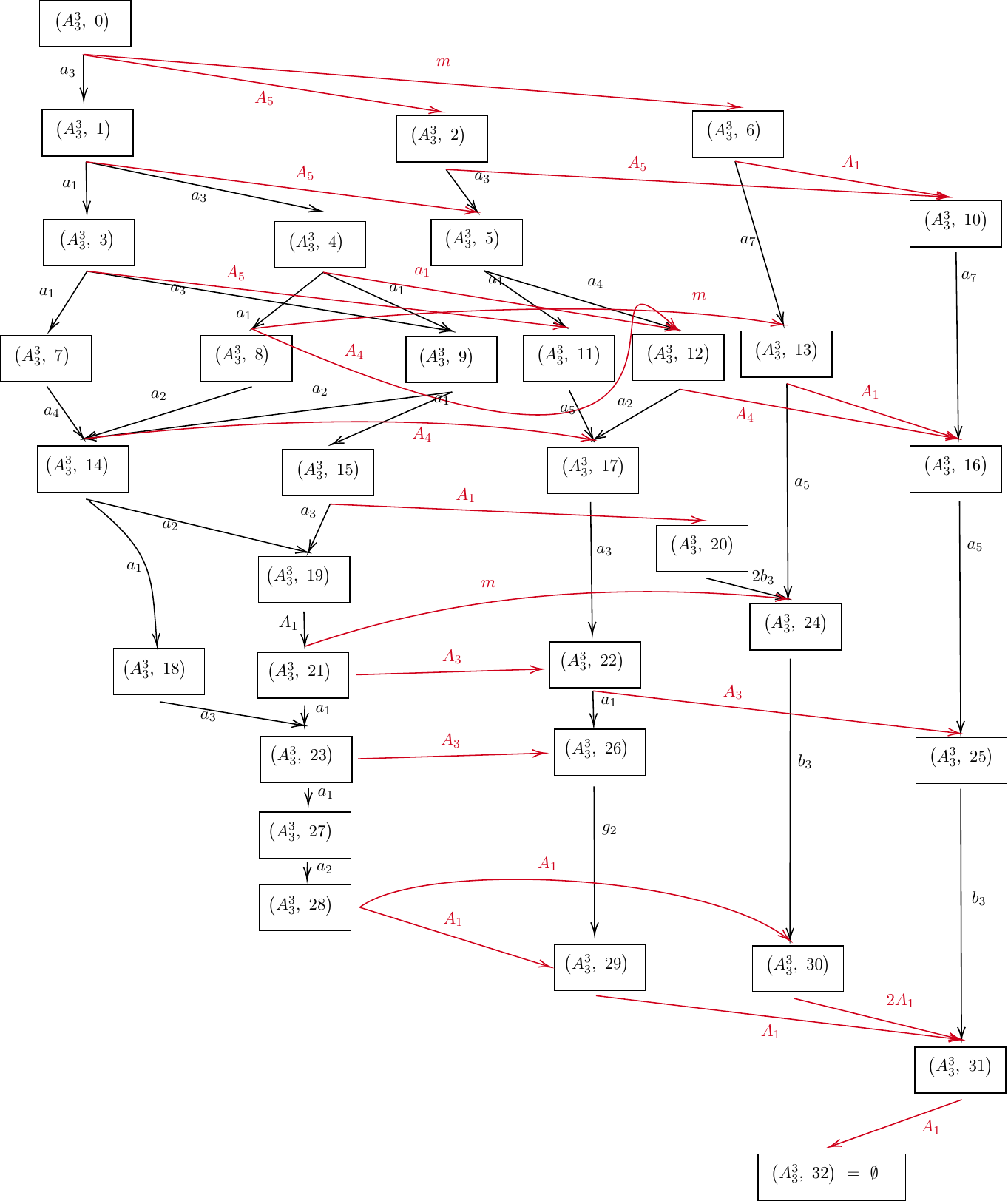}
    \caption{Atomic Hasse diagram of the UV theory \eqref{eqn:A3rank3Hasse_UV}. The nodes are given in Table \ref{A3rank3Hasse_nodes}. The red arrows correspond to the endpoint-changing flows.}\label{A3rank3Hasse}
\end{figure}

Let us start with an $A_3$ $\mathcal{N}=(2,0)$ theory and decorate each $-2$ curve with a $\ksu(4)$ gauge symmetry:
\be
    [\SU(4)] \ \ \overset{\ksu(4)}{2} \ \ \overset{\ksu(4)}{2} \ \ \overset{\ksu(4)}{2} \ \ [\SU(4)].\label{eqn:A3rank3Hasse_UV}
\ee
The Hasse diagram is given in Figure \ref{A3rank3Hasse}, where the 6d SCFT encoded by each node is listed in a separate table in Table \ref{A3rank3Hasse_nodes} to avoid clutter.

\begin{longtable}{|c|c|c|c|}
\hline
$A_3^3$ label &  Tensor branch description & $d_{\mathbb{H}}$ & (descendant \#; flow type) \\ \hline \hline
$(A_3^3, 0)$ & $[\SU(4)] \ \ \overset{\ksu(4)}{2} \ \ \overset{\ksu(4)}{2} \ \ \overset{\ksu(4)}{2} \ \ [\SU(4)]$ & 19 & $(1; a_3),\ (2; A_5),\ (6; m)$ \\ \hline
$(A_3^3, 1)$ & $[\SU(2)] \ \ {\overset{\ksu(3)}{2}}\ \ \underset{[N_f = 1]}{\overset{\ksu(4)}{2}}\ \ \overset{\ksu(4)}{2} \ \ [\SU(4)]$ & 16 & $(3; a_1),\ (4; a_3),\ (5;A_5)$ \\ \hline
$(A_3^3, 2)$ & $[\SU(4)] \ \ \overset{\ksu(4)}{2} \ \ \overset{\ksu(4)}{2} \ \ [\SU(4)]$ & 18 & $(5; a_3),\ (10; A_5)$ \\ \hline
$(A_3^3, 3)$ & $\overset{\ksu(2)}{2}\ \ \underset{[\SU(2)]}{\overset{\ksu(4)}{2}}\ \ \overset{\ksu(4)}{2} \ \ [\SU(4)]$ & 15 & $(7; a_1),\ (9; a_3),\ (11; A_5)$ \\ \hline
$(A_3^3, 4)$ & $[\SU(2)] \ \ {\overset{\ksu(3)}{2}}\ \ \underset{[\SU(2)]}{\overset{\ksu(4)}{2}}\ \ \overset{\ksu(3)}{2} \ \ [\SU(2)]$ & 13 & $(8; a_1),\ (9, a_1),\ (12, a_1)$ \\ \hline
$(A_3^3, 5)$ & $[\SU(2)] \ \ {\overset{\ksu(3)}{2}}\ \ \overset{\ksu(4)}{2}\ \ [\SU(5)]$ & 15 & $(11; a_1),\ (12; a_4)$ \\ \hline
$(A_3^3, 6)$ & $\overset{\ksu(4)}{2} \ \ [\SU(8)] \ \ \sqcup \ \ 2$ & 18 & $(10; A_1),\ (13, a_7)$ \\ \hline
$(A_3^3, 7)$ & $[N_f = 1] \ \ \overset{\ksu(2)}{2}\ \ \overset{\ksu(3)}{2}\ \ \overset{\ksu(4)}{2} \ \ [\SU(5)]$ & 14 & $(14; a_4)$ \\ \hline
$(A_3^3, 8)$ & $[\SU(3)] \ \ {\overset{\ksu(3)}{2}}\ \ {\overset{\ksu(3)}{2}}\ \ \overset{\ksu(3)}{2} \ \ [\SU(3)]$ & 12 & $(12;A_4),\ (13,m),\ (14; a_2)$ \\ \hline
$(A_3^3, 9)$ & ${\overset{\ksu(2)}{2}}\ \ \underset{[\SU(3)]}{\overset{\ksu(4)}{2}}\ \ \overset{\ksu(3)}{2} \ \ [\SU(2)]$ & 12 & $(14; a_2),\ (15; a_1)$ \\ \hline
$(A_3^3, 10)$ & $\overset{\ksu(4)}{2} \ \ [\SU(8)]$ & 17 & $(16; a_7)$ \\ \hline
$(A_3^3, 11)$ & $\overset{\ksu(2)}{2}\ \ {\overset{\ksu(4)}{2}}\ \ [\SU(6)]$ & 14 & $(17; a_5)$ \\ \hline
$(A_3^3, 12)$ & $[\SU(3)] \ \ {\overset{\ksu(3)}{2}}\ \ {\overset{\ksu(3)}{2}}\ \ [\SU(3)]$ & 11 & $(17; a_2),\ (16; A_4)$ \\ \hline
$(A_3^3, 13)$ & $\overset{\ksu(3)}{2} \ \ [\SU(6)] \ \ \sqcup \ \ 2$ & 11 & $(16; A_1),\ (24, a_5)$ \\ \hline
$(A_3^3, 14)$ & $[N_f = 1]\ \ {\overset{\ksu(2)}{2}}\ \ \underset{[N_f = 1]}{\overset{\ksu(3)}{2}}\ \ \overset{\ksu(3)}{2} \ \ [\SU(3)]$ & 10 & $(17; A_4),\ (18; a_1),\ (19; a_2)$ \\ \hline
$(A_3^3, 15)$ & ${\overset{\ksu(2)}{2}}\ \ \underset{[\SU(4)]}{\overset{\ksu(4)}{2}}\ \ \overset{\ksu(2)}{2}$ & 11 & $(19; a_3),\, (20; A_1)$ \\ \hline
$(A_3^3, 16)$ & $\overset{\ksu(3)}{2} \ \ [\SU(6)]$ & 10 & $(25; A_5)$ \\ \hline
$(A_3^3, 17)$ & $[N_f = 1] \ \ {\overset{\ksu(2)}{2}}\ \ {\overset{\ksu(3)}{2}}\ \ [\SU(4)]$ & 9 & $(22; a_3)$ \\ \hline
$(A_3^3, 18)$ & $2 \ \ \underset{[N_f = 1/2]}{\overset{\ksu(2)}{2}} \ \ \overset{\ksu(3)}{2} \ \  [\SU(4)]$ & 9 & $(23; a_3)$ \\ \hline
$(A_3^3, 19)$ & $[N_f = 1] \ \ \overset{\ksu(2)}{2} \ \ \underset{[\SU(2)]}{\overset{\ksu(3)}{2}} \ \  \overset{\ksu(2)}{2}\ \ [N_f = 1]$ & 8 & $(21; A_1)$ \\ \hline
$(A_3^3, 20)$ & $\overset{\ksu(2)}{2} \ \ [\SO(7)] \ \ \sqcup \ \ \overset{\ksu(2)}{2} \ \ [\SO(7)]$ & 10 & $(24; 2b_3)$ \\ \hline
$(A_3^3, 21)$ & $[\SU(2)] \ \ \overset{\ksu(2)}{2} \ \ \overset{\ksu(2)}{2} \ \  \overset{\ksu(2)}{2}\ \ [\SU(2)]$ & 7 & $(22; A_3),\ (23; a_1),\ (24,m)$ \\ \hline
$(A_3^3, 22)$ & $[\SU(2)] \ \ {\overset{\ksu(2)}{2}}\ \ {\overset{\ksu(2)}{2}}\ \ [\SU(2)]$ & 6 & $(25; A_3),\ (26, a_1)$ \\ \hline
$(A_3^3, 23)$ & $2 \ \ \underset{[N_f = 3/2]}{\overset{\ksu(2)}{2}} \ \ \overset{\ksu(2)}{2} \ \  [\SU(2)]$ & 6 & $(26; A_3),\ (27; a_1)$ \\ \hline
$(A_3^3, 24)$ & $\overset{\ksu(2)}{2} \ \ [\SO(7)] \ \ \sqcup \ \ 2$ & 6 & $(30; b_3)$ \\ \hline
$(A_3^3, 25)$ & $\overset{\ksu(2)}{2} \ \ [\SO(7)]$ & 5 & $(31; b_3)$ \\ \hline
$(A_3^3, 26)$ & $2 \ \ {\overset{\ksu(2)}{2}} \ \ [G_2]$ & 5 & $(29; g_2)$ \\ \hline
$(A_3^3, 27)$ & $2 \ \ \underset{[\SU(3)]}{\overset{\ksu(2)}{2}} \ \ 2 $ & 5 & $(28; a_2)$ \\ \hline
$(A_3^3, 28)$ & $2 \ \ 2 \ \ 2$ & 3 & $(29; A_1),\ (30; A_1)$ \\ \hline
$(A_3^3, 29)$ & $2 \ \ 2$ & 2 & $(31; A_1)$ \\ \hline
$(A_3^3, 30)$ & $2 \ \ \sqcup \ \ 2$ & 2 & $(31; 2A_1)$ \\ \hline
$(A_3^3, 31)$ & $2$ & 1 & $(32; A_1)$ \\ \hline
$(A_3^3, 32)$ & $\varnothing$ & 0 & IR Theory \\ \hline
\caption{The descendant theories of the rank 3 $A_3$ theory \eqref{eqn:A3rank3Hasse_UV} with the Hasse diagram in Figure \ref{A3rank3Hasse}. In the table, besides the labels and the tensor branch descriptions of the theories, we also give the (quaternionic) dimensions of the Higgs branches and the slices of the flows.}\label{A3rank3Hasse_nodes}
\end{longtable}

As mentioned above, the complete Hasse diagrams generically contain more theories than the Hasse diagrams for the nilpotent hierarchies. For this particular UV theory, the extra descendant theories come from either decorating the A-type $(2,0)$ theories or performing the minimal plateau Higgsings.

Notice that the A-type theories on $-2$ curves have both A-type $(2,0)$ endpoints and A-type fibre decorations. With these features, the Hasse diagrams of this family can be completely reproduced via magnetic quivers as will be illustrated in \S\ref{SUtypeMQs}. Readers are also referred to \cite{Lawrie:2024zon} for a careful analysis of the theories still with A-type fibre decorations but with DE-type bases, which we shall not repeat.

Later in \S\ref{Dtypes}, we will see that this Hasse diagram would reappear in its exact form as part of the complete Hasse diagram of a D-type conformal matter theory. This demonstrates the iterativeness property that we promised earlier. In this sense, it unifies an A-type nilpotent hierarchy with a D-type one, both treated explicitly in \cite{Hassler:2019eso}, into a single complete Hasse diagram.

\subsection{The 6D Conformal Matter Theories}\label{conformalmatters}
Now, let us analyze the Higgsings of the 6d $(G, G)$ conformal matter theories \cite{DelZotto:2014hpa}, where $G$ is a Lie group of DE-type. A large subset of Higgsing such theories by an arbitrary nilpotent VEV into a single flavour symmetry has been found to fall into the nilpotent hierarchies \cite{Heckman:2016ssk}. If one turns on both of the flavour symmetries, then the IR theory can be determined by either the string junctions for D-types \cite{Hassler:2019eso} or the $T^2$ compactification down to 4d class $\mathcal{S}$ theories of arbitrary types in \cite{Baume:2021qho}.

However, some Higgsings are not covered by the nilpotent hierarchies. First of all, there are RG flows that change the IR $\mathcal{N}=(2,0)$ theories under the nilpotent VEVs, namely changing the ranks of the conformal matters. They would connect different nilpotent hierarchies. In addition, for cases with a pair of T-brane VEVs, there are flavour symmetries in the middle of the quivers. Their VEVs can be used to generate new theories. We shall incorporate both of these cases in our larger Hasse diagrams.

\subsubsection{D-Types}\label{Dtypes}
We begin by considering the full Higgsing Hasse diagram for the $(D_4, D_4)$ conformal matter. Let us first take the short quiver of rank 3. This theory with the nilpotent hierarchy was treated explicitly in \cite[Figure 45]{Hassler:2019eso}, but here we would obtain more theories. In particular, there exists an intermediate theory that is also the $(1,0)$ theory given by decorating the $A_3$ $(2,0)$ theory with $\ksu(4)$ flavour symmetry, and thus further incorporating the SU(4) short quiver case in \cite[Figure 42]{Hassler:2019eso}.

\begin{figure}[H]
    \centering
    \includegraphics[width=13cm]{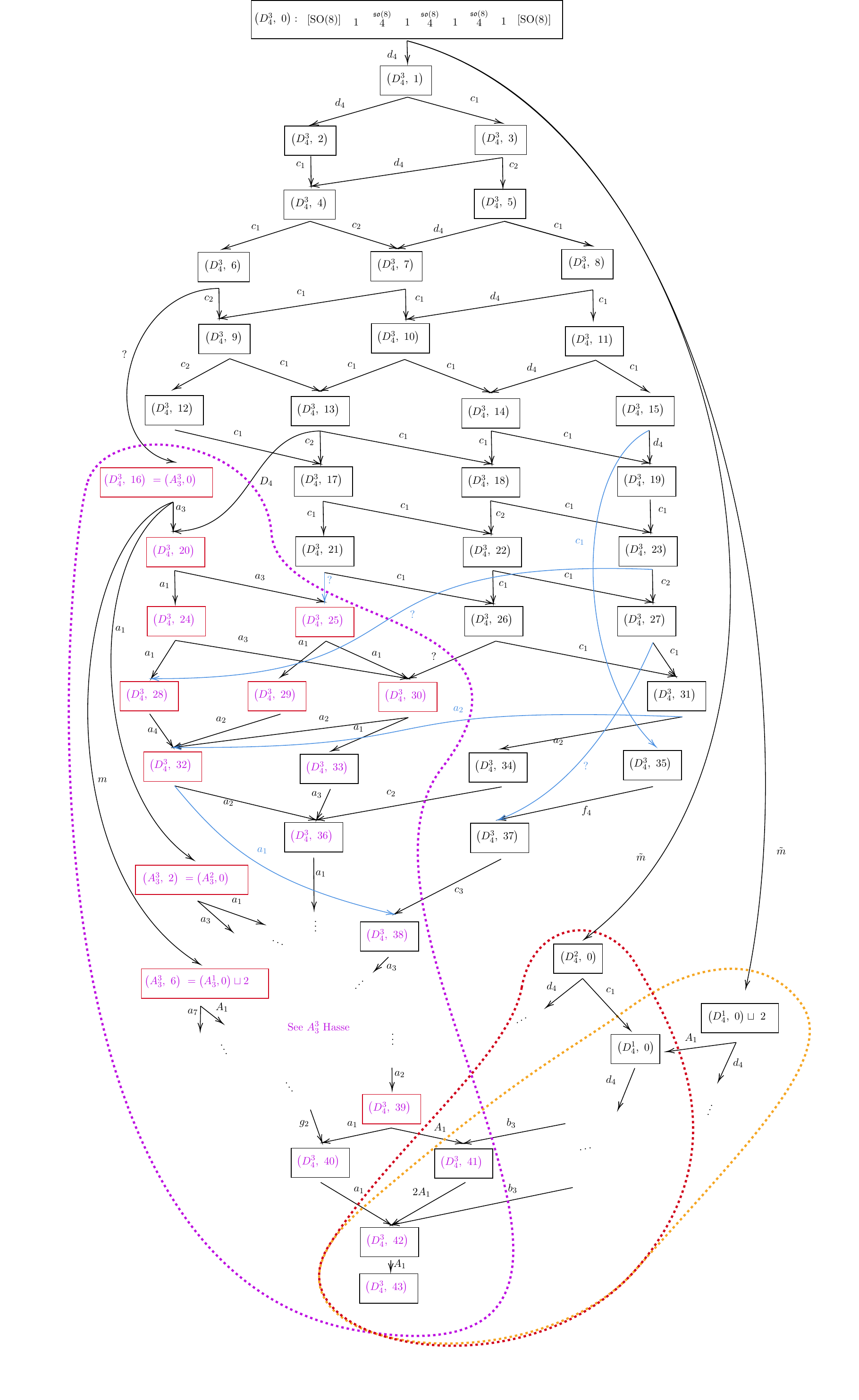}
    \caption{The complete Hasse diagram for $(D_4, D_4)$ conformal matter theory of rank 3 with each node given in Table \ref{D4rank3_table}. We emphasize that the table is self-contained, whereas the figure is more to illustrate the overall landscape of possible Higgsings. To avoid cluttering of the figure, we leave parts of the figure implicit once we Higgs into (the Hasse diagrams of) either the $(D_4, D_4)$ theories of lower ranks or the $(A_3, A_3)$ theory of rank 3. }\label{D4rank3Hasse}
\end{figure}

The Hasse diagram is depicted in Figure \ref{D4rank3Hasse}, with the nodes labelled in Table \ref{D4rank3_table}. It turns out that the complete Hasse diagram also contains the Higgsings of the $(D_4,D_4)$ conformal matter theories of ranks 1 and 2, as well as the $A_3^3$ theory discussed in the previous subsection. To avoid clutter, they are illustrated as dashed blobs in Figure \ref{D4rank3Hasse}. The nodes from Higgsing the rank 1 and rank 2 theories are listed in Tables \ref{D4rank2_table} and \ref{D4rank1_table}. Moreover, in the tables, we leave the end-point changing RG flows implicit but only list the endpoint-preserving flows.

\begin{longtable}{|c|c|c|c|c|}
\hline
$D_4^3$ label &  Tensor branch description & $d_{\mathbb{H}}$ & (descendant \#; flow type) & $A_3^3$ label \\ \hline \hline
$(D_4^3, 0)$ & $\underset{[\SO(8)]}{1} \ \ \overset{\kso(8)}{4} \ \ 1 \ \ \overset{\kso(8)}{4} \ \ 1 \ \ \overset{\kso(8)}{4} \ \ \underset{[\SO(8)]}{1}$ & 32 & $(1; d_4)$ & - \\ \hline
$(D_4^3, 1)$ & $\underset{[\Sp(1)^3]}{\overset{\kso(8)}{3}} \ \ 1 \ \ \overset{\kso(8)}{4} \ \ 1 \ \ \overset{\kso(8)}{4} \ \ \underset{[\SO(8)]}{1}$ & 27 & $(2; d_4),\ (3; c_1)$ & - \\ \hline
$(D_4^3, 2)$ & $\underset{[\Sp(1)^3]}{\overset{\kso(8)}{3}} \ \ 1 \ \ \overset{\kso(8)}{4} \ \ 1 \ \ \underset{[\Sp(1)^3]}{\overset{\kso(8)}{3}}$ & 22 & $(4; c_1)$ & - \\ \hline
$(D_4^3, 3)$ & $\underset{[\Sp(2)]}{\overset{\kso(7)}{3}} \ \ 1 \ \ \overset{\kso(8)}{4} \ \ 1 \ \ \overset{\kso(8)}{4} \ \ \underset{[\SO(8)]}{1}$ & 26 & $(4; d_4),\ (5; c_2)$ & - \\ \hline
$(D_4^3, 4)$ & $\underset{[\Sp(2)]}{\overset{\kso(7)}{3}} \ \ 1 \ \ \overset{\kso(8)}{4} \ \ 1 \ \ \underset{[\Sp(1)^3]}{\overset{\kso(8)}{3}}$ & 21 & $(6; c_1),\ (7; c_2)$ & - \\ \hline
$(D_4^3, 5)$ & $\underset{[\Sp(1)]}{\overset{\kg_2}{3}} \ \ 1 \ \ \overset{\kso(8)}{4} \ \ 1 \ \ \overset{\kso(8)}{4} \ \ \underset{[\SO(8)]}{1}$ & 24 & $(7; d_4),\ (8; c_1)$ & - \\ \hline
$(D_4^3, 6)$ & $\underset{[\Sp(2)]}{\overset{\kso(7)}{3}} \ \ 1 \ \ \overset{\kso(8)}{4} \ \ 1 \ \ \underset{[\Sp(2)]}{\overset{\kso(7)}{3}}$ & 20 & $(9; c_2),\ (16; ?)$ & - \\ \hline
$(D_4^3, 7)$ & $\underset{[\Sp(1)]}{\overset{\kg_2}{3}} \ \ 1 \ \ \overset{\kso(8)}{4} \ \ 1 \ \ \underset{[\Sp(1)^3]}{\overset{\kso(8)}{3}}$ & 19 & $(9; c_1),\ (10; c_1)$ & - \\ \hline
$(D_4^3, 8)$ & $\overset{\ksu(3)}{3} \ \ 1 \ \ \overset{\kso(8)}{4} \ \ 1 \ \ \overset{\kso(8)}{4} \ \ \underset{[\SO(8)]}{1}$ & 23 & $(10; d_4),\ (11; c_1)$ & - \\ \hline
$(D_4^3, 9)$ & $\underset{[\Sp(1)]}{\overset{\kg_2}{3}} \ \ 1 \ \ \overset{\kso(8)}{4} \ \ 1 \ \ \underset{[\Sp(2)]}{\overset{\kso(7)}{3}}$ & 28 & $(12; c_2),\ (13; c_1)$ & - \\ \hline
$(D_4^3, 10)$ & $\overset{\ksu(3)}{3} \ \ 1 \ \ \overset{\kso(8)}{4} \ \ 1 \ \ \underset{[\Sp(1)^3]}{\overset{\kso(8)}{3}}$ & 28 & $(13; c_1),\ (14; c_1)$ & - \\ \hline
$(D_4^3, 11)$ & $\overset{\ksu(2)}{2} \ \ \underset{[\Sp(1)]}{\overset{\kso(7)}{3}} \ \ 1 \ \ \overset{\kso(8)}{4} \ \ \underset{[\SO(8)]}{1}$ & 22 & $(14; d_4),\ (15; c_1)$ & - \\ \hline
$(D_4^3, 12)$ & $\underset{[\Sp(1)]}{\overset{\kg_2}{3}} \ \ 1 \ \ \overset{\kso(8)}{4} \ \ 1 \ \ \underset{[\Sp(1)]}{\overset{\kg_2}{3}}$ & 16 & $(17; c_1)$ & - \\ \hline
$(D_4^3, 13)$ & $\overset{\ksu(3)}{3} \ \ 1 \ \ \overset{\kso(8)}{4} \ \ 1 \ \ \underset{[\Sp(2)]}{\overset{\kso(7)}{3}}$ & 17 & $(17; c_2),\ (18; c_1)$ & - \\ \hline
$(D_4^3, 14)$ & $\overset{\ksu(2)}{2} \ \ \underset{[\Sp(1)]}{\overset{\kso(7)}{3}} \ \ 1 \ \ \underset{[\Sp(1)^3]}{\overset{\kso(8)}{3}}$ & 17 & $(18; c_1),\ (19; c_1)$ & - \\ \hline
$(D_4^3, 15)$ & $\overset{\ksu(2)}{2} \ \ \overset{\kg_2}{3} \ \ 1 \ \ \overset{\kso(8)}{4} \ \ \underset{[\SO(8)]}{1}$ & 21 & $(19; d_4),\ (35; c_1)$ & - \\ \hline
$(D_4^3, 16)$ & $\underset{[\SU(4)]}{\overset{\ksu(4)}{2}}\ \ \overset{\ksu(4)}{2}\ \ \underset{[\SU(4)]}{\overset{\ksu(4)}{2}}$ & 19 & $(20; a_3)$ & $(A_3^3,0)$ \\ \hline
$(D_4^3, 17)$ & $\overset{\ksu(3)}{3} \ \ 1 \ \ \overset{\kso(8)}{4} \ \ 1 \ \ \overset{\kg_2}{3}  \ \ \underset{[\Sp(1)]}{1}$ & 15 & $(21; c_1),\ (22; c_1)$ & - \\ \hline
$(D_4^3, 18)$ & $\overset{\ksu(2)}{2} \ \ \underset{[\Sp(1)]}{\overset{\kso(7)}{3}} \ \ 1 \ \ \underset{[\Sp(2)]}{\overset{\kso(7)}{3}}$ & 16 & $(22; c_2),\ (23; c_1)$ & - \\ \hline
$(D_4^3, 19)$ & $\overset{\ksu(2)}{2} \ \ \overset{\kg_2}{3} \ \ 1 \ \ \underset{[\Sp(1)^3]}{\overset{\kso(8)}{3}}$ & 16 & $(23; c_1)$ & - \\ \hline
$(D_4^3, 20)$ & $\underset{[\SU(2)]}{{\overset{\ksu(3)}{2}}}\ \ \underset{[N_f = 1]}{\overset{\ksu(4)}{2}}\ \ \underset{[\SU(4)]}{\overset{\ksu(4)}{2}}$ & 19 & $(24; a_1),\ (25; a_3)$ & $(A_3^3,1)$ \\ \hline
$(D_4^3, 21)$ & $\overset{\ksu(3)}{3} \ \ 1 \ \ \overset{\kso(8)}{4} \ \ 1 \ \ \overset{\ksu(3)}{3}$ & 14 & $(25; ?),\ (26; c_1)$ & - \\ \hline
$(D_4^3, 22)$ & $\overset{\ksu(2)}{2} \ \ \underset{[\Sp(1)]}{\overset{\kso(7)}{3}} \ \ 1 \ \ \overset{\kg_2}{3}  \ \ \underset{[\Sp(1)]}{1}$ & 14 & $(26; c_1),\ (27; c_1)$ & - \\ \hline
$(D_4^3, 23)$ & $\overset{\ksu(2)}{2} \ \ \overset{\kg_2}{3} \ \ 1 \ \ \underset{[\Sp(2)]}{\overset{\kso(7)}{3}}$ & 15 & $(27; c_2),\ (28; ?)$ & - \\ \hline
$(D_4^3, 24)$ & $\overset{\ksu(2)}{2}\ \ \underset{[\SU(2)]}{\overset{\ksu(4)}{2}}\ \ \underset{[\SU(4)]}{\overset{\ksu(4)}{2}}$ & 15 & $(28; a_1),\ (30; a_3)$ & $(A_3^3,3)$ \\ \hline
$(D_4^3, 25)$ & $\underset{[\SU(2)]}{{\overset{\ksu(3)}{2}}}\ \ \underset{[\SU(2)]}{\overset{\ksu(4)}{2}}\ \ \underset{[\SU(2)]}{\overset{\ksu(3)}{2}}$ & 13 & $(29; a_1),\ (30; a_1)$ & $(A_3^3,4)$ \\ \hline
$(D_4^3, 26)$ & $\overset{\ksu(2)}{2} \ \ \underset{[\Sp(1)]}{\overset{\kso(7)}{3}} \ \ \underset{[\SU(2)]}{1} \ \ \overset{\ksu(3)}{3}$ & 13 & $(30; ?),\ (31; c_1)$ & - \\ \hline
$(D_4^3, 27)$ & $\overset{\ksu(2)}{2} \ \ \overset{\kg_2}{3} \ \ \underset{[\SU(2)]}{1} \ \ \underset{[\Sp(1)]}{\overset{\kg_2}{3}}$ & 13 & $(31; c_1),\ (37; ?)$ & - \\ \hline
$(D_4^3, 28)$ & $\underset{[N_f = 1]}{\overset{\ksu(2)}{2}}\ \ \overset{\ksu(3)}{2}\ \ \underset{[\SU(5)]}{\overset{\ksu(4)}{2}}$ & 14 & $(32; a_4)$ & $(A_3^3,7)$ \\ \hline
$(D_4^3, 29)$ & $\underset{[\SU(3)]}{{\overset{\ksu(3)}{2}}}\ \ {\overset{\ksu(3)}{2}}\ \ \underset{[\SU(3)]}{\overset{\ksu(3)}{2}}$ & 12 & $(32; a_2)$ & $(A_3^3,8)$ \\ \hline
$(D_4^3, 30)$ & ${\overset{\ksu(2)}{2}}\ \ \underset{[\SU(3)]}{\overset{\ksu(4)}{2}}\ \ \underset{[\SU(2)]}{\overset{\ksu(3)}{2}}$ & 12 & $(32; a_2),\ (34; a_1)$ & $(A_3^3,9)$ \\ \hline
$(D_4^3, 31)$ & $\overset{\ksu(2)}{2} \ \ \overset{\kg_2}{3} \ \ \underset{[\SU(3)]}{1} \ \ \overset{\ksu(3)}{3}$ & 12 & $(32; a_2),\ (34; a_2)$ & - \\ \hline
$(D_4^3, 32)$ & $\underset{[N_f = 1]}{{\overset{\ksu(2)}{2}}} \ \ \underset{[N_f = 1]}{\overset{\ksu(3)}{2}}\ \ \underset{[\SU(3)]}{\overset{\ksu(3)}{2}}$ & 10 & $(36; a_2)$ & $(A_3^3,14)$ \\ \hline
$(D_4^3, 33)$ & ${\overset{\ksu(2)}{2}}\ \ \underset{[\SU(4)]}{\overset{\ksu(4)}{2}}\ \ \overset{\ksu(2)}{2}$ & 11 & $(36; a_3)$ & $(A_3^3,15)$ \\ \hline
$(D_4^3, 34)$ & $\underset{[N_f = 1/2]}{\overset{\ksu(2)}{2}} \ \ \underset{[\Sp(2)]}{\overset{\kg_2}{2}} \ \ \underset{[N_f=1/2]}{\overset{\ksu(2)}{2}}$ & 10 & $(33; c_1),\ (36; c_2)$ & - \\ \hline
$(D_4^3, 35)$ & $2 \ \ \overset{\ksu(2)}{2} \ \ \overset{\kg_2}{3} \ \ \underset{[F_4]}{1}$ & 20 & $(37; f_4)$ & - \\ \hline
$(D_4^3, 36)$ & $\underset{[N_f = 1]}{\overset{\ksu(2)}{2}} \ \ \underset{[\SU(2)]}{\overset{\ksu(3)}{2}} \ \ \underset{[N_f=1]}{\overset{\ksu(2)}{2}}$ & 8 & see $A_3^3$ Hasse & $(A_3^3,19)$ \\ \hline
$(D_4^3, 37)$ & $2 \ \ \overset{\ksu(2)}{2} \ \ \underset{[\Sp(3)]}{\overset{\kg_2}{2}}$ & 12 & $(38; c_3)$ & - \\ \hline
$(D_4^3, 38)$ & $2 \ \ \underset{[N_f = 1/2]}{\overset{\ksu(2)}{2}} \ \ \underset{[\SU(4)]}{\overset{\ksu(3)}{2}}$ & 9 & see $A_3^3$ Hasse & $(A_3^3,18)$ \\ \hline
$(D_4^3, 39)$ & $2 \ \ 2 \ \ 2$ & 3 & $(40,A_1),\, (41,A_1)$ & $(A_3^3,28)$ \\ \hline
$(D_4^3, 40)$ & $2 \ \ 2$ & 2 & $(42,A_1)$ & $(A_3^3,29)$ \\ \hline
$(D_4^3, 41)$ & $2 \ \ \sqcup \ \ 2$ & 2 & $(42,2A_1)$ & $(A_3^3,30)$ \\ \hline
$(D_4^3, 42)$ & $2$ & 1 & $(43,A_1)$ & $(A_3^3,31)$ \\ \hline
$(D_4^3, 43)$ & $\varnothing$ & 0 & IR Theory & $(A_3^3,32)$ \\ \hline
\caption{Theories corresponding to the nodes in the Hasse diagram of the rank 3 $(D_4, D_4)$ conformal matter. Endpoint-changing flows are left implicit in the table, which has the following two scenarios. A theory with a ``41-tail'' on one end admits a flow that loses this tail. Moreover, a theory with a ``4141-tail'' on one end admits another endpoint-changing flow that drops this tail but gains a disjoint $A_1$ $(2,0)$ theory. In this table, we have RG flow types labelled by question marks - these are the unknown cases discussed in \S\ref{unknownslices}.}\label{D4rank3_table}
\end{longtable}

\begin{longtable}{|c|c|c|c|}
\hline
$D_4^1$ label &  Tensor branch description & $d_{\mathbb{H}}$ & (descendant \#; flow type) \\ \hline \hline
$(D_4^1, 0)$ & $[\SO(8)] \ \ 1 \ \ \overset{\kso(8)}{4} \ \ 1 \ \  \ \ [\SO(8)]$ & 30 & $(1; d_4)$ \\ \hline
$(D_4^1, 1)$ & $[\SO(8)] \ \ 1 \ \ \overset{\kso(8)}{3} \ \  [\Sp(1)^3]$ & 25 & $(2; d_4), (3; c_1)$ \\ \hline
$(D_4^1, 2)$ & $\overset{\kso(8)}{2} \ \  [\Sp(2)^3]$ & 20 & $(4; c_2)$ \\ \hline
$(D_4^1, 3)$ & $[\SO(9)] \ \ 1 \ \ \overset{\kso(7)}{3} \ \  [\Sp(2)]$ & 24 & $(4; b_4), (5;  c_2)$ \\ \hline
$(D_4^1, 4)$ & $\overset{\kso(7)}{2} \ \  [\Sp(4) \times\Sp(1)]$ & 18 & $(6; c_1), (7; c_4)$ \\ \hline
$(D_4^1, 5)$ & $[F_4] \ \ 1 \ \ \overset{\kg_2}{3} \ \  [\Sp(1)]$ & 22 & $(7; f_4), (8; c_1)$ \\ \hline
$(D_4^1, 6)$ & $\overset{\ksu(4)}{2} \ \  [\SU(8)]$ & 17 & $(9; a_7)$ \\ \hline
$(D_4^1, 7)$ & $\overset{\kg_2}{2} \ \  [\Sp(4)]$ & 14 & $(9; c_4)$ \\ \hline
$(D_4^1, 8)$ & $[E_6] \ \ 1 \ \ \overset{\ksu(3)}{3} $ & 21 & $(9; e_6)$ \\ \hline
$(D_4^1, 9)$ & $\overset{\ksu(3)}{2} \ \  [\SU(6)] $ & 10 & $(10; a_5)$ \\ \hline
$(D_4^1, 10)$ & $\overset{\ksu(2)}{2} \ \  [\SO(7)] $ & 5 & $(11; b_3)$ \\ \hline
$(D_4^1, 11)$ & $2$ & 1 & $(12; A_1)$ \\ \hline
$(D_4^1, 12)$ & $\varnothing$ & 0 & IR theory \\ \hline
\caption{Theories corresponding to the nodes in the Hasse diagram of the rank 1 $(D_4, D_4)$ conformal matter.}\label{D4rank1_table}
\end{longtable}

\begin{longtable}{|c|c|c|c|c|}
\hline
$D_4^2$ label &  Tensor branch description & $d_{\mathbb{H}}$ & (descendant \#; flow type) & $A_3^3$ label \\ \hline \hline
$(D_4^2, 0)$ & $[\SO(8)] \ \ 1 \ \ \overset{\kso(8)}{4} \ \ 1 \ \ \overset{\kso(8)}{4} \ \ 1 \ \ [\SO(8)]$ & 31 & $(1; d_4)$ & -  \\ \hline
$(D_4^2, 1)$ & $[\SO(8)] \ \ 1 \ \ \overset{\kso(8)}{4} \ \ 1 \ \ \overset{\kso(8)}{3} \ \ [\Sp(1)^3]$ & 26 & $(2; d_4),\ (3; c_1)$ & -  \\ \hline
$(D_4^2, 2)$ & $[\Sp(1)^3] \ \ \overset{\kso(8)}{3} \ \ 1 \ \ \overset{\kso(8)}{3} \ \ [\Sp(1)^3]$ & 21 & $(4; c_1)$ & -  \\ \hline
$(D_4^2, 3)$ & $[\SO(8)] \ \ 1 \ \ \overset{\kso(8)}{4} \ \ 1 \ \ \overset{\kso(7)}{3} \ \ [\Sp(2)]$ & 25 & $(4; d_4),\ (5; c_2)$ & -  \\ \hline
$(D_4^2, 4)$ & $[\Sp(1)^3] \ \ \overset{\kso(8)}{3} \ \ 1 \ \ \overset{\kso(7)}{3} \ \ [\Sp(2)]$ & 20 & $(6; c_1),\ (7; c_2)$ & -  \\ \hline
$(D_4^2, 5)$ & $[\SO(8)] \ \ 1 \ \ \overset{\kso(8)}{4} \ \ 1 \ \ \overset{\kg_2}{3} \ \ [\Sp(1)]$ & 23 & $(7; d_4),\ (8; c_1)$ & -  \\ \hline
$(D_4^2, 6)$ & $[\Sp(2)] \ \ \overset{\kso(7)}{3} \ \ 1 \ \ \overset{\kso(7)}{3} \ \ [\Sp(2)]$ & 19 & $(9; A_1),\ (10; c_2)$ & -  \\ \hline
$(D_4^2, 7)$ & $[\Sp(1)^3] \ \ \overset{\kso(8)}{3} \ \ 1 \ \ \overset{\kg_2}{3} \ \ [\Sp(1)]$ & 18 & $(10; c_1),\ (11; c_1)$ & -  \\ \hline
$(D_4^2, 8)$ & $[\SO(8)] \ \ 1 \ \ \overset{\kso(8)}{4} \ \ 1 \ \ \overset{\ksu(3)}{3}$ & 22 & $(11; d_4),\ (12; A_1)$ & -  \\ \hline
$(D_4^2, 9)$ & $[\SU(4)] \ \ \overset{\ksu(4)}{2} \ \ \overset{\ksu(4)}{2} \ \ [\SU(4)]$ & 18 & $(17; a_3)$ & $(A_3^3, 2)$  \\ \hline
$(D_4^2, 10)$ & $[\Sp(2)] \ \ \overset{\kso(7)}{3} \ \ 1 \ \ \overset{\kg_2}{3} \ \ [\Sp(1)]$ & 17 & $(13; c_2),\ (14; c_1)$ & -  \\ \hline
$(D_4^2, 11)$ & $[\Sp(1)^3] \ \ \overset{\kso(8)}{3} \ \ 1 \ \ \overset{\ksu(3)}{3}$ & 17 & $(14; c_1),\ (15; ?)$ & -  \\ \hline
$(D_4^2, 12)$ & $[\SO(9)] \ \ 1 \ \ \underset{[\Sp(1)]}{\overset{\kso(7)}{3}} \ \ \overset{\ksu(2)}{2}$ & 21 & $(15; b_4),\ (16; c_1)$ & -  \\ \hline
$(D_4^2, 13)$ & $[\Sp(1)] \ \ \overset{\kg_2}{3} \ \ 1 \ \ \overset{\kg_2}{3} \ \ [\Sp(1)]$ & 15 & $(18; c_1)$ & -  \\ \hline
$(D_4^2, 14)$ & $[\Sp(2)] \ \ \overset{\kso(7)}{3} \ \ 1 \ \ \overset{\ksu(3)}{3}$ & 16 & $(17; ?),\ (18; c_2),\ (21; ??)$ & -  \\ \hline
$(D_4^2, 15)$ & $[\Sp(3)\times\Sp(1)] \ \ \overset{\kso(7)}{2} \ \ \overset{\ksu(2)}{2} \ \ [N_f = 1/2]$ & 15 & $(21; c_1),\ (22; c_3)$ & -  \\ \hline
$(D_4^2, 16)$ & $[F_4] \ \ 1 \ \ {\overset{\kg_2}{3}} \ \ \overset{\ksu(2)}{2} \ \ [N_f = 1/2]$ & 20 & $(22; f_4)$ & -  \\ \hline
$(D_4^2, 17)$ & $[\SU(5)] \ \ \overset{\ksu(4)}{2} \ \ \overset{\ksu(3)}{2} \ \ [\SU(2)]$ & 15 & $(20; a_4),\ (21; a_1)$ & $(A_3^3, 5)$  \\ \hline
$(D_4^2, 18)$ & $[\Sp(1)] \ \ \overset{\kg_2}{3} \ \ 1 \ \ \overset{\ksu(3)}{3}$ & 14 & $(19; c_1)$ & -  \\ \hline
$(D_4^2, 19)$ & $ \overset{\ksu(3)}{3} \ \ 1 \ \ \overset{\ksu(3)}{3}$ & 13 & $(20; c_2)$ & -  \\ \hline
$(D_4^2, 20)$ & $[\SU(3)] \ \ {\overset{\ksu(3)}{2}}\ \ {\overset{\ksu(3)}{2}}\ \ [\SU(3)]$ & 11 & $(23; a_2)$ & $(A_3^3, 12)$  \\ \hline
$(D_4^2, 21)$ & $[\SU(6)] \ \ {\overset{\ksu(4)}{2}} \ \ \overset{\ksu(2)}{2}$ & 14 & $(23; a_5)$ & $(A_3^3, 11)$  \\ \hline
$(D_4^2, 22)$ & $[\Sp(3)] \ \ {\overset{\kg_2}{2}} \ \ \overset{\ksu(2)}{2} \ \ [N_f = 1/2]$ & 12 & $(23; c_3)$ & -  \\ \hline
$(D_4^2, 23)$ & $[\SU(4)] \ \ {\overset{\ksu(3)}{2}}\ \ {\overset{\ksu(2)}{2}}\ \ [N_f = 1]$ & 9 & $(24; a_3)$ & $(A_3^3, 17)$  \\ \hline
$(D_4^2, 24)$ & $[\SU(2)] \ \ {\overset{\ksu(2)}{2}}\ \ {\overset{\ksu(2)}{2}}\ \ [\SU(2)]$ & 6 & $(25; a_1)$ & $(A_3^3, 22)$  \\ \hline
$(D_4^2, 25)$ & $[G_2] \ \ {\overset{\ksu(2)}{2}} \ \ 2$ & 5 & $(26; g_2)$ & $(A_3^3, 26)$  \\ \hline
$(D_4^2, 26)$ & $2 \ \ 2$ & 2 & $(27,A_1)$ & $(A_3^3,29)$  \\ \hline
$(D_4^3, 27)$ & $2$ & 1 & $(28,A_1)$ & $(A_3^3,31)$ \\ \hline
$(D_4^3, 28)$ & $\varnothing$ & 0 & IR Theory & $(A_3^3,32)$ \\ \hline
\caption{Theories corresponding to the nodes in the Hasse diagram of the rank 2 $(D_4, D_4)$ conformal matter. Endpoint-changing flows can be read off following the caption of Table \ref{D4rank3_table}.}\label{D4rank2_table}
\end{longtable}

To determine the endpoint-changing flows in the $D_4^r$ families for arbitrary $r$, we only need to examine the theories with one end in the form of
\be
    [\SO(8)] \ \ 1 \ \ \overset{\kso(8)}{4} \ \ \dots \ \ 1 \ \ \overset{\kso(8)}{4} \ \ \dots,
\ee
where there are $n$ pieces of the $(-1)\text{-}(-4)$ curve configurations on the tail. Then, all possible endpoint-changing flows amount to removing $k$ pieces of the $(-1)\text{-}(-4)$ curve configurations ($k \leq n$) and adding a disjoint $(2,0)$ theory of $A_{k-1}$ type (which is trivial for $k = 1$). For example, the $(D_4^3, 1)$ theory admits the following hierarchy of the endpoint-changing flows:
\begin{equation}
    \includegraphics[width=10cm]{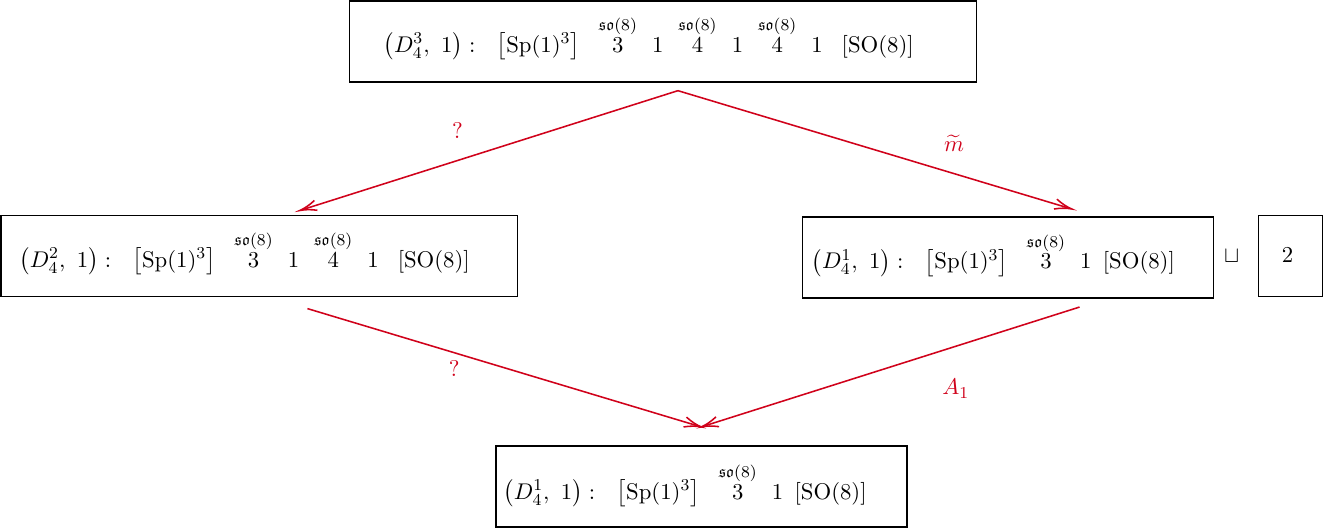}.
\end{equation}
We expect the unknown 1-dimensional slices to be similar to the $A_n$ singularities in \S\ref{nilporbslices} which arise from reducing $\U(k)$ nodes to $\U(k-1)$ nodes in terms of magnetic quivers. Here, one reduces the rank of the $\mathfrak{c}_k$ node ($k=4\rightarrow3\rightarrow2$) with an antisymmetric in the orthosymplectic magnetic quiver. For theories with both ends Higgsed, they do not admit any \textit{minimal} endpoint-changing flows until further Higgsed down to a theory with all $-2$ curves and $\ksu$-type gauge symmetries.

\paragraph{Long quivers} Let us now discuss the long quivers for the $(D_4,D_4)$ conformal matters. It turns out that all the Higgsed theories without the endpoint-changing flows involved belong to the nilpotent hierarchy. Therefore, there is a subdiagram following the structure of the partial ordering of the nilpotent orbits in $\kso(8)$, which can be found for example in \cite[Figure 4]{Heckman:2016ssk}, and we shall not repeat this here. For the endpoint-changing flows, we find that they are the same as the cases for the short quivers. In other words, the complete Hasse diagram can be obtained by adding the theories from the endpoint-changing Higgsings on the ``41\dots41-tails'' as described in the previous paragraph to the nilpotent hierarchy.

Therefore, when the rank $r$ of $D_4$ conformal matter theory goes to infinity, we may say that the Hasse diagram is ``quasi-finite'' in the following sense. The endpoint-preserving flows always produce a finite diagram that coincides with the nilpotent hierarchy. Then the endpoint-changing flows simply reduce the rank $r$ or separate disjoint $(2,0)$ theories of A-type under Higgsings.

\subsubsection{E-Types}\label{Etypes}
We now give a few examples of the E-type conformal matter theories. The simplest example is the rank 0 $(E_6, E_6)$ conformal matter theory. The full Hasse diagram turns out to be identical to the nilpotent hierarchy, which is obtained in \cite{Baume:2021qho} by comparing to the 4d theories. However, this is no longer the case if we either unHiggs the gauge group to the rank 0 $(E_7, E_7)$ conformal matter or increase $r$ to the rank 1 $(E_6, E_6)$ conformal matter. These will be the examples that we explore in this part.

\begin{figure}[H]
    \centering
    \includegraphics[width=0.8\linewidth]{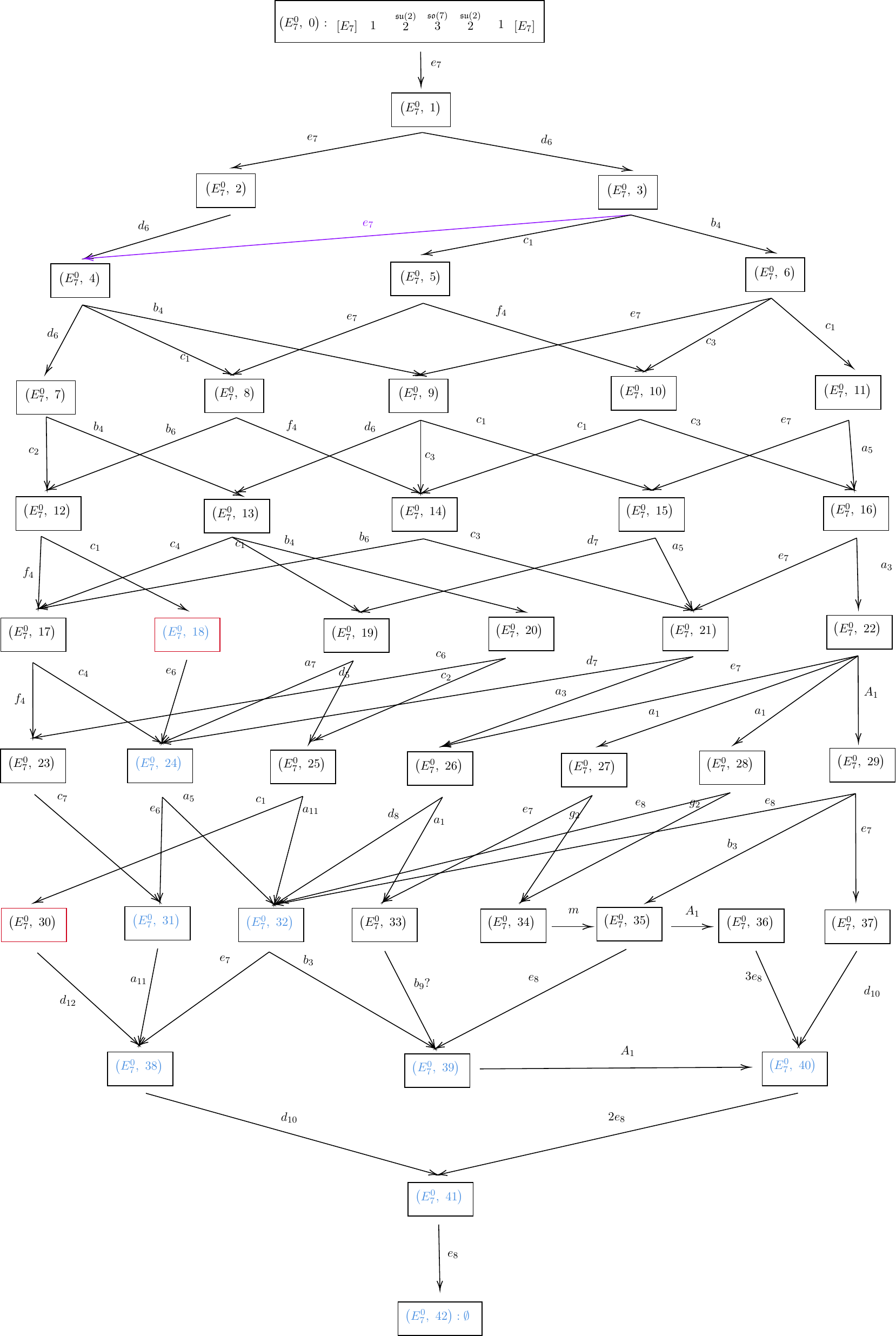}
    \caption{The Hasse diagram of the rank 0 $(E_7,E_7)$ conformal matter. The purple frames (resp.~arrows) indicate the theories (resp.~flows) that are not detected by the $(E_7,E_7)$ nilpotent hierarchy. The blue labels correspond to the Hasse diagram of the rank 0 $(E_6,E_6)$ conformal matter.}\label{E7rank0Hasse}
\end{figure}

\paragraph{Rank 0 $(E_7,E_7)$ conformal matter} We begin by analyzing the rank 0 $(E_7, E_7)$ conformal matter, with the complete Hasse diagram presented in Figure \ref{E7rank0Hasse} and the nodes labelled in Table \ref{E7rank0_table}. As indicated in blue in Figure \ref{E7rank0Hasse}, it contains the whole Hasse diagram for the rank 0 $(E_6,E_6)$ conformal matter as a subdiagram. We again see the unifying power of the atomic Higgsing approach, in that the ``lower" conformal matter theory can always be found in the case of a ``higher'' one. Similarly, without working out the full details of the Hasse diagram of the rank 0 $(E_8, E_8)$ conformal matter, one can identify an explicit path from it down to the rank 0 $(E_7, E_7)$ conformal matter:
\begin{align}
    & \left(E_8^0, 0\right): \quad [E_8] \ \ 1 \ \ 2 \ \ \overset{\ksu(2)}{2} \ \ \overset{\kg_2}{3} \ \ 1 \ \ \overset{\kf_4}{5} \ \ 1 \ \ \overset{\kg_2}{3} \ \ \overset{\ksu(2)}{2}\ \ 2 \ \ 1 \ \ [E_8]  \\
    \xrightarrow{(\calO_L, \calO_R)\ =\ (A_2, A_2)} & \ \  [E_6] \ \ 1 \ \ \overset{\ksu(3)}{3} \ \ 1 \ \ \overset{\kf_4}{5} \ \ 1 \ \ \overset{\ksu(3)}{3} \ \ 1 \ \ [E_6]  \\
    \xrightarrow{~~\quad\text{combo flow}\quad~~} & \ \ \left(E_7^0, 0\right): \quad [E_7] \ \  1 \ \ \overset{\ksu(2)}{2} \ \ \overset{\kso(7)}{3} \ \ \overset{\ksu(2)}{2} \ \ 1 \ \ [E_7].
\end{align}

\begin{longtable}{|c|c|c|c|c|}
\hline
$E_7^0$ label &  Tensor branch description & $d_{\mathbb{H}}$ & (descendant \#; flow type) & ($E_6^0$ label)\\ \hline \hline
$(E_7^0, 0)$ & $[E_7] \ \ 1 \ \ \overset{\ksu(2)}{2} \ \ \overset{\kso(7)}{3} \ \ \overset{\ksu(2)}{2} \ \ 1 \ \ [E_7]$ & 134 & $(1; e_7)$ & - \\ \hline
$(E_7^0, 1)$ & $[E_7] \ \ 1 \ \ \overset{\ksu(2)}{2} \ \ \overset{\kso(7)}{3} \ \ \overset{\ksu(2)}{1} \ \    [\SO(12)]$ & 117 & $(2; e_7),\ (3; d_6)$  & - \\ \hline
$(E_7^0, 2)$ & $[\SO(12)]    \ \ \overset{\ksu(2)}{1} \ \ \overset{\kso(7)}{3} \ \ \overset{\ksu(2)}{2} \ \ 1 \ \ [E_7]$ & 100 & $(4; d_6)$ &  - \\ \hline
$(E_7^0, 3)$ & $[E_7] \ \ 1 \ \ \overset{\ksu(2)}{2} \ \ \underset{[\Sp(1)]}{\overset{\kso(7)}{3}} \ \   1 \ \ [\SO(9)]$ & 108 & $(4; e_7),\ (5; c_1),\ (6; b_4)$  & - \\ \hline
$(E_7^0, 4)$ & $[\SO(12)]    \ \ \overset{\ksu(2)}{1} \ \ \underset{[\Sp(1)]}{\overset{\kso(7)}{3}} \ \  1 \ \ [\SO(9)]$ & 91 & $(7; d_6),\ (8; c_1),\ (9; b_4)$ & - \\ \hline
$(E_7^0, 5)$ & $[E_7] \ \ 1 \ \ \overset{\ksu(2)}{2} \ \ {\overset{\kg_2}{3}} \ \  \ \ 1 \ \ [F_4]$ & 107 & $(8; e_7),\ (10; f_4)$ & - \\ \hline
$(E_7^0, 6)$ & $\underset{[E_7]}{1} \ \ \overset{\ksu(2)}{2} \ \ \underset{[\Sp(3)\times\Sp(1)]}{\overset{\kso(7)}{2}}$ & 102 & $(9; e_7),\ (10; c_3),\ (11; c_1)$ &  - \\ \hline
$(E_7^0, 7)$ & $[\SO(9)] \ \ 1 \ \ \underset{[\Sp(2)]}{\overset{\kso(7)}{3}} \ \ 1 \ \ [\SO(9)]$ & 82 &  $(12; c_2),\ (13; b_4)$ & - \\ \hline
$(E_7^0, 8)$ & $[\SO(13)] \ \ \overset{\ksu(2)}{1} \ \ \overset{\kg_2}{3} \ \  \ \ 1 \ \ [F_4]$ & 90 & $(12; b_6),\ (14; f_4)$ & - \\ \hline
$(E_7^0, 9)$ & $\underset{[\SO(12)]}{\overset{\ksu(2)}{1}} \ \ \underset{[\Sp(3)\times\Sp(1)]}{\overset{\kso(7)}{2}}$ & 85 & $(13; d_6),\ (14; c_3),\ (15; c_1)$ &- \\ \hline
$(E_7^0, 10)$ & $[E_7] \ \ 1 \ \ \overset{\ksu(2)}{2} \ \ \overset{\kg_2}{2} \ \ [\Sp(3)]$ & 99 & $(14; e_7),\ (16; c_3)$ & - \\ \hline
$(E_7^0, 11)$ & $[E_7] \ \ 1 \ \ \overset{\ksu(2)}{2} \ \ \overset{\ksu(4)}{2} \ \ [\SU(6)]$ & 101 & $(15; e_7),\ (16; a_5)$ & - \\ \hline
$(E_7^0, 12)$ & $[F_4] \ \ 1 \ \ \underset{[\Sp(1)]}{\overset{\kg_2}{3}} \ \ 1 \ \ [F_4]$ & 80 & $(17; f_4),\ (18; c_1)$ & - \\ \hline
$(E_7^0, 13)$ & $\underset{[\SO(9)]}{1} \ \ {\overset{\kso(7)}{2}} \ \ \underset{[\Sp(4)\times\Sp(1)]}{1}$ & 76 & $(17; c_4),\ (19; c_1),\ (20; b_4)$ & - \\ \hline
$(E_7^0, 14)$ & $[\SO(13)] \ \  \overset{\ksu(2)}{1} \ \ \overset{\kg_2}{2} \ \ [\Sp(3)]$ & 82 & $(17; b_6),\ (21; c_3)$ & - \\ \hline
$(E_7^0, 15)$ & $[\SO(14)]  \ \ \overset{\ksu(2)}{1} \ \ \overset{\ksu(4)}{2} \ \ [\SU(6)]$ & 84 & $(19; d_7),\ (21; a_6)$ & - \\ \hline
$(E_7^0, 16)$ & $[E_7] \ \ 1 \ \ \overset{\ksu(2)}{2} \ \ \overset{\ksu(3)}{2} \ \ [\SU(4)]$ & 96 & $(21; e_7),\ (22; a_3)$ & - \\ \hline
$(E_7^0, 17)$ & $[F_4] \ \ 1 \ \ {\overset{\kg_2}{2}} \ \ [\Sp(4)]$ & 72 & $(23; f_4),\ (24; c_4)$ & - \\ \hline
$(E_7^0, 18)$ & $[E_6] \ \ 1 \ \ {\overset{\ksu(3)}{3}} \ \ 1 \ \ [E_6]$ & 79 & $(24; e_6)$ & $(E_6^0, 0)$ \\ \hline
$(E_7^0, 19)$ & $[\SO(10)] \ \ 1 \ \ {\overset{\ksu(4)}{2}} \ \ 1 \ \ [\SU(4)]$ & 75 & $(24; a_7),\ (25; d_5)$ & - \\ \hline
$(E_7^0, 20)$ & ${\overset{\kso(7)}{1}} \ \ 1 \ \ [\Sp(6) \times \Sp(2)]$ & 70 & $(23; c_6),\ (25; c_2)$ & - \\ \hline
$(E_7^0, 21)$ & $[\SO(14)] \ \ \overset{\ksu(2)}{1} \ \ \overset{\ksu(3)}{2} \ \ [\SU(4)]$ & 79 & $(24; d_7),\ (26; a_3)$ & - \\ \hline
 &  &  & $(26; e_7),\ (27; a_1)$ &  \\
$(E_7^0, 22)$ & $[E_7] \ \ 1 \ \ \underset{[\SU(2)]}{\overset{\ksu(2)}{2}} \ \ \overset{\ksu(2)}{2} \ \ [\SU(2)]$ & 93 & $(28; a_1),\ (29; m)$ & - \\ \hline
$(E_7^0, 23)$ & $ {\overset{\kg_2}{1}} \ \ [\Sp(7)]$ & 64 & $(31; c_7)$ & - \\ \hline
$(E_7^0, 24)$ & $[E_6] \ \ 1 \ \ {\overset{\ksu(3)}{2}} \ \ [\SU(6)]$ & 68 & $(31; e_6),\ (32; a_5)$ & $(E_6^0, 1)$ \\ \hline
$(E_7^0, 25)$ & ${\overset{\ksu(4)}{1}} \ \ 1 \ \ [\SU(12) \times \Sp(1)]$ & 68 & $(30; a_1),\ (32; a_{11})$ & - \\ \hline
$(E_7^0, 26)$ & $[\SO(16)] \ \  {\overset{\ksu(2)}{1}} \ \ \overset{\ksu(2)}{2} \ \ [\SU(2)]$ & 76 & $(32; d_8),\ (33; a_1)$ & - \\ \hline
$(E_7^0, 27)$ & $[E_7] \ \ 1 \ \ \underset{[G_2]}{\overset{\ksu(2)}{2}} \ \ \overset{I_1}{2} \ \ $ & 92 & $(33; e_7),\ (34; g_2)$ & - \\ \hline
$(E_7^0, 28)$ & $[E_8] \ \ 1 \ \ {\overset{I_1}{2}} \ \ \overset{\ksu(2)}{2} \ \ [G_2]$ & 92 & $(32; e_8),\ (34; g_2)$ & - \\ \hline
$(E_7^0, 29)$ & $[E_7] \ \ 1 \ \ {\overset{\ksu(2)}{2}} [\SO(7)] \ \ \sqcup \ \ 1 \ \ [E_8]$ & 92 & $(32; e_8),\ (35; b_3),\ (37; e_7)$ & - \\ \hline
$(E_7^0, 30)$ & ${\overset{\ksp(2)}{1}} \ \  [\SO(24)]$ & 67 & $(38; d_{12})$ & - \\ \hline
$(E_7^0, 31)$ & ${\overset{\ksu(3)}{1}} \ \  [\SU(12)]$ & 57 & $(38; a_{11})$ & $(E_6^0, 2)$ \\ \hline
$(E_7^0, 32)$ & $[E_7] \ \ 1 \ \ {\overset{\ksu(2)}{2}} \ \ [\SO(7)]$ & 63 & $(38; e_7),\ (39; b_3)$ & $(E_6^0, 3)$ \\ \hline
$(E_7^0, 33)$ & $[\SO(19)]\ \ {\overset{\ksu(2)}{1}} \ \ \overset{I_1}{2} \ \ $ & 75 & $(39; d_{10})$ & - \\ \hline
$(E_7^0, 34)$ & $[E_8] \ \ 1 \ \ {2} \ \ {2} \ \ [\SU(2)] $ & 89 & $(35; m)$ & - \\ \hline
$(E_7^0, 35)$ & $[E_8] \ \ 1 \ \ {2} \ \ [\SU(2)] \ \ \sqcup \ \ 1 \ \ [E_8] $ & 88 & $(36; A_1),\ (39; e_8)$ & - \\ \hline
$(E_7^0, 36)$ & $[E_8] \ \ 1 \ \ \sqcup \ \ 1 \ \ [E_8] \ \ \sqcup \ \ 1 \ \ [E_8] $ & 87 & $(40; 3e_8)$ & - \\ \hline
$(E_7^0, 37)$ & $ {\overset{\ksu(2)}{1}} [\SO(20)] \ \ \sqcup \ \ 1 \ \ [E_8]$ & 75 & $(40; d_{10})$ & - \\ \hline
$(E_7^0, 38)$ & $ {\overset{\ksu(2)}{1}} \ \ [\SO(20)]$ & 46 & $(41; d_{10})$ & $(E_6^0, 4)$ \\ \hline
$(E_7^0, 39)$ & $[E_8] \ \ 1 \ \ 2 \ \ [\SU(2)]$ & 59 & $(40; A_1)$ & $(E_6^0, 5)$ \\ \hline
$(E_7^0, 40)$ & $[E_8] \ \ 1 \ \ \sqcup \ \ 1 \ \ [E_8]$ & 58 & $(41; 2e_8)$ & $(E_6^0, 6)$ \\ \hline
$(E_7^0, 41)$ & $1 \ \ [E_8]$ & 29 & $(42; e_8)$ & $(E_6^0, 7)$ \\ \hline
$(E_7^0, 42)$ & $\varnothing$ & 0 & IR theory & $(E_6^0, 8)$ \\ \hline
\caption{Theories corresponding to the nodes in the Hasse diagram of the rank 0 $(E_7, E_7)$ conformal matter.}\label{E7rank0_table}
\end{longtable}

\begin{figure}[H]
    \centering
    \includegraphics[width=0.6\linewidth]{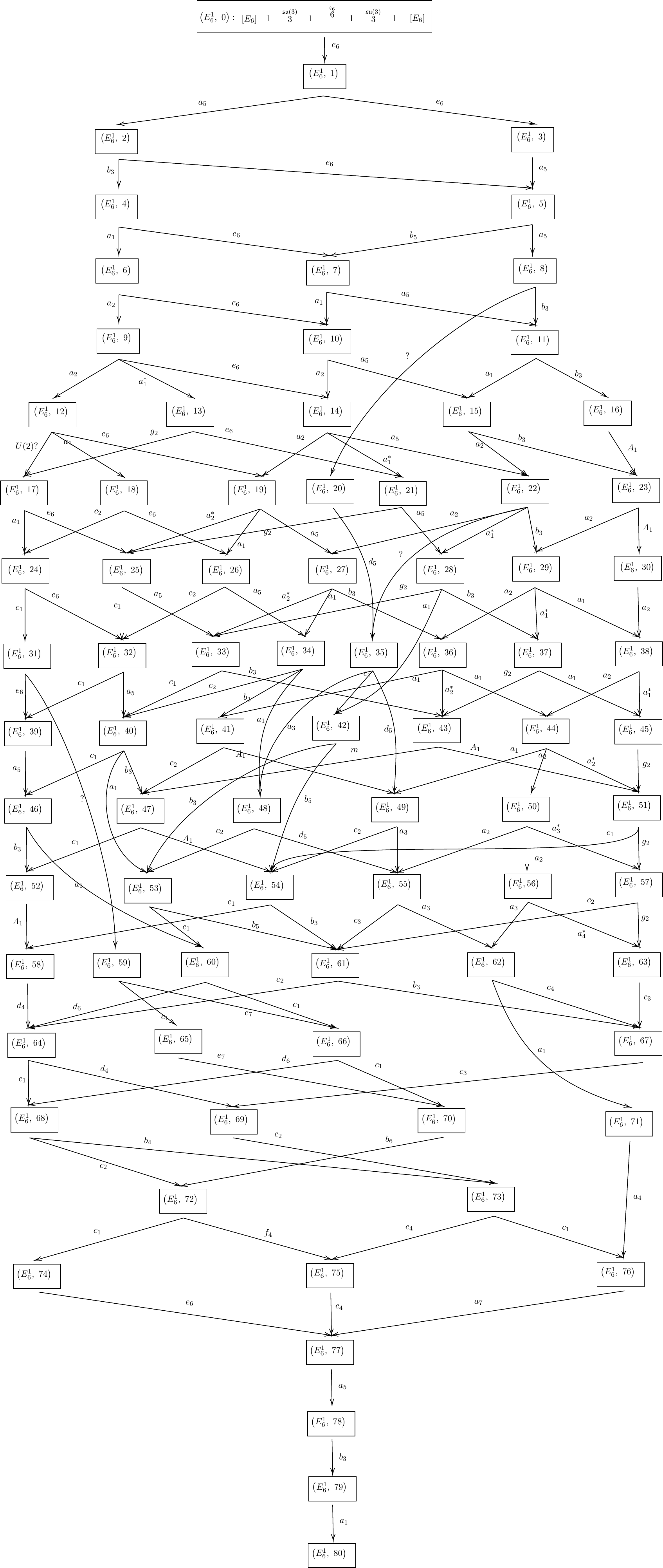}
    \caption{The Hasse diagram of the rank 1 $(E_6,E_6)$ conformal matter. The notation $a_n^*$ of the slices are explained in \S\ref{unknownslices}.}\label{E6rank1Hasse}
\end{figure}

\paragraph{Rank 1 $(E_6,E_6)$ conformal matter} We conclude our exploration of the conformal matter theory by considering an example of higher rank. The simplest such case would be the $(E_6, E_6)$ conformal matter of rank 1. The Hasse diagram is given in Figure \ref{E6rank1Hasse}, with the nodes labelled in Table \ref{E6rank1_table}. One can see that the rank 0 $(D_5, D_5)$ nilpotent hierarchy is embedded in the rank 0 $(E_6, E_6)$ nilpotent hierarchy.

Here, we have some slices denoted as $a_n^*$ of dimension $n$ from Higgsing the $\U(n)$ flavour symmetries. By checking the Hasse diagram of the nilpotent orbits in the $E_6$ case in \cite{fu2017generic}, we may identify some of them as certain known slices. From the Higgsing $(E_6^1,9)\rightarrow(E_6^1,13)$, we learn that $a_1^*$ is the slice $A_2$. From the Higgsing $(E_6^1,12)\rightarrow(E_6^1,17)$, we learn that $a_2^*$ is the slice called $\tau$ in \cite{fu2017generic}.

\begin{longtable}{|c|c|c|c|}
\hline
$E_6^1$ label &  Tensor branch description & $d_{\mathbb{H}}$ & (descendant \#; flow type) \\ \hline \hline
$(E_6^1, 0)$ & $[E_6] \ \ 1 \ \ {\overset{\ksu(3)}{3}} \ \ 1 \ \ \overset{\ke_6}{6} \ \ 1 \ \ {\overset{\ksu(3)}{3}} \ \ 1 \ \ [E_6]$ & 80 & $(1; e_6)$ \\ \hline
$(E_6^1, 1)$ & $[\SU(6)] \ \ {\overset{\ksu(3)}{2}} \ \ 1 \ \ \overset{\ke_6}{6} \ \ 1 \ \ {\overset{\ksu(3)}{3}} \ \ 1 \ \ [E_6]$ & 69  & $(2; a_5),\ (3; e_6)$ \\ \hline
$(E_6^1, 2)$ & $[\SO(7)]  \ \ {\overset{\ksu(2)}{2}} \ \ 1 \ \ \overset{\ke_6}{6} \ \ 1 \ \ {\overset{\ksu(3)}{3}} \ \ 1 \ \ [E_6]$ & 64  & $(4; b_3),\ (5; e_6)$ \\ \hline
$(E_6^1, 3)$ & $[\SU(6)] \ \ {\overset{\ksu(3)}{2}} \ \ 1 \ \ \overset{\ke_6}{6} \ \ 1 \ \ {\overset{\ksu(3)}{2}} \ \ [\SU(6)]$ & 58  & $(5; a_5)$ \\ \hline
$(E_6^1, 4)$ & $[\SU(2)]  \ \ 2 \ \ 1 \ \ \overset{\ke_6}{6} \ \ 1 \ \ {\overset{\ksu(3)}{3}} \ \ 1 \ \ [E_6]$ & 60  & $(6; A_1),\ (7; e_6)$ \\ \hline
$(E_6^1, 5)$ & $[\SO(7)]  \ \ {\overset{\ksu(2)}{2}} \ \ 1 \ \ \overset{\ke_6}{6} \ \ 1 \ \ {\overset{\ksu(3)}{2}} \ \ [\SU(6)]$ & 53  & $(7; b_3),\ (8; a_5)$ \\ \hline
$(E_6^1, 6)$ & $\big([\SU(3)] \ \ 1 \big)^{\otimes 2} \ \ \overset{\ke_6}{6} \ \ 1 \ \ {\overset{\ksu(3)}{3}} \ \ 1 \ \ [E_6]$ & 59  & $(9; [2a_2]^+),\ (10; e_6)$ \\ \hline
$(E_6^1, 7)$ & $[\SU(2)]  \ \ 2 \ \ 1 \ \ \overset{\ke_6}{6} \ \ 1 \ \ {\overset{\ksu(3)}{2}} \ \ [\SU(6)]$ & 49  & $(10; a_1),\ (11; a_5)$ \\ \hline
$(E_6^1, 8)$ & $[\SO(7)]  \ \ {\overset{\ksu(2)}{2}} \ \ 1 \ \ \overset{\ke_6}{6} \ \ 1 \ \ {\overset{\ksu(2)}{2}} \ \ [\SO(7)]$ & 48  & $(11; b_3),\ (20; E_6)$ \\ \hline
$(E_6^1, 9)$ & $[\SU(3)] \ \ 1  \ \ \underset{[\U(1)]}{\overset{\ke_6}{5}} \ \ 1 \ \ {\overset{\ksu(3)}{3}} \ \ 1 \ \ [E_6]$ & 59  & $(12; a_2),\ (13; a_1^*),\ (14; e_6)$ \\ \hline
$(E_6^1, 10)$ & $\big([\SU(3)] \ \ 1 \big)^{\otimes 2} \ \ \overset{\ke_6}{6} \ \ 1 \ \ {\overset{\ksu(3)}{2}} \ \ [\SU(6)]$ & 48  & $(14; a_2),\ (15; a_5)$ \\ \hline
$(E_6^1, 11)$ & $[\SU(2)]  \ \ 2 \ \ 1 \ \ \overset{\ke_6}{6} \ \ 1 \ \ {\overset{\ksu(2)}{2}} \ \ [\SO(7)]$ & 44  & $(15; a_1),\ (16; b_3)$ \\ \hline
$(E_6^1, 12)$ & $[\U(2)] \ \ {\overset{\ke_6}{4}} \ \ 1 \ \ {\overset{\ksu(3)}{3}} \ \ 1 \ \ [E_6]$ & 55 & $(17; a_2^*),\ (18; a_1),\ (19; e_6)$ \\ \hline
$(E_6^1, 13)$ & $[G_2] \ \ 1  \ \ {\overset{\kf_4}{5}} \ \ 1 \ \ {\overset{\ksu(3)}{3}} \ \ 1 \ \ [E_6]$ & 56  & $(17; g_2),\ (21; e_6)$ \\ \hline
$(E_6^1, 14)$ & $[\SU(3)] \ \ 1  \ \ \underset{[\U(1)]}{\overset{\ke_6}{5}} \ \ 1 \ \ {\overset{\ksu(3)}{2}} \ \ [\SU(6)]$ & 46  & $(19; a_2),\ (21; a_1^*),\ (22; a_5)$ \\ \hline
$(E_6^1, 15)$ & $[\SU(3)] \ \ 1  \ \ \underset{[\U(1)]}{\overset{\ke_6}{5}} \ \ 1 \ \ {\overset{\ksu(2)}{2}} \ \ [\SO(7)]$ & 43  & $(22; a_2),\ (23; b_3)$ \\ \hline
$(E_6^1, 16)$ & $[\SU(2)]  \ \ 2 \ \ 1 \ \ \overset{\ke_6}{6} \ \ 1 \ \ 2 \ \ [\SU(2)]$ & 40  & $(23; A_1)$ \\ \hline
$(E_6^1, 17)$ & $[\Sp(1)] \ \ {\overset{\kf_4}{4}} \ \ 1 \ \ {\overset{\ksu(3)}{3}} \ \ 1 \ \ [E_6]$ & 53 & $(24; a_1),\ (25; e_6)$ \\ \hline
$(E_6^1, 18)$ & $[\Sp(2)] \ \ {\overset{\kso(10)}{4}} \ \ 1 \ \ {\overset{\ksu(3)}{3}} \ \ 1 \ \ [E_6]$ & 54 & $(24; c_2),\ (26; e_6)$ \\ \hline
$(E_6^1, 19)$ & $[\U(2)] \ \ {\overset{\ke_6}{4}} \ \ 1 \ \ \ \ {\overset{\ksu(3)}{2}} \ \ [\SU(6)]$ & 44 & $(25; a_2^*),\ (26; a_1),\ (27; a_5)$ \\ \hline
$(E_6^1, 20)$ & $[\SO(10)] \ \ \overset{\ksp(1)}{1} \ \ \overset{\kso(10)}{4} \ \ \overset{\ksp(1)}{1} \ \ [\SO(10)]$ & 47 & $(35; d_5)$ \\ \hline
$(E_6^1, 21)$ & $[G_2] \ \ 1  \ \ {\overset{\kf_4}{5}} \ \ 1 \ \ {\overset{\ksu(3)}{2}} \ \ [\SU(6)]$ & 45  & $(25; g_2),\ (28; a_5)$ \\ \hline
$(E_6^1, 22)$ & $[\SU(3)] \ \ 1  \ \ \underset{[\U(1)]}{\overset{\ke_6}{5}} \ \ 1 \ \ {\overset{\ksu(2)}{2}} \ \ [\SO(7)]$ & 41  & $(27; a_2),\ (28; a_1^*),\ (29; b_3),\ (35;?)$ \\ \hline
$(E_6^1, 23)$ & $\big([\SU(3)] \ \ 1 \big)^{\otimes 2} \ \ \overset{\ke_6}{6} \ \ 1 \ \ 2 \ \ [\SU(2)]$ & 39  & $(29; a_1),\ (30; A_1)$ \\ \hline
$(E_6^1, 24)$ & $[\Sp(1)] \ \ \overset{\kso(9)}{4} \ \ 1 \ \ {\overset{\ksu(3)}{3}} \ \ 1 \ \ [E_6]$ & 52 & $(31; c_1),\ (32; e_6)$ \\ \hline
$(E_6^1, 25)$ & $[\Sp(1)] \ \ \overset{\kf_4}{4} \ \ 1 \ \ {\overset{\ksu(3)}{2}} \ \ [\SU(6)]$ & 42 & $(31; c_1),\ (33; a_5)$ \\ \hline
$(E_6^1, 26)$ & $[\Sp(2)] \ \ \overset{\kso(10)}{4} \ \ \underset{[\U(1)]}{1} \ \ \ \ {\overset{\ksu(3)}{2}} \ \ [\SU(6)]$ & 43 & $(32; c_2),\ (34; a_5)$ \\ \hline
$(E_6^1, 27)$ & $[\U(2)] \ \ {\overset{\ke_6}{4}} \ \ 1 \ \ \ \ {\overset{\ksu(2)}{2}} \ \ [\SO(7)]$ & 39 & $(33; a_2^*),\ (34; a_1),\ (35; b_3)$ \\ \hline
$(E_6^1, 28)$ & $[G_2] \ \ 1  \ \ \overset{\kf_4}{5} \ \ \underset{[\SU(2)]}{1} \ \ {\overset{\ksu(2)}{2}} \ \ [\SO(7)]$ & 40  & $(33; g_2),\ (37; b_3),\ (42; a_1)$ \\ \hline
$(E_6^1, 29)$ & $[\SU(3)] \ \ 1  \ \ \underset{[\U(1)]}{\overset{\ke_6}{5}} \ \ 1 \ \ 2 \ \ [\SU(2)]$ & 37  & $(36; a_2),\ (37; a_1^*),\ (38; a_1)$ \\ \hline
$(E_6^1, 30)$ & $\big([\SU(3)] \ \ 1 \big)^{\otimes 4} \ \ \overset{\ke_6}{6} $ & 38  & $(38; a_2)$ \\ \hline
$(E_6^1, 31)$ & $\overset{\kso(8)}{4} \ \ 1 \ \ {\overset{\ksu(3)}{3}} \ \ 1 \ \ [E_6]$ & 51 & $(39; e_6),\ (59; D_4)$ \\ \hline
$(E_6^1, 32)$ & $[\Sp(1)] \ \ \overset{\kso(9)}{4} \ \ 1 \ \ {\overset{\ksu(3)}{2}} \ \ [\SU(6)]$ & 41 & $(39; c_1),\ (40; a_5)$ \\ \hline
$(E_6^1, 33)$ & $[\Sp(1)] \ \ {\overset{\kf_4}{4}} \ \ 1 \ \ \ \ {\overset{\ksu(2)}{2}} \ \ [\SO(7)]$ & 37 & $(40; c_1),\ (43; b_3)$ \\ \hline
$(E_6^1, 34)$ & $[\Sp(2)] \ \ {\overset{\kso(10)}{4}} \ \ \underset{[\SU(2)]}{1} \ \ \ \ {\overset{\ksu(2)}{2}} \ \ [\SO(7)]$ & 38 & $(40; c_2),\ (41; b_3),\ (48; a_1)$ \\ \hline
$(E_6^1, 35)$ & $[\SU(4)] \ \ 1 \ \ \underset{[\Sp(1)]}{\overset{\kso(10)}{4}} \ \ \overset{\ksp(1)}{1} \ \ [\SO(10)]$ & 40 & $(42; c_1),\ (48; a_3),\ (49; d_5)$ \\ \hline
$(E_6^1, 36)$ & $[\U(2)]\ \  \overset{\ke_6}{4} \ \ 1 \ \ 2 \ \ [\SU(2)]$ & 35  & $(41; a_1),\ (43; a_2^*),\ (44; a_1)$ \\ \hline
$(E_6^1, 37)$ & $[G_2] \ \ 1  \ \ \overset{\kf_4}{5} \ \ 1 \ \ 2 \ \ [\SU(2)]$ & 36  & $(43; g_2),\ (45; a_1)$ \\ \hline
$(E_6^1, 38)$ & $\big([\SU(3)] \ \ 1 \big)^{\otimes 3} \ \ \overset{\ke_6}{5} \ \ [\U(1)] $ & 36  & $(44; a_2),\ (45; a_1^*)$ \\ \hline
$(E_6^1, 39)$ & $\overset{\kso(8)}{4} \ \ 1 \ \ {\overset{\ksu(3)}{2}} \ \ [\SU(6)]$ & 40 & $(46; a_5)$ \\ \hline
$(E_6^1, 40)$ & $[\Sp(1)] \ \ \overset{\kso(9)}{4} \ \ \underset{[\SU(2)^2]}{1} \ \ {\overset{\ksu(2)}{2}} \ \ [\SO(7)]$ & 36 & $(46; c_1),\ (47; b_3),\ (53; a_1)$ \\ \hline
$(E_6^1, 41)$ & $[\Sp(2)] \ \ {\overset{\kso(10)}{4}} \ \ 1 \ \ \ \ 2 \ \ [\SU(2)]$ & 34 & $(47; c_2),\ (49; A_1)$ \\ \hline
$(E_6^1, 42)$ & $[\SO(7)] \ \ 1 \ \ \overset{\kso(9)}{4} \ \ \overset{\ksp(1)}{1} \ \ [\SO(11)]$ & 39 & $(53; b_3),\ (54; b_5)$ \\ \hline
$(E_6^1, 43)$ & $[\Sp(1)]\ \  \overset{\kf_4}{4} \ \ 1 \ \ 2 \ \ [\SU(2)]$ & 33  & $(47; m),\ (51; A_1)$ \\ \hline
$(E_6^1, 44)$ & $\big([\SU(3)] \ \ 1 \big)^{\otimes 2}\ \  \overset{\ke_6}{4}\ \ [\U(2)]$ & 34  & $(49; a_1),\ (50; a_2),\ (51; a_2^*)$ \\ \hline
$(E_6^1, 45)$ & $\big([G_2] \ \ 1 \big)^{\otimes 3} \ \ \overset{\kf_4}{5} $ & 35  & $(51; g_2)$ \\ \hline
$(E_6^1, 46)$ & $\overset{\kso(8)}{4} \ \ \underset{[\SU(2)^3]}{1} \ \ {\overset{\ksu(2)}{2}} \ \ [\SO(7)]$ & 35 & $(52; b_3),\ (60; a_1)$ \\ \hline
$(E_6^1, 47)$ & $[\Sp(1)] \ \ \overset{\kso(9)}{4} \ \ 1 \ \ 2 \ \ [\SU(2)]$ & 32 & $(52; c_1)\ (54; A_1)$ \\ \hline
$(E_6^1, 48)$ & $[\Sp(2) \times \U(1)] \ \ {\overset{\kso(10)}{3}} \ \ {\overset{\ksp(1)}{1}} \ \ [\SO(10)]$ & 37 & $(53; c_2),\ (55; d_5)$ \\ \hline
$(E_6^1, 49)$ & $[\SU(4)] \ \ 1 \ \ \underset{[\Sp(2)]}{\overset{\kso(10)}{4}} \ \ 1 \ \  [\SU(4)]$ & 33 & $(54; c_2),\ (55; a_3)$ \\ \hline
$(E_6^1, 50)$ & $[\SU(3)] \ \ 1 \ \  \overset{\ke_6}{3}\ \ [\U(3)]$ & 32  & $(55; a_2),\ (56; a_2),\ (57; a_3^*)$ \\ \hline
$(E_6^1, 51)$ & $\big([G_2] \ \ 1 \big)^{\otimes 2}\ \  \overset{\kf_4}{4}\ \ [\Sp(1)]$ & 33  & $(54; c_1),\ (57; g_2)$ \\ \hline
$(E_6^1, 52)$ & $\overset{\kso(8)}{4} \ \ 1 \ \ 2  \ \ [\SU(2)]$ & 31 & $(58; A_1)$ \\ \hline
$(E_6^1, 53)$ & $[\Sp(1) \times \Sp(1)] \ \ {\overset{\kso(9)}{3}} \ \ {\overset{\ksp(1)}{1}} \ \ [\SO(11)]$ & 35 & $(60; c_1),\ (61; b_5)$ \\ \hline
$(E_6^1, 54)$ & $[\SO(7)] \ \ 1 \ \ \underset{[\Sp(1)]}{\overset{\kso(9)}{4}} \ \ 1 \ \  [\SO(7)]$ & 31 & $(58; c_1),\ (61; b_3)$ \\ \hline
$(E_6^1, 55)$ & $[\SU(4)] \ \ 1 \ \ {\overset{\kso(10)}{3}} \ \ [\Sp(3) \times \U(1)]$ & 30 & $(61; c_3),\ (62; a_2)$ \\ \hline
$(E_6^1, 56)$ & $ \overset{\ke_6}{2}\ \ [\U(4)]$ & 30  & $(62; a_3),\ (63; a_4^*)$ \\ \hline
$(E_6^1, 57)$ & $[G_2] \ \ 1 \ \  \overset{\kf_4}{3}\ \ [\Sp(2)]$ & 29  & $(61; c_2),\ (63; g_2)$ \\ \hline
$(E_6^1, 58)$ & $[\SO(8)] \ \ 1 \ \ \overset{\kso(8)}{4} \ \ 1 \ \  [\SO(8)]$ & 30 & $(64; d_4)$ \\ \hline
$(E_6^1, 59)$ & $[\Sp(1)] \ \ \overset{\kso(7)}{3} \ \ {\overset{\ksu(2)}{2}} \ \ 1 \ \ [E_7]$ & 50 & $(65; c_1),\ (66; e_7)$ \\ \hline
$(E_6^1, 60)$ & $[\Sp(1) \times \Sp(1)] \ \ {\overset{\kso(8)}{3}} \ \ {\overset{\ksp(1)}{1}} \ \ [\SO(12)]$ & 34 & $(64; d_6),\ (66; c_1)$ \\ \hline
$(E_6^1, 61)$ & $[\SU(4)] \ \ 1 \ \ {\overset{\kso(9)}{3}} \ \ [\Sp(2) \times \Sp(1)]$ & 27 & $(64; c_2),\ (67; b_3)$ \\ \hline
$(E_6^1, 62)$ & $\overset{\kso(10)}{2} \ \ [\Sp(4) \times \U(2)]$ & 27 & $(67; c_4),\ (71; a_1)$ \\ \hline
$(E_6^1, 63)$ & $ \overset{\kf_4}{2}\ \ [\Sp(3)]$ & 26  & $(67; c_3)$ \\ \hline
$(E_6^1, 64)$ & $[\Sp(1)^3] \ \ \overset{\kso(8)}{3} \ \ 1 \ \  [\SO(8)]$ & 25 & $(68; c_1),\ (69; d_4)$ \\ \hline
$(E_6^1, 65)$ & $\overset{\kg_2}{3} \ \ \overset{\ksu(2)}{2} \ \ 1 \ \ [E_7]$ & 49 & $(70; e_7)$ \\ \hline
$(E_6^1, 66)$ & $[\Sp(1)] \ \ {\overset{\kso(7)}{3}} \ \ {\overset{\ksp(1)}{1}} \ \ [\SO(12)]$ & 33 & $(68; d_6),\ (70; c_1)$ \\ \hline
$(E_6^1, 67)$ & $\overset{\kso(9)}{2} \ \ [\Sp(3) \times \Sp(2)]$ & 23 & $(67; c_4),\ (71; a_1)$ \\ \hline
$(E_6^1, 68)$ & $[\Sp(2)] \ \ \overset{\kso(7)}{3} \ \ 1 \ \  [\SO(9)]$ & 24 & $(72; c_2),\ (73; b_4)$ \\ \hline
$(E_6^1, 69)$ & $\overset{\kso(8)}{2}\ \ [\Sp(2)^3] $ & 20 & $(73; c_2)$ \\ \hline
$(E_6^1, 70)$ & ${\overset{\kg_2}{3}} \ \ {\overset{\ksp(1)}{1}} \ \ [\SO(13)]$ & 32 & $(72; b_6)$ \\ \hline
$(E_6^1, 71)$ & $\overset{\ksu(5)}{2} \ \ [\SU(10)]$ & 26 & $(76; a_4)$ \\ \hline
$(E_6^1, 72)$ & $[\Sp(1)] \ \ \overset{\kg_2}{3} \ \ 1 \ \  [F_4]$ & 22 & $(74; c_1),\ (75; f_4)$ \\ \hline
$(E_6^1, 73)$ & $[\Sp(4) \times \Sp(1)] \ \ \overset{\kso(7)}{2} $ & 18 & $(75; c_4),\ (76; c_1)$ \\ \hline
$(E_6^1, 74)$ & $\overset{\ksu(3)}{3} \ \ 1 \ \  [E_6]$ & 21 & $(77; e_6)$ \\ \hline
$(E_6^1, 75)$ & $\overset{\kg_2}{2} \ \ [\Sp(4)]$ & 14 & $(77; c_4)$ \\ \hline
$(E_6^1, 76)$ & $\overset{\ksu(4)}{2} \ \ [\SU(8)]$ & 17 & $(77; a_7)$ \\ \hline
$(E_6^1, 77)$ & $\overset{\ksu(3)}{2} \ \ [\SU(6)]$ & 10 & $(78; a_5)$ \\ \hline
$(E_6^1, 78)$ & $\overset{\ksu(2)}{2} \ \ [\SO(7)]$ & 5 & $(79; b_3)$ \\ \hline
$(E_6^1, 79)$ & $2$ & 1 & $(80; a_1)$ \\ \hline
$(E_6^1, 80)$ & $\varnothing$ & 0 & IR Theory \\ \hline
\caption{Theories corresponding to the nodes in the Hasse diagram of the rank 1 $(E_6, E_6)$ conformal matter.}\label{E6rank1_table}
\end{longtable}

\subsection{Orbi-Instanton Theories}\label{orbi-instanton}
In this subsection, we shall consider the orbi-instanton theories. For an A-type orbi-instanton theory engineered by $k$ M5 brane probing an M9 wall, the Hasse diagram can be completely obtained by the magnetic quiver techniques (see \cite{Bourget:2024mgn} for the $k=4$ example). Here, we demonstrate that our approach also reproduces the same result by working out the case of $k = 5$,
\be
    [E_8] \ \ 1 \ \ 2 \ \ \overset{\ksu(2)}{2} \ \ \overset{\ksu(3)}{2} \ \ \overset{\ksu(4)}{2} \ \ [\SU(5)],
\ee
in full detail, which can be compared with the magnetic quiver description in \S\ref{orbiinstantonMQs}.

However, our approach applies not only to the A-type but also to the DE-type orbi-instanton theories. To illustrate this point, we work out the Hasse diagram starting from a descendant theory of the D-type orbi-instanton:
\begin{align}
    [\SO(8)] \ \ 1 \ \ &\overset{\kso(8)}{4} \ \ 1 \ \ [\SO(8)]\nonumber\\
    \widehat{D}_4^1:\quad\quad\quad\quad\quad\quad\quad&\ \ 1\\ 
    [&\SO(8)]\nonumber.
\end{align}
This theory can be seen by taking the rank 1 $(D_4, D_4)$ conformal matter (``141'') and affinizing it via sticking another $-1$ curve to the $-4$ curve. This is why we label it by adding a hat to $D_4^1$. This $\widehat{D}_4^1$ theory can be obtained by performing various Higgsings on the D-type orbi-instanton theory:
\be
    [E_8] \ \ 1 \ \ 2 \ \ \overset{\ksu(2)}{2} \ \ \overset{\ksu(3)}{3} \ \ 1 \ \ \overset{\kso(8)}{4} \ \ 1 \ \ [\SO(8)] \label{eqn:D4_orbi_instanton},
\ee
and we choose to the $\widehat{D}_4^1$ theory for simplicity. In \S\ref{sec:D4hatrank1MQ}, we will give the magnetic quivers for this $\widehat{D}_4^1$ theory and some of its descendants, and verify them to exhibit decay-type flows whenever the orthosymplectic magnetic quivers for the parent theory and the child theory are both applicable.

\subsubsection{The A-Type Orbi-Instanton Theories}\label{Aorbi-instanton}
\begin{figure}[H]
    \centering
    \includegraphics[width=1.3\linewidth]{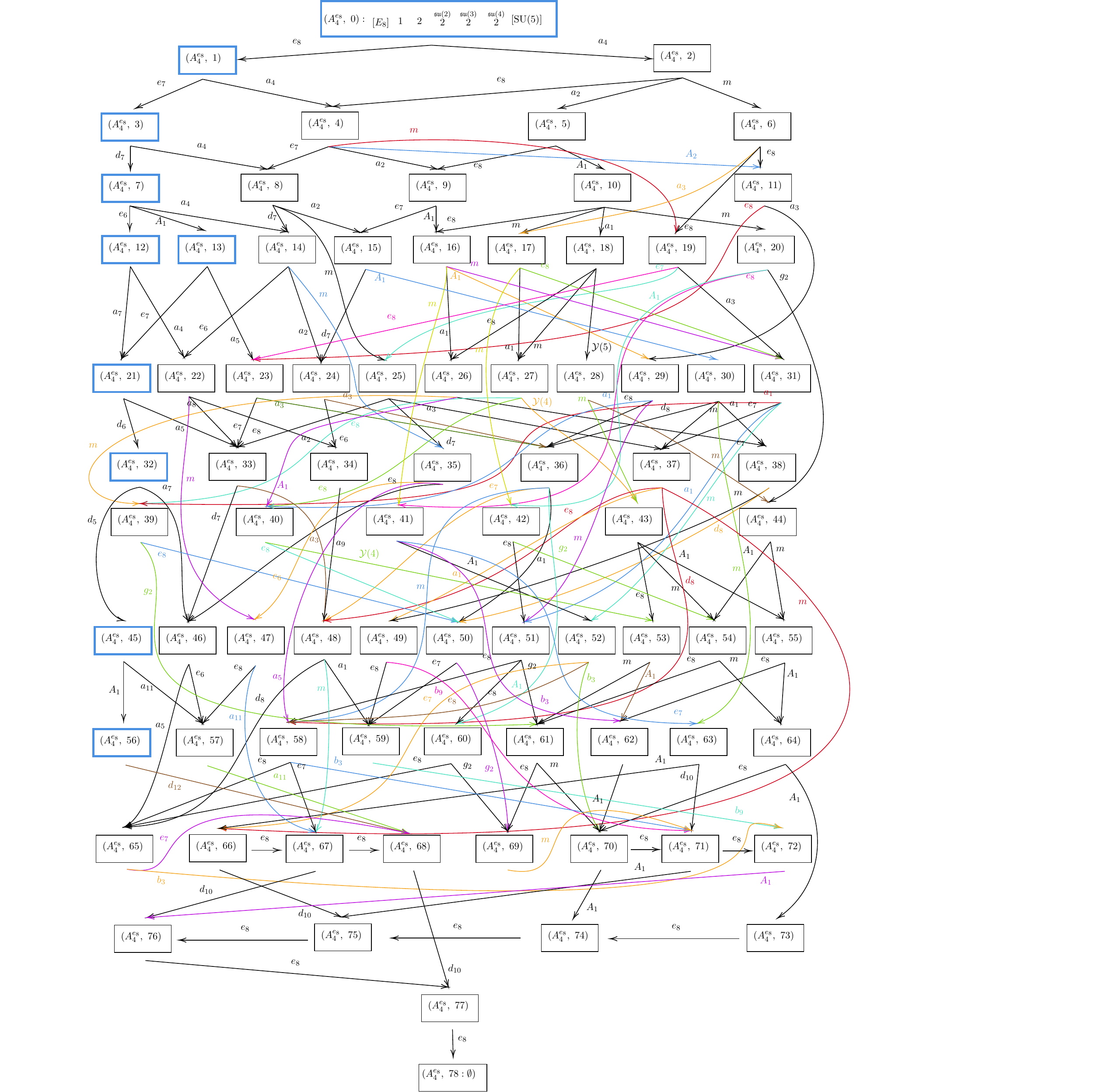}
    \caption{Complete Hasse diagram of the A-type orbi-instanton theories with flavour symmetry $E_8\times\SU(5)$. The nodes with all the flows are catalogued in Table \ref{A4_orbi-instanton_table}.}\label{A4orbi-instantonHasse}
\end{figure}

Let us start with the Higgsings of an A-type orbi-instanton theory. Such theories have known unitary magnetic quiver descriptions that match the decay and fission algorithms available. Therefore, it is an ideal place to cross-validate our algorithm on the 6d tensor branches. In \cite{Bourget:2024mgn}, the Higgsings of the A-type orbi-instanton theory with flavour symmetry $E_8\times\SU(4)$ was performed. Here, we would increase one tensor so that the flavour symmetry becomes $E_8\times\SU(5)$. The full Hasse diagram is given in Figure \ref{A4orbi-instantonHasse}, with the nodes and the RG flows catalogued in Table \ref{A4_orbi-instanton_table}. One may check that this agrees with the result from the decay and fission algorithm for the magnetic quivers in \S\ref{orbiinstantonMQs}.

Notice that our Hasse diagram automatically includes all theories that can be obtained by performing a $\bbZ_k \rightarrow E_8$ discrete homomorphism Higgsing from the orbi-instanton theories. Such theories are labelled with blue frames in Figure \ref{A4orbi-instantonHasse}.

\paragraph{Affinizations} It is worth remarking that the orbi-instanton theories can be related to the theories with A-type $(2,0)$ endpoints by (de)affinizations. More concretely, taking a fission-type flow in the small instanton family and removing all the $-1$ curves, one would get another endpoint-changing flow among theories with $(2,0)$ endpoints. Conversely, for each endpoint-changing flow among theories with $(2,0)$ endpoints, it is always possible to decorate them so that one gets a fission-type flow among heterotic strings.

As an example, the above procedure relates the following flow from $(A_4^{e_8}; 10)$ to $(A_4^{e_8}, 20)$:
\be
    \underset{[E_8]}{1} \ \ 2 \ \ \overset{\ksu(2)}{2} \ \ {\overset{\ksu(2)}{2}} \ \ \underset{[\SU(2) \times \SU(2)_L]}{\overset{\ksu(2)}{2}} \ \ \rightarrow \ \ [E_8] \ \ 1 \ \ 2 \ \ \underset{[G_2]}{\overset{\ksu(2)}{2}} \ \ \sqcup \ \ \underset{[E_8]}{1} \ \ 2.
\ee
If we remove all the $-1$ curves from the above theories, we can get another endpoint-changing flow:
\be
    2 \ \ \overset{\ksu(2)}{2} \ \ {\overset{\ksu(2)}{2}} \ \ \underset{[\SU(2) \times \SU(2)_L]}{\overset{\ksu(2)}{2}} \ \ \rightarrow \ \  \ \ 2 \ \ {\overset{\ksu(2)}{2}} \ \ [G_2] \ \ \sqcup  \ \ 2.
\ee
In this way, we can take our knowledge of all the endpoint-changing flows to understand the fission-type flows for theories without endpoints.

\begin{longtable}{|c|c|c|c|}
\hline
$A_4^{e_8}$ label &  Tensor branch description & $d_{\mathbb{H}}$ & (descendant \#; flow type) \\ \hline \hline
$(A_4^{e_8}, 0)$ & $[E_8] \ \ 1 \ \ 2 \ \ \overset{\ksu(2)}{2} \ \ \overset{\ksu(3)}{2} \ \ \overset{\ksu(4)}{2} \ \ [\SU(5)]$ & 159 & $(1; e_8),\ (2; a_4)$ \\ \hline
$(A_4^{e_8}, 1)$ & $[E_7] \ \ 1 \ \ \underset{[N_f = 1]}{\overset{\ksu(2)}{2}} \ \ \overset{\ksu(3)}{2} \ \ \overset{\ksu(4)}{2} \ \ [\SU(5)]$ & 130 & $(3; e_7),\ (4; a_4)$ \\ \hline
$(A_4^{e_8}, 2)$ & $[E_8] \ \ 1 \ \ 2 \ \ \overset{\ksu(2)}{2} \ \ \underset{[N_f = 1]}{\overset{\ksu(3)}{2}} \ \ \overset{\ksu(3)}{2} \ \ [\SU(3)]$ & 155 & $(4; e_8),\ (5; a_2),\ (6; m)$ \\ \hline
$(A_4^{e_8}, 3)$ & $[\SO(14)] \ \ \overset{\ksu(2)}{1} \ \ \overset{\ksu(3)}{2} \ \ \overset{\ksu(4)}{2} \ \ [\SU(5)]$ & 113 & $(7; d_7),\ (8; a_4)$ \\ \hline
$(A_4^{e_8}, 4)$ & $[E_7] \ \ 1 \ \ \overset{\ksu(2)}{2} \ \ \underset{[N_f = 1]}{\overset{\ksu(3)}{2}} \ \ \overset{\ksu(3)}{2} \ \ [\SU(3)]$ & 126 & $(8; e_7),\ (9; a_2),\ (11; A_2),\ (19, m)$ \\ \hline
$(A_4^{e_8}, 5)$ & $[E_8] \ \ 1 \ \ 2 \ \ \overset{\ksu(2)}{2} \ \ \underset{[\SU(2)]}{\overset{\ksu(3)}{2}} \ \ \overset{\ksu(2)}{2} \ \ [N_f = 1]$ & 153 & $(9; e_8),\ (10; A_1)$ \\ \hline
$(A_4^{e_8}, 6)$ & $[E_8] \ \ 1 \ \ 2 \ \ \underset{[N_f = 1/2]}{\overset{\ksu(2)}{2}} \ \ \underset{[\SU(4)]}{\overset{\ksu(3)}{2}} \ \ \sqcup \ \ \underset{[E_8]}{1}$ & 154 & $(11; e_8),\ (17; a_3),\ (19, e_8)$ \\ \hline
$(A_4^{e_8}, 7)$ & $[E_6] \ \ 1 \ \ \underset{[\SU(2)]}{\overset{\ksu(3)}{2}} \ \ \overset{\ksu(4)}{2} \ \ [\SU(5)]$ & 102 & $(12; e_6),\ (13; A_1),\ (14; a_4)$ \\ \hline
$(A_4^{e_8}, 8)$ & $[\SO(14)] \ \ \overset{\ksu(2)}{1} \ \ \underset{[N_f = 1]}{\overset{\ksu(3)}{2}} \ \ \overset{\ksu(3)}{2} \ \ [\SU(3)]$  & 109 & $(14; d_7),\ (15; a_2),\ (25; m)$ \\ \hline
$(A_4^{e_8}, 9)$ & $[E_7] \ \ 1 \ \ \underset{[N_f = 1]}{\overset{\ksu(2)}{2}} \ \ \underset{[\SU(3)]}{\overset{\ksu(3)}{2}} \ \ \overset{\ksu(2)}{2} \ \ [N_f = 1]$ & 124 & $(15; e_7),\ (16; A_1)$ \\ \hline
$(A_4^{e_8}, 10)$ & $\underset{[E_8]}{1} \ \ 2 \ \ \overset{\ksu(2)}{2} \ \ {\overset{\ksu(2)}{2}} \ \ \underset{[\SU(2)]}{\overset{\ksu(2)}{2}}$ & 152 & $(16; e_8),\ (17; m),\ (18; a_1),\ (20, m)$ \\ \hline
$(A_4^{e_8}, 11)$ & $[E_8] \ \ 1 \ \ 2 \ \ \underset{[N_f = 1/2]}{\overset{\ksu(2)}{2}} \ \ \overset{\ksu(3)}{2} \ \ [\SU(4)]$ & 125 & $(23; e_8),\ (29; a_3)$ \\ \hline
$(A_4^{e_8}, 12)$ & $[\SU(8)] \ \ {\overset{\ksu(3)}{1}} \ \ \overset{\ksu(4)}{2} \ \ [\SU(5)]$ & 91 & $(21; a_7),\ (22; a_4)$ \\ \hline
$(A_4^{e_8}, 13)$ & $[E_6] \ \ 1 \ \ \overset{\ksu(2)}{2} \ \ \overset{\ksu(4)}{2} \ \ [\SU(6)]$ & 101 & $(21; e_7),\ (23; a_5)$ \\ \hline
$(A_4^{e_8}, 14)$ & $[\SO(10)] \ \ 1 \ \ \underset{[\SU(3)]}{\overset{\ksu(3)}{2}} \ \ \overset{\ksu(3)}{2} \ \ [\SU(3)]$ & 98 & $(22; e_6),\ (24; a_2),\ (35; m)$ \\ \hline
$(A_4^{e_8}, 15)$ & $[\SO(14)] \ \ \overset{\ksu(2)}{1} \ \ \underset{[\SU(2)]}{\overset{\ksu(3)}{2}} \ \ \overset{\ksu(2)}{2} \ \ [N_f = 1]$ & 107 & $(24; d_7),\ (30; A_1)$ \\ \hline
$(A_4^{e_8}, 16)$ & $\underset{[E_7]}{1} \ \ \underset{[\SU(2)]}{\overset{\ksu(2)}{2}} \ \ {\overset{\ksu(2)}{2}} \ \ \underset{[\SU(2)]}{\overset{\ksu(2)}{2}}$ & 123 & $(26; a_1),\ (29; A_1),\ (31; m),\ (41; m)$ \\ \hline
$(A_4^{e_8}, 17)$ & $\underset{[E_8]}{1} \ \ 2 \ \ \underset{[N_f = 1/2]}{\overset{\ksu(2)}{2}} \ \ \underset{[\SU(2)]}{\overset{\ksu(2)}{2}} \ \ \sqcup \ \ \underset{[E_8]}{1} $ & 151 & $(27; a_1),\ (31; e_8),\ (42; m)$ \\ \hline
$(A_4^{e_8}, 18)$ & $[E_8] \ \ 1 \ \ 2 \ \ \underset{[N_f = 3/2]}{\overset{\ksu(2)}{2}} \ \ \underset{[N_f = 3/2]}{\overset{\ksu(2)}{2}} \ \ 2$ & 151 & $(26; e_8),\ (27; m),\ (28; \calY(5))$ \\ \hline
$(A_4^{e_8}, 19)$ & $[E_7] \ \ 1 \ \ \underset{[N_f = 1/2]}{\overset{\ksu(2)}{2}} \ \ \underset{[\SU(4)]}{\overset{\ksu(3)}{2}} \ \ \sqcup \ \ \underset{[E_8]}{1}$ & 125 & $(25; e_7),\ (31; a_3),\ (23; e_8)$ \\ \hline
$(A_4^{e_8}, 20)$ & $[E_8] \ \ 1 \ \ 2 \ \ \underset{[G_2]}{\overset{\ksu(2)}{2}} \ \ \sqcup \ \ \underset{[E_8]}{1} \ \ 2$ & 151 & $(41; e_8),\ (42; A_1),\ (44; g_2)$ \\ \hline
$(A_4^{e_8}, 21)$ & $[\SO(12)] \ \ {\overset{\ksu(2)}{1}} \ \ \overset{\ksu(4)}{2} \ \ [\SU(6)]$ & 84 & $(32; d_6),\ (33; a_5)$ \\ \hline
$(A_4^{e_8}, 22)$ & $[\SU(9)] \ \ {\overset{\ksu(3)}{1}} \ \ \overset{\ksu(3)}{2} \ \ [\SU(3)]$ & 87 & $(33; a_8),\ (34; a_2)$ \\ \hline
$(A_4^{e_8}, 23)$ & $[E_7] \ \ 1 \ \ \underset{[N_f = 1/2]}{\overset{\ksu(2)}{2}} \ \ \overset{\ksu(3)}{2} \ \ [\SU(4)]$ & 96 & $(33; e_7),\ (36; a_3)$ \\ \hline
$(A_4^{e_8}, 24)$ & $[E_6] \ \ 1 \ \ \underset{[\SU(4)]}{\overset{\ksu(3)}{2}} \ \ \overset{\ksu(2)}{2} \ \ [N_f = 1]$ & 96 & $(34; e_6),\ (36; a_3)$ \\ \hline
$(A_4^{e_8}, 25)$ & $[\SO(14)] \ \ {\overset{\ksu(2)}{1}} \ \ \underset{\SU(4)}{\overset{\ksu(3)}{2}}  \ \ \sqcup \ \ \underset{[E_8]}{1}$ & 108 & $(33; e_8),\ (35; d_7),\ (37; a_3)$ \\ \hline
$(A_4^{e_8}, 26)$ & $[E_7] \ \ 1 \ \ \underset{[\SU(2)]}{\overset{\ksu(2)}{2}} \ \ \underset{[N_f = 3/2]}{\overset{\ksu(2)}{2}} \ \ 2$ & 122 & $(38; e_7),\ (40; A_1)$ \\ \hline
$(A_4^{e_8}, 27)$ & $[E_8] \ \ 1 \ \ 2 \ \ \underset{[\SU(3)]}{\overset{\ksu(2)}{2}} \ \ 2 \ \ \sqcup \ \ \underset{[E_8]}{1}$ & 150 & $(39; e_8),\ (40; e_8),\ (43; \calY(4))$ \\ \hline
$(A_4^{e_8}, 28)$ & $[E_8] \ \ 1 \ \ 2 \ \ 2 \ \ 2 \ \ 2$ & 149 & $(43; m),\ (44; m)$ \\ \hline
$(A_4^{e_8}, 29)$ & $\underset{[E_8]}{1} \ \ 2 \ \ \underset{[N_f = 3/2]}{\overset{\ksu(2)}{2}} \ \ \underset{[\SU(2)]}{\overset{\ksu(2)}{2}}$ & 122 & $(36; e_8),\ (40; a_1),\ (51; m)$ \\ \hline
$(A_4^{e_8}, 30)$ & $[\SO(16)] \ \ \overset{\ksu(2)}{1} \ \ {\overset{\ksu(2)}{2}} \ \ \overset{\ksu(2)}{2} \ \ [\SU(2)]$ & 106 & $(36; d_8),\ (37; m),\ (38; a_1), (63; m)$ \\ \hline
$(A_4^{e_8}, 31)$ & $[E_7] \ \ 1 \ \ \underset{[N_f = 3/2]}{\overset{\ksu(2)}{2}} \ \ \underset{[\SU(2)]}{\overset{\ksu(2)}{2}} \ \ \sqcup \ \ \underset{[E_8]}{1}$ & 122 & $(37; e_7),\ (39; a_1),\ (51; a_1),\ (52; m)$ \\ \hline
$(A_4^{e_8}, 32)$ & $[\SO(10)] \ \ 1 \ \ \overset{\ksu(4)}{2} \ \ [\SU(8)]$ & 75 & $(45; d_5),\ (46; a_7)$ \\ \hline
$(A_4^{e_8}, 33)$ & $[\SO(14)] \ \ \overset{\ksu(2)}{1} \ \ \overset{\ksu(3)}{2} \ \ [\SU(4)]$ & 79 & $(46; d_7),\ (48; a_3)$ \\ \hline
$(A_4^{e_8}, 34)$ & $[\SU(10)]  {\overset{\ksu(3)}{1}} \ \ \overset{\ksu(2)}{2} \ \ [N_f = 1]$ & 85 & $(48; a_9)$ \\ \hline
$(A_4^{e_8}, 35)$ & $[E_6] \ \ 1 \ \ \underset{[\SU(6)]}{\overset{\ksu(3)}{2}} \ \ \sqcup \ \ \underset{[E_8]}{1}$ & 97 & $(46; e_8),\ (47; e_6), (58; a_5)$ \\ \hline
$(A_4^{e_8}, 36)$ & $\underset{[E_7]}{1} \ \ \underset{[\SU(2)]}{\overset{\ksu(2)}{2}} \ \ \underset{[\SU(2)]}{\overset{\ksu(2)}{2}}$ & 93 & $(48; e_7),\ (50; a_1),\ (58; m),\ (60; A_1)$ \\ \hline
$(A_4^{e_8}, 37)$ & $\underset{[\SO(14)]}{{\overset{\ksu(2)}{1}}} \ \ \underset{[\SU(2)]}{\overset{\ksu(2)}{2}} \ \ \sqcup \ \ \underset{[E_8]}{1}$ & 105 & $(48; e_8),\ (49; a_1),\ (58; d_8),\ (66; m)$ \\ \hline
$(A_4^{e_8}, 38)$ & $[\SO(16)] \ \ {\overset{\ksu(2)}{1}} \ \ \underset{[N_f = 3/2]}{\overset{\ksu(2)}{2}} \ \ 2$ & 105 & $(49; m),\ (50; d_8)$ \\ \hline
$(A_4^{e_8}, 39)$ & $[E_7] \ \ 1 \ \ \underset{[G_2]}{\overset{\ksu(2)}{2}} \ \ 2 \ \ \sqcup \ \ 1$ & 121 & $(50; e_8),\ (61; g_2)$ \\ \hline
$(A_4^{e_8}, 40)$ & $[E_8] \ \ 1 \ \ 2 \ \ \underset{[\SU(3)]}{\overset{\ksu(2)}{2}} \ \ 2$ & 121 & $(50; e_8),\ (53; \calY(4))$ \\ \hline
$(A_4^{e_8}, 41)$ & $[E_7] \ \ 1 \ \ \underset{[\SO(7)]}{\overset{\ksu(2)}{2}} \ \ \sqcup \ \ \underset{[E_8]}{1} \ \ 2$ & 122 & $(52; A_1),\ (62; b_3),\ (63; e_7)$ \\ \hline
$(A_4^{e_8}, 42)$ & $[E_8] \ \ 1 \ \ 2 \ \ \underset{[G_2]}{\overset{\ksu(2)}{2}} \ \ \sqcup \ \ \left\{\underset{[E_8]}{1}\right\}^{\otimes2} $ & 150 & $(51; e_8),\ (54; g_2)$ \\ \hline
$(A_4^{e_8}, 43)$ & $[E_8] \ \ 1 \ \ 2 \ \ 2 \ \ 2 \ \ \sqcup \ \ \underset{[E_8]}{1}$ & 148 & $(53; e_8),\ (54; m),\ (55, A_1)$ \\ \hline
$(A_4^{e_8}, 44)$ & $[E_8] \ \ 1 \ \ 2 \ \ 2 \ \ \sqcup \ \ \underset{[E_8]}{1} \ \ 2$ & 148 & $(54; A_1),\ (55; m)$ \\ \hline
$(A_4^{e_8}, 45)$ & $\underset{[\SU(12) \times \Sp(1)]}{\overset{\ksu(4)}{1}}$ & 68 & $(56; A_1),\ (57; a_{11})$ \\ \hline
$(A_4^{e_8}, 46)$ & $[E_6] \ \ 1 \ \ \overset{\ksu(3)}{2} \ \ [\SU(6)]$ & 68 & $(57; e_6),\ (65; a_5)$ \\ \hline
$(A_4^{e_8}, 47)$ & $\underset{[\SU(12)]}{\overset{\ksu(3)}{1}} \ \ \sqcup \ \ \underset{[E_8]}{1}$ & 85 & $(57; e_8),\ (67; a_{11})$ \\ \hline
$(A_4^{e_8}, 48)$ & $[\SO(16)] \ \ {\overset{\ksu(2)}{1}} \ \ \overset{\ksu(2)}{2} \ \ [\SU(2)]$ & 76 & $(59; a_1),\ (65; d_8)$ \\ \hline
$(A_4^{e_8}, 49)$ & $[\SO(19)] \ \ {\overset{\ksu(2)}{1}} \ \ 2  \ \ \sqcup \ \ \underset{[E_8]}{1} $ & 104 & $(59; e_8),\ (71; b_9)$ \\ \hline
$(A_4^{e_8}, 50)$ & $[E_7] \ \ 1 \ \ \underset{[G_2]}{\overset{\ksu(2)}{2}} \ \ 2$ & 92 & $(59; e_7),\ (69; g_2)$ \\ \hline
$(A_4^{e_8}, 51)$ & $[E_8] \ \ 1 \ \ 2 \ \ \underset{[G_2]}{\overset{\ksu(2)}{2}} \ \ \sqcup \ \ \underset{[E_8]}{1}$ & 121 & $(58; e_8),\ (60; e_8),\ (61; g_2)$ \\ \hline
$(A_4^{e_8}, 52)$ & $[E_7] \ \ 1 \ \ \underset{[\SO(7)]}{\overset{\ksu(2)}{2}} \ \ \sqcup \ \ \left\{\underset{[E_8]}{1}\right\}^{\otimes2} $ & 121 & $(58; e_8),\ (66; e_7),\ (70; b_3)$ \\ \hline
$(A_4^{e_8}, 53)$ & $[E_8] \ \ 1 \ \ 2 \ \ 2 \ \ 2$ & 119 & $(61; m),\ (62; a_1)$ \\ \hline
$(A_4^{e_8}, 54)$ & $[E_8] \ \ 1 \ \ 2 \ \ 2 \ \ \sqcup \ \ \left\{\underset{[E_8]}{1}\right\}^{\otimes2}$ & 147 & $(61; e_8),\ (64; m)$ \\ \hline
$(A_4^{e_8}, 55)$ & $\{[E_8] \ \ 1 \ \ 2\}^{\otimes2} \ \ \sqcup \ \ \underset{[E_8]}{1} $ & 147 & $(62; e_8),\ (64; A_1)$ \\ \hline
$(A_4^{e_8}, 56)$ & $\underset{[\SO(24)]}{\overset{\ksp(2)}{1}}$ & 67 & $(68; d_{12})$ \\ \hline
$(A_4^{e_8}, 57)$ & $ \underset{[\SU(12)]}{\overset{\ksu(3)}{1}}$ & 57 & $(68; a_{11})$ \\ \hline
$(A_4^{e_8}, 58)$ & $[E_7] \ \ 1 \ \ \underset{[\SO(7)]}{\overset{\ksu(2)}{2}} \ \ \sqcup \ \ \underset{[E_8]}{1} $ & 92 & $(65; e_8),\ (67; e_7),\ (71; b_3)$ \\ \hline
$(A_4^{e_8}, 59)$ & $[\SO(19)] \ \ {\overset{\ksu(2)}{1}} \ \ 2$ & 75 & $(72; b_9)$ \\ \hline
$(A_4^{e_8}, 60)$ & $[E_8] \ \ 1 \ \ 2 \ \ \overset{\ksu(2)}{2} \ \ [G_2]$ & 92 & $(65; e_8),\ (69; g_2)$ \\ \hline
$(A_4^{e_8}, 61)$ & $[E_8] \ \ 1 \ \ 2 \ \ 2 \ \ \sqcup \ \ \underset{[E_8]}{1} $ & 118 & $(69; e_8),\ (70; m)$ \\ \hline
$(A_4^{e_8}, 62)$ & $\{[E_8] \ \ 1 \ \ 2\}^{\otimes2}$ & 118 & $(70; A_1)$ \\ \hline
$(A_4^{e_8}, 63)$ & $\underset{[\SO(20)]}{\overset{\ksu(2)}{1}} \ \ \sqcup \ \ \underset{[E_8]}{1} \ \ 2$ & 105 & $(66; A_1),\ (71; d_{10})$ \\ \hline
$(A_4^{e_8}, 64)$ & $[E_8] \ \ 1 \ \ 2 \ \ \sqcup \ \ \left\{\underset{[E_8]}{1}\right\}^{\otimes3} $ & 146 & $(70; e_8),\ (73; A_1)$ \\ \hline
$(A_4^{e_8}, 65)$ & $[E_7] \ \ 1 \ \ \overset{\ksu(2)}{2} \ \ [\SO(7)]$ & 63 & $(68; e_7),\ (72; b_3)$ \\ \hline
$(A_4^{e_8}, 66)$ & $\underset{[\SO(20)]}{\overset{\ksu(2)}{1}} \ \ \sqcup \ \ \left\{\underset{[E_8]}{1}\right\}^{\otimes2} $ & 104 & $(67; e_8),\ (75; d_{10})$ \\ \hline
$(A_4^{e_8}, 67)$ & $\underset{[\SO(20)]}{\overset{\ksu(2)}{1}} \ \ \sqcup \ \ \underset{[E_8]}{1} $ & 159 & $(68; e_8),\ (76; d_{10})$ \\ \hline
$(A_4^{e_8}, 68)$ & $\underset{[\SO(20)]}{\overset{\ksu(2)}{1}} $ & 46 & $(77; d_{10})$ \\ \hline
$(A_4^{e_8}, 69)$ & $[E_8] \ \ 1 \ \ 2  \ \ 2$ & 89 & $(71; m)$ \\ \hline
$(A_4^{e_8}, 70)$ & $[E_8] \ \ 1 \ \ 2 \ \ \sqcup \ \ \left\{\underset{[E_8]}{1}\right\}^{\otimes2}$ & 117 & $(71; e_8),\ (74; a_1)$ \\ \hline
$(A_4^{e_8}, 71)$ & $[E_8] \ \ 1 \ \ 2 \ \ \sqcup \underset{[E_8]}{1}$ & 88 & $(72; e_8),\ (75, A_1)$ \\ \hline
$(A_4^{e_8}, 72)$ & $[E_8] \ \ 1 \ \ 2$ & 59 & $(76; A_1)$ \\ \hline
$(A_4^{e_8}, 73)$ & $\{[E_8] \ \ 1\}^{\otimes5}$ & 145 & $(74; e_8)$ \\ \hline
$(A_4^{e_8}, 74)$ & $\{[E_8] \ \ 1\}^{\otimes4}$ & 116 & $(75; e_8)$ \\ \hline
$(A_4^{e_8}, 75)$ & $\{[E_8] \ \ 1\}^{\otimes3}$ & 87 & $(76; e_8)$ \\ \hline
$(A_4^{e_8}, 76)$ & $\{[E_8] \ \ 1\}^{\otimes2}$ & 58 & $(77; e_8)$ \\ \hline
$(A_4^{e_8}, 77)$ & $\underset{[E_8]}{1}$ & 29 & $(78; e_8)$ \\ \hline
$(A_4^{e_8}, 78)$ & $\varnothing$ & 0 & IR theory \\ \hline
\caption{Theories corresponding to the nodes in the Hasse diagram of the $A_4$-type orbi-instanton theory.}\label{A4_orbi-instanton_table}
\end{longtable}

\subsubsection{The Tri-Leg \texorpdfstring{$D_4$}{D4} Theory}\label{D4hatrank1}
\begin{figure}[h]
    \centering
    \includegraphics[width=7cm]{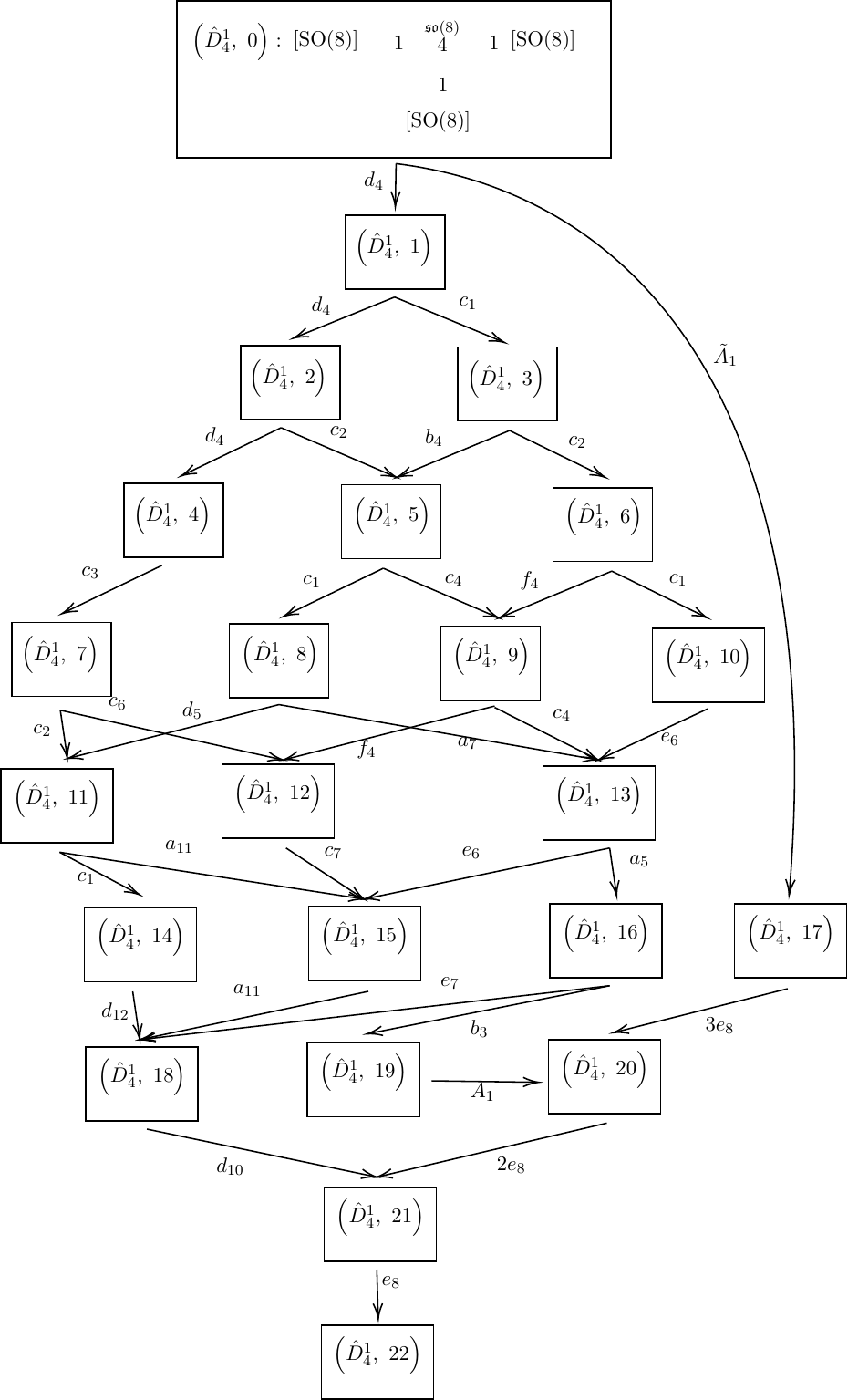}
    \caption{Complete Hasse diagram of the rank 1 $\widehat{D}_4$ theory. The nodes can be found in Table \ref{D4hatrank1_table}.}\label{D4hatrank1Hasse}
\end{figure}

Now, let us consider the $\widehat{D}_4^1$ theory. The full Hasse diagram is given in Figure \ref{D4hatrank1Hasse}, with the nodes and the RG flows catalogued in Table \ref{D4hatrank1_table}. Notice that the Hasse diagram contains the full Hasse diagram of the rank 0 $(E_6,E_6)$ conformal matter theory as a subdiagram.

\vspace{10mm}

\begin{longtable}{|c|c|c|c|c|c|} \hline
$\widehat{D}_4^1$ label &  Tensor branch description & $d_{\mathbb{H}}$ & (descendant \#; flow type) & $E_6^0$ label & $A_4^{e_8}$ label \\ \hline \hline
$\left(\widehat{D}_4^1, 0\right)$ & $ {\overset{\kso(8)}{4}} \ \ (1 \ \ [\SO(8)])^{\otimes 3}$ & 88 & $(1; d_4), (17; \widetilde{A}_1)$ & - & - \\ \hline
$\left(\widehat{D}_4^1, 1\right)$ & $\underset{[\SO(8)]}{1} \ \ \underset{[\Sp(1)^3]}{\overset{\kso(8)}{3}} \ \ \underset{[\SO(8)]}{1}$ & 83 & $(2; d_4), (3; c_1)$ & - & - \\ \hline
$\left(\widehat{D}_4^1, 2\right)$ & $ \underset{[\Sp(2)^3]}{\overset{\kso(8)}{2}} \ \ 1 \ \ [\SO(8)]$ & 78 & $(4; d_4), (5; c_2)$ & - & - \\ \hline
$\left(\widehat{D}_4^1, 3\right)$ & $\underset{[\SO(9)]}{1} \ \ \underset{[\Sp(2)]}{\overset{\kso(7)}{3}} \ \ \underset{[\SO(9)]}{1}$ & 82 & $(5; b_4), (6; c_2)$ & - & - \\ \hline
$\left(\widehat{D}_4^1, 4\right)$ & ${\overset{\kso(8)}{1}} \ \ [\Sp(3)^3]$ & 73 & $(7; c_3)$ & - & - \\ \hline
$\left(\widehat{D}_4^1, 5\right)$ & $\underset{[\SO(9)]}{1} \ \ \underset{[\Sp(4)\times\Sp(1)]}{\overset{\kso(7)}{2}}$ & 76 & $(7; b_4), (8; c_1), (9, c_4)$ & - & - \\ \hline
$\left(\widehat{D}_4^1, 6\right)$ & $[F_4] \ \ 1 \ \ \underset{[\Sp(1)]}{\overset{\kg_2}{3}} \ \ 1 \ \ [F_4]$ & 80 & $(9; f_4), (10; c_1)$ & - & - \\ \hline
$\left(\widehat{D}_4^1, 7\right)$ & $ {\overset{\kso(7)}{1}} \ \ [\Sp(6) \times \Sp(2)]$ & 70 & $(11; c_2), (12; c_6)$ & - & - \\ \hline
$\left(\widehat{D}_4^1, 8\right)$ & $\underset{[\SO(10)]}{1} \ \ \underset{[\SU(8)]}{\overset{\ksu(4)}{2}}$ & 75 & $(11; d_5), (13; a_7)$ & - & $(A_4^{e_8},32)$ \\ \hline
$\left(\widehat{D}_4^1, 9\right)$ & $[F_4] \ \ 1 \ \ {\overset{\kg_2}{2}} \ \ [\Sp(4)]$ & 72 & $(12; f_4), (13; c_4)$ & - & - \\ \hline
$\left(\widehat{D}_4^1, 10\right)$ & $[E_6] \ \ 1 \ \ {\overset{\ksu(3)}{3}} \ \ 1 \ \ [E_6]$ & 79 & $(13; e_6)$ & $(E_6^0, 0)$ & - \\ \hline
$\left(\widehat{D}_4^1, 11\right)$ & $ {\overset{\ksu(4)}{1}} \ \ [\Sp(6) \times \Sp(2)]$ & 68 & $(14; c_1), (15; a_{11})$ & - & $(A_4^{e_8},45)$ \\ \hline
$\left(\widehat{D}_4^1, 12\right)$ & $ {\overset{\kg_2}{1}} \ \ [\Sp(7)]$ & 64 & $(15; c_7)$ & - & - \\ \hline
$\left(\widehat{D}_4^1, 13\right)$ & $[E_6] \ \ 1 \ \ {\overset{\ksu(3)}{2}} \ \ [\SU(6)]$ & 68 & $(15; e_6),\ (16; a_5)$ & $(E_6^0, 1)$ & $(A_4^{e_8},46)$ \\ \hline
$\left(\widehat{D}_4^1, 14\right)$ & $ {\overset{\ksp(2)}{1}} \ \ [\SO(24)]$ & 67 & $(18; d_{12})$ & - & $(A_4^{e_8},56)$ \\ \hline
$\left(\widehat{D}_4^1, 15\right)$ & ${\overset{\ksu(3)}{1}} \ \  [\SU(12)]$ & 57 & $(18; a_{11})$ & $(E_6^0, 2)$ & $(A_4^{e_8},57)$ \\ \hline
$\left(\widehat{D}_4^1, 16\right)$ & $[E_7] \ \ 1 \ \ {\overset{\ksu(2)}{2}} \ \ [\SO(7)]$ & 63 & $(18; e_7),\ (19; b_3)$ & $(E_6^0, 3)$ & $(A_4^{e_8},65)$ \\ \hline
$\left(\widehat{D}_4^1, 17\right)$ & $\{1 \ \ [E_8]\}^{\otimes3}$ & 87 & $(20; 3e_8)$ & - & $(A_4^{e_8},75)$ \\ \hline
$\left(\widehat{D}_4^1, 18\right)$ & $ {\overset{\ksu(2)}{1}} \ \ [\SO(20)]$ & 46 & $(21; d_{10})$ & $(E_6^0, 4)$ & $(A_4^{e_8},68)$ \\ \hline
$\left(\widehat{D}_4^1, 19\right)$ & $[E_8] \ \ 1 \ \ 2 \ \ [\SU(2)]$ & 59 & $(20; A_1)$ & $(E_6^0, 5)$ & $(A_4^{e_8},72)$ \\ \hline
$\left(\widehat{D}_4^1, 20\right)$ & $[E_8] \ \ 1 \ \ \sqcup \ \ 1 \ \ [E_8]$ & 58 & $(21; 2e_8)$ & $(E_6^0, 6)$ & $(A_4^{e_8},76)$ \\ \hline
$\left(\widehat{D}_4^1, 21\right)$ & $1 \ \ [E_8]$ & 29 & $(22; e_8)$ & $(E_6^0, 7)$ & $(A_4^{e_8},77)$ \\ \hline
$\left(\widehat{D}_4^1, 22\right)$ & $\varnothing$ & 0 & IR theory & $(E_6^0, 8)$ & $(A_4^{e_8},78)$ \\ \hline
\caption{Theories corresponding to the nodes in the Hasse diagram of the rank 1 $\widehat{D}_4$ theory.}\label{D4hatrank1_table}
\end{longtable}

Let us make a comment on the $\widetilde{A}_1$ flow. We have checked that the $a$ central charge decreases along this flow. On the other hand, there could be some subtleties if one applies quiver subtractions \cite{Cabrera:2018ann} (whose magnetic quivers will be given in \S\ref{sec:D4hatrank1MQ}) to this process. Despite the subtleties, it is tempting to identify this $\widetilde{A}_1$ slice with the Kleinian singularity $D_4$ from quiver subtractions.


\section{Magnetic Phases}\label{sec:magneticquivers}
As mentioned above, the magnetic quivers, whose Coulomb branches/moduli spaces of dressed monopole operators are the same as the Higgs branches in the corresponding 6d theories, provide a powerful tool to study the Higgsing structure. In this section, we shall consider some examples from the viewpoint of the magnetic quivers. In these examples, we have the Type IIA constructions (possibly with negatively charged branes allowed), and the magnetic quivers can thus be obtained by moving to the magnetic phases of the systems. The generalized/electric quivers and the magnetic quivers often encode different information, and it would be worth comparing them. For instance, from the electric side, we can see that the minimal nilpotent orbit $\calO_{\mathrm{min}}$ in non-abelian $G$ is physically encoded in the $\calO_{\mathrm{min}}$ T-brane deformation as recalled in \S\ref{Tbranes}.

When we look at the magnetic quivers, there are already systematic ways to obtain the complete Hasse diagrams, at least for unitary ones. From the brane systems, some magnetic quivers were constructed for example in \cite{Cabrera:2019izd,Mekareeya:2017jgc,Cabrera:2018jxt,VanBeest:2020kxw,DelZotto:2023nrb,Lawrie:2023uiu,Cabrera:2019dob}. However, in general, given a theory, the magnetic quiver description may not be known. For unitary magnetic quivers, we shall compare the Hasse diagrams with those obtained from our algorithm, and verify that they would coincide. For orthosymplectic magnetic quivers, many techniques are still under development. We hope that the Hasse diagrams obtained from our algorithm will also shed light on the study of the orthosymplectic magnetic quivers.

\subsection{Unitary Magnetic Quivers}\label{unitaryMQs}
For magnetic quivers with only unitary nodes, the Higgs mechanism and the structure of the symplectic singularities can be obtained in a rather straightforward manner thanks to techniques such as quiver subtractions \cite{Cabrera:2018ann,Bourget:2019aer} and quiver decays and fissions \cite{Bourget:2023dkj,Bourget:2024mgn}. Since we are focusing on the Higgsed theories under the RG flows, we shall mainly be considering the quiver decays and fissions. For minimal nilpotent orbit, minimal plateau, and combo Higgsings, they correspond to quiver decays. For endpoint-changing flows, they can have either quiver decays or quiver fissions as their incarnation on the magnetic side.

\subsubsection{Rank 0 D-Type Conformal Matters}\label{rank0DkuMQs}
\begin{figure}[h]
    \centering
    \includegraphics[width=0.13\linewidth]{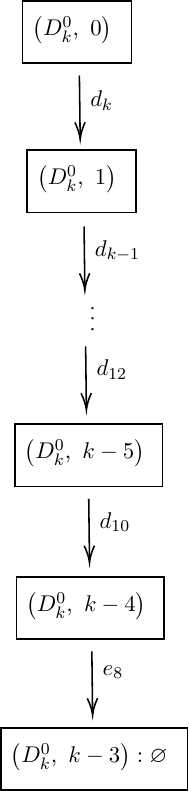}
    \caption{The Hasse diagram of the rank 0 $(\SO(2k),\SO(2k))$ conformal matter.}\label{Dkrank0Hasse}
\end{figure}

As a warm-up, let us first consider the rank 0 D-type conformal matter theories:
\begin{equation}
    \underset{[G]}{\overset{\mathfrak{sp}(k-4)}{1}},\label{Dkrank0}
\end{equation}
where the flavour symmetry $\SO(2k)\times\SO(2k)$ is enhanced to $G$, with $G=E_8$ and $G=\SO(4k)$ for $k=4$ and $k>4$ respectively. The dimension of the Higgs branch is $(2k^2-k+1)$. The Hasse diagram has a rather simple structure as given in Figure \ref{Dkrank0Hasse}.

For $0\leq j\leq k-4$, the $(D_k^0,j)$ theory is just the rank 0 $(\SO(2k-2j),\SO(2k-2j))$ conformal matter theory. The magnetic quiver for the rank 0 $D_k$ theory is
\begin{equation}
    \includegraphics[width=5cm]{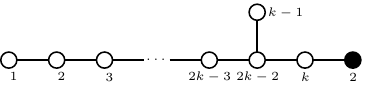}.
\end{equation}
In particular, when $k=4$, node 2 on the right end also becomes balanced, reflecting the symmetry enhancement to $E_8$. The elementary slices are simply the closure of the minimal nilpotent orbit of the corresponding (affine) Dynkin diagrams (given by the differences of the adjacent magnetic quivers in the Hasse diagram). Later, we shall compare this with the orthosymplectic magnetic quivers in \S\ref{rank0DkospMQs}.

\subsubsection{The A-type Orbi-Instanton Theories}\label{orbiinstantonMQs}
Our next example would be the orbi-instanton theory $A_4^{e_8}$ whose generalized quiver reads
\begin{equation}
    [E_8] \ \ 1 \ \ 2 \ \ \overset{\ksu(2)}{2} \ \ \overset{\ksu(3)}{2} \ \ \overset{\ksu(4)}{2} \ \ [\SU(5)].
\end{equation}
The magnetic quiver is \cite{Cabrera:2019izd,Mekareeya:2017jgc}
\begin{equation}
    \includegraphics[width=8cm]{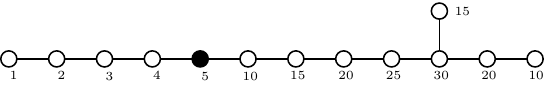}.
\end{equation}
Using the quiver decay and fission algorithm \cite{Bourget:2023dkj,Bourget:2024mgn}, one may check that the Hasse diagram agrees with the one obtained in Figure \ref{A4orbi-instantonHasse}. Let us list the magnetic quivers in Table \ref{A4_orbi-instanton_MQs_table}.
\begin{longtable}{|c|c|c|c|}
\hline
$A_4^{e_8}$ label &  Magnetic quiver & $d_{\mathbb{H}}$ & (descendant \#; flow type) \\ \hline \hline
\textcolor{blue}{$(A_4^{e_8}, 0)$} & \includegraphics[width=5cm]{figures/A4orbi-instantonMQ.pdf} & 159 & $(1; e_8),\ (2; a_4)$ \\ \hline
\textcolor{blue}{$(A_4^{e_8}, 1)$} & \includegraphics[width=5cm]{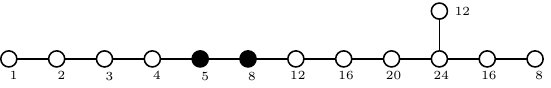} & 130 & $(3; e_7),\ (4; a_4)$ \\ \hline
$(A_4^{e_8}, 2)$ & \includegraphics[width=5cm]{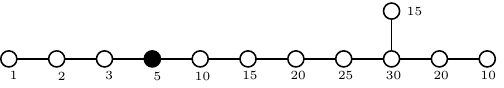} & 155 & $(4; e_8),\ (5; a_2),\ (6; m)$ \\ \hline
\textcolor{blue}{$(A_4^{e_8}, 3)$} & \includegraphics[width=5cm]{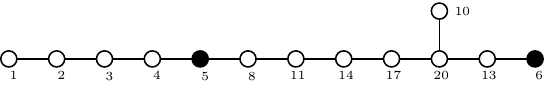} & 113 & $(7; d_7),\ (8; a_4)$ \\ \hline
$(A_4^{e_8}, 4)$ & \includegraphics[width=5cm]{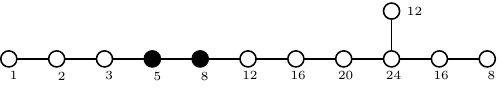} & 126 & $(8; e_7),\ (9; a_2),\ (11; A_2),\ (19, m)$ \\ \hline
$(A_4^{e_8}, 5)$ & \includegraphics[width=5cm]{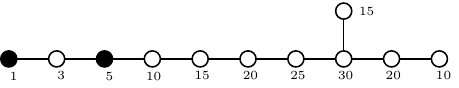} & 153 & $(9; e_8),\ (10; A_1)$ \\ \hline
$(A_4^{e_8}, 6)$ & \includegraphics[width=5cm]{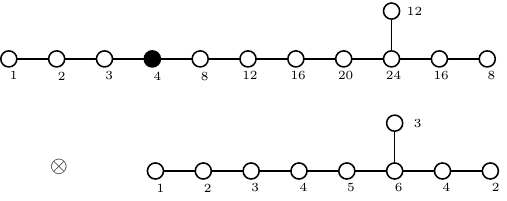} & 154 & $(11; e_8),\ (17; a_3),\ (19, e_8)$ \\ \hline
\textcolor{blue}{$(A_4^{e_8}, 7)$} & \includegraphics[width=5cm]{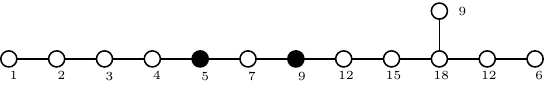} & 102 & $(12; e_6),\ (13; A_1),\ (14; a_4)$ \\ \hline
$(A_4^{e_8}, 8)$ & \includegraphics[width=5cm]{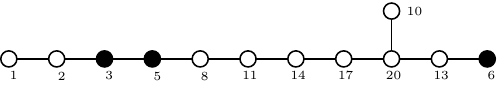} & 109 & $(14; d_7),\ (15; a_2),\ (25; m)$ \\ \hline
$(A_4^{e_8}, 9)$ & \includegraphics[width=5cm]{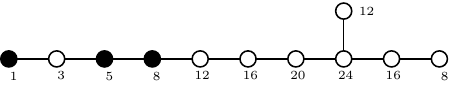} & 124 & $(15; e_7),\ (16; A_1)$ \\ \hline
$(A_4^{e_8}, 10)$ & \includegraphics[width=5cm]{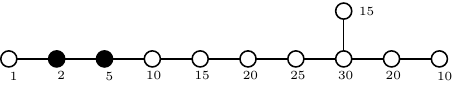} & 152 & $(16; e_8),\ (17; m),\ (18; a_1),\ (20, m)$ \\ \hline
$(A_4^{e_8}, 11)$ & \includegraphics[width=5cm]{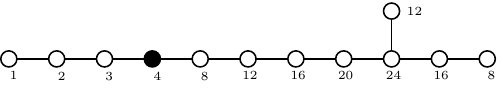} & 125 & $(23; e_8),\ (29; a_3)$ \\ \hline
\textcolor{blue}{$(A_4^{e_8}, 12)$} & \includegraphics[width=5cm]{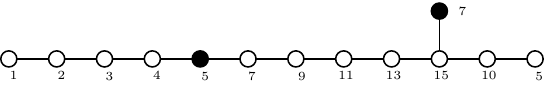} & 91 & $(21; a_7),\ (22; a_4)$ \\ \hline
\textcolor{blue}{$(A_4^{e_8}, 13)$} & \includegraphics[width=5cm]{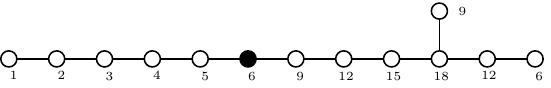} & 101 & $(21; e_7),\ (23; a_5)$ \\ \hline
$(A_4^{e_8}, 14)$ & \includegraphics[width=5cm]{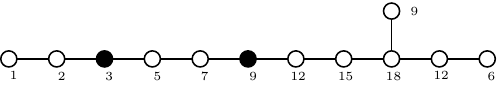} & 98 & $(22; e_6),\ (24; a_2),\ (35; m)$ \\ \hline
$(A_4^{e_8}, 15)$ & \includegraphics[width=5cm]{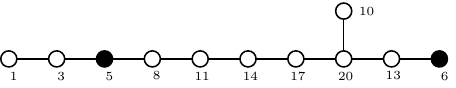} & 107 & $(24; d_7),\ (30; A_1)$ \\ \hline
$(A_4^{e_8}, 16)$ & \includegraphics[width=5cm]{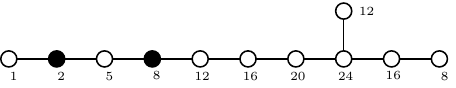} & 123 & $(26; a_1),\ (29; A_1),\ (31; m),\ (41; m)$ \\ \hline
$(A_4^{e_8}, 17)$ & \includegraphics[width=5cm]{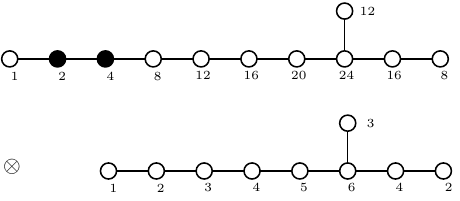} & 151 & $(27; a_1),\ (31; e_8),\ (42; m)$ \\ \hline
$(A_4^{e_8}, 18)$ & \includegraphics[width=5cm]{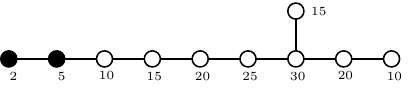} & 151 & $(26; e_8),\ (27; m),\ (28; \calY(5))$ \\ \hline
$(A_4^{e_8}, 19)$ & \includegraphics[width=5cm]{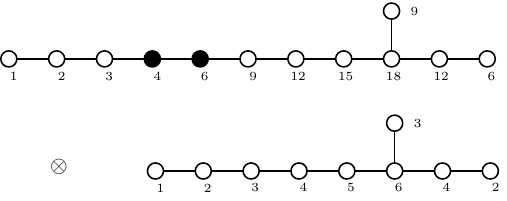} & 125 & $(25; e_7),\ (31; a_3),\ (23; e_8)$ \\ \hline
$(A_4^{e_8}, 20)$ & \includegraphics[width=5cm]{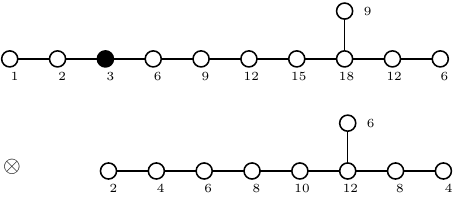} & 151 & $(41; e_8),\ (42; A_1),\ (44; g_2)$ \\ \hline
\textcolor{blue}{$(A_4^{e_8}, 21)$} & \includegraphics[width=5cm]{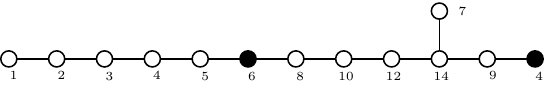} & 84 & $(32; d_6),\ (33; a_5)$ \\ \hline
$(A_4^{e_8}, 22)$ & \includegraphics[width=5cm]{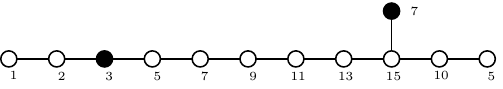} & 87 & $(33; a_8),\ (34; a_2)$ \\ \hline
$(A_4^{e_8}, 23)$ & \includegraphics[width=5cm]{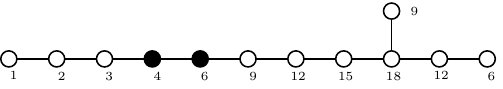} & 96 & $(33; e_7),\ (36; a_3)$ \\ \hline
$(A_4^{e_8}, 24)$ & \includegraphics[width=5cm]{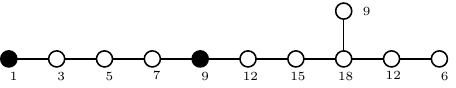} & 96 & $(34; e_6),\ (36; a_3)$ \\ \hline
$(A_4^{e_8}, 25)$ & \includegraphics[width=5cm]{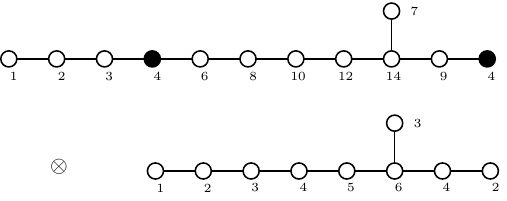} & 108 & $(33; e_8),\ (35; d_7),\ (37; a_3)$ \\ \hline
$(A_4^{e_8}, 26)$ & \includegraphics[width=5cm]{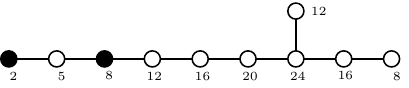} & 122 & $(38; e_7),\ (40; A_1)$ \\ \hline
$(A_4^{e_8}, 27)$ & \includegraphics[width=5cm]{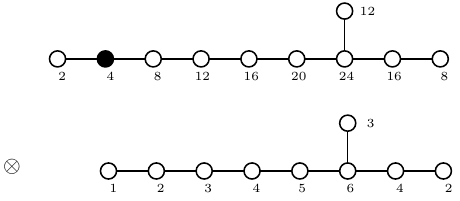} & 150 & $(39; e_8),\ (40; e_8),\ (43; \calY(4))$ \\ \hline
$(A_4^{e_8}, 28)$ & \includegraphics[width=5cm]{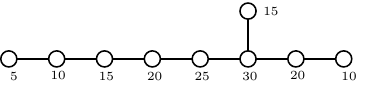} & 149 & $(43; m),\ (44; m)$ \\ \hline
$(A_4^{e_8}, 29)$ & \includegraphics[width=5cm]{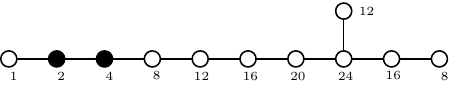} & 122 & $(36; e_8),\ (40; a_1),\ (51; m)$ \\ \hline
$(A_4^{e_8}, 30)$ & \includegraphics[width=5cm]{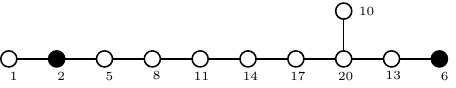} & 106 & $(36; d_8),\ (37; m),\ (38; a_1), (63; m)$ \\ \hline
$(A_4^{e_8}, 31)$ & \includegraphics[width=5cm]{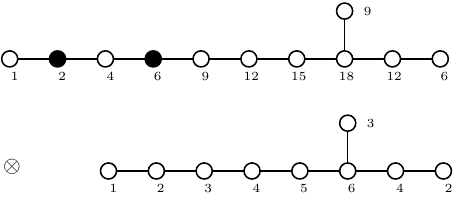} & 122 & $(37; e_7),\ (39; a_1),\ (51; a_1),\ (52; m)$ \\ \hline
\textcolor{blue}{$(A_4^{e_8}, 32)$} & \includegraphics[width=5cm]{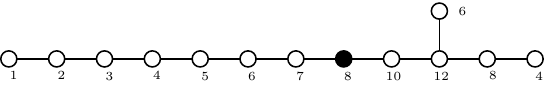} & 75 & $(45; d_5),\ (46; a_7)$ \\ \hline
$(A_4^{e_8}, 33)$ & \includegraphics[width=5cm]{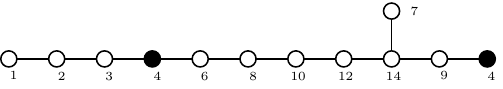} & 79 & $(46; d_7),\ (48; a_3)$ \\ \hline
$(A_4^{e_8}, 34)$ & \includegraphics[width=5cm]{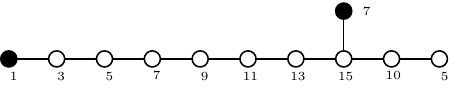} & 85 & $(48; a_9)$ \\ \hline
$(A_4^{e_8}, 35)$ & \includegraphics[width=5cm]{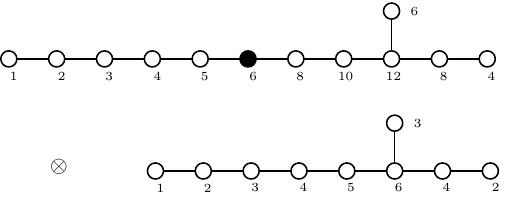} & 97 & $(46; e_8),\ (47; e_6), (58; a_5)$ \\ \hline
$(A_4^{e_8}, 36)$ & \includegraphics[width=5cm]{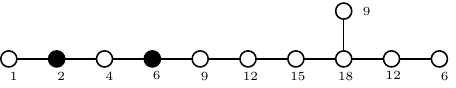} & 93 & $(48; e_7),\ (50; a_1),\ (58; m),\ (60; A_1)$ \\ \hline
$(A_4^{e_8}, 37)$ & \includegraphics[width=5cm]{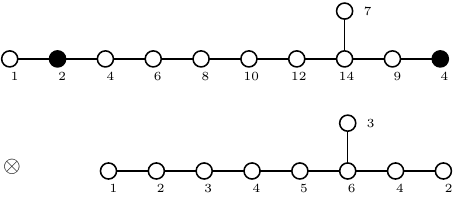} & 105 & $(48; e_8),\ (49; a_1),\ (58; d_8),\ (66; m)$ \\ \hline
$(A_4^{e_8}, 38)$ & \includegraphics[width=5cm]{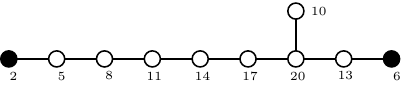} & 105 & $(49; m),\ (50; d_8)$ \\ \hline
$(A_4^{e_8}, 39)$ & \includegraphics[width=5cm]{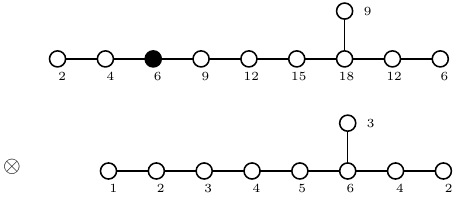} & 121 & $(50; e_8),\ (61; g_2)$ \\ \hline
$(A_4^{e_8}, 40)$ & \includegraphics[width=5cm]{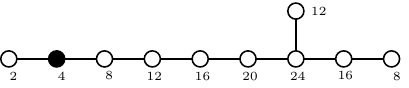} & 121 & $(50; e_8),\ (53; \calY(4))$ \\ \hline
$(A_4^{e_8}, 41)$ & \includegraphics[width=5cm]{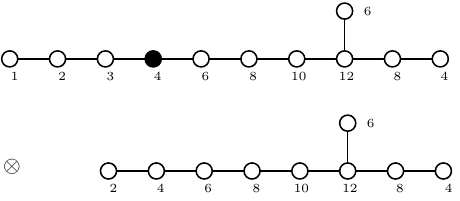} & 122 & $(52; A_1),\ (62; b_3),\ (63; e_7)$ \\ \hline
$(A_4^{e_8}, 42)$ & \includegraphics[width=5cm]{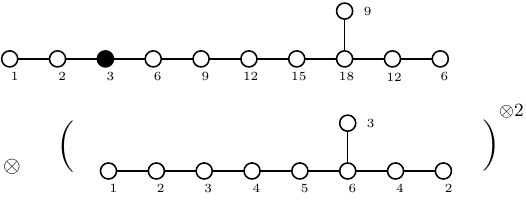} & 150 & $(51; e_8),\ (54; g_2)$ \\ \hline
$(A_4^{e_8}, 43)$ & \includegraphics[width=5cm]{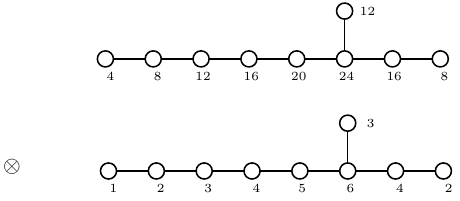} & 148 & $(53; e_8),\ (54; m),\ (55, A_1)$ \\ \hline
$(A_4^{e_8}, 44)$ & \includegraphics[width=5cm]{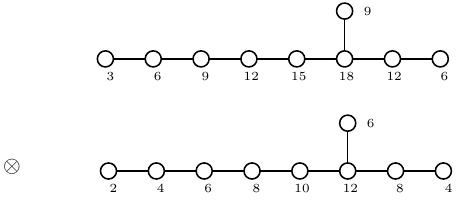} & 148 & $(54; A_1),\ (55; m)$ \\ \hline
\textcolor{blue}{$(A_4^{e_8}, 45)$} & \includegraphics[width=5cm]{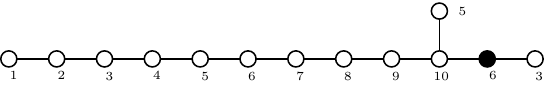} & 68 & $(56; A_1),\ (57; a_{11})$ \\ \hline
$(A_4^{e_8}, 46)$ & \includegraphics[width=5cm]{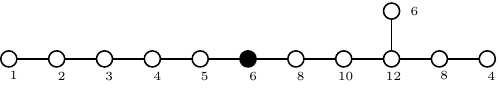} & 68 & $(57; e_6),\ (65; a_5)$ \\ \hline
$(A_4^{e_8}, 47)$ & \includegraphics[width=5cm]{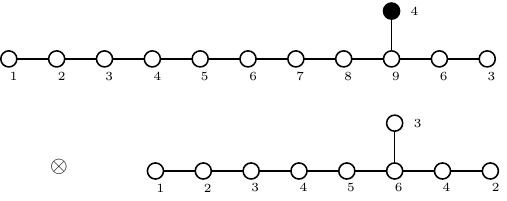} & 85 & $(57; e_8),\ (67; a_{11})$ \\ \hline
$(A_4^{e_8}, 48)$ & \includegraphics[width=5cm]{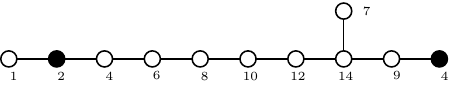} & 76 & $(59; a_1),\ (65; d_8)$ \\ \hline
$(A_4^{e_8}, 49)$ & \includegraphics[width=5cm]{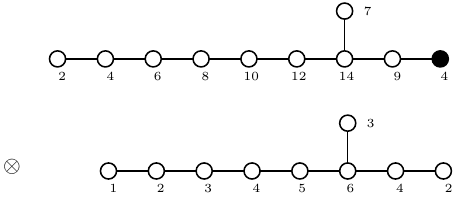} & 104 & $(59; e_8),\ (71; b_9)$ \\ \hline
$(A_4^{e_8}, 50)$ & \includegraphics[width=5cm]{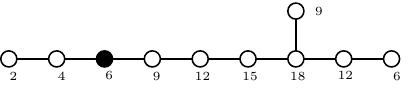} & 92 & $(59; e_7),\ (69; g_2)$ \\ \hline
$(A_4^{e_8}, 51)$ & \includegraphics[width=5cm]{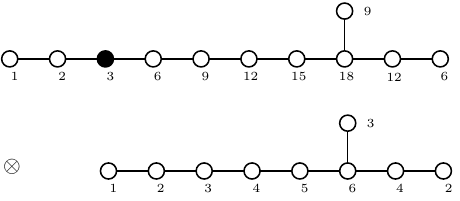} & 121 & $(58; e_8),\ (60; e_8),\ (61; g_2)$ \\ \hline
$(A_4^{e_8}, 52)$ & \includegraphics[width=5cm]{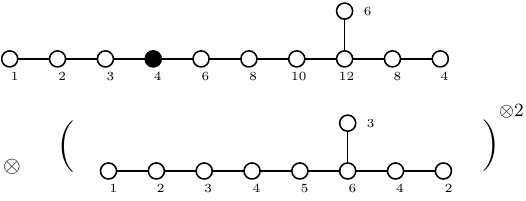} & 121 & $(58; e_8),\ (66; e_7),\ (70; b_3)$ \\ \hline
$(A_4^{e_8}, 53)$ & \includegraphics[width=5cm]{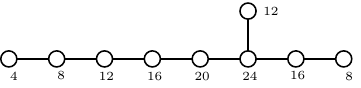} & 119 & $(61; m),\ (62; a_1)$ \\ \hline
$(A_4^{e_8}, 54)$ & \includegraphics[width=5cm]{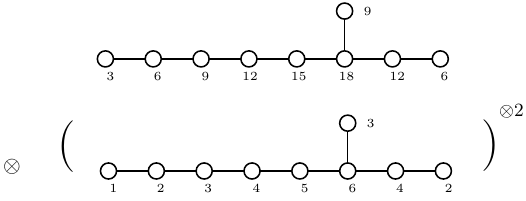} & 147 & $(61; e_8),\ (64; m)$ \\ \hline
$(A_4^{e_8}, 55)$ & \includegraphics[width=5cm]{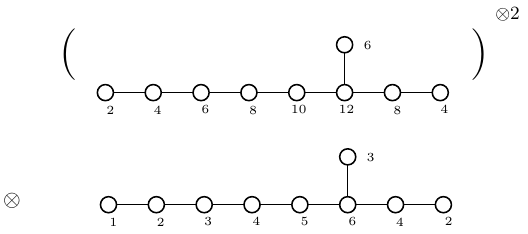} & 147 & $(62; e_8),\ (64; A_1)$ \\ \hline
\textcolor{blue}{$(A_4^{e_8}, 56)$} & \includegraphics[width=5cm]{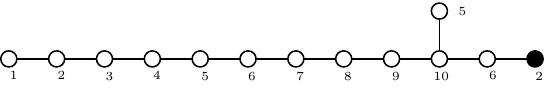} & 67 & $(68; d_{12})$ \\ \hline
$(A_4^{e_8}, 57)$ & \includegraphics[width=5cm]{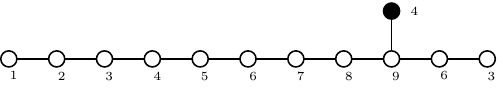} & 57 & $(68; a_{11})$ \\ \hline
$(A_4^{e_8}, 58)$ & \includegraphics[width=5cm]{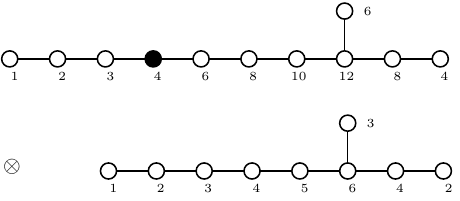} & 92 & $(65; e_8),\ (67; e_7),\ (71; b_3)$ \\ \hline
$(A_4^{e_8}, 59)$ & \includegraphics[width=5cm]{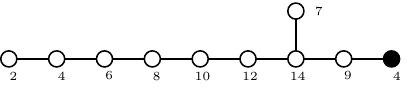} & 75 & $(72; b_9)$ \\ \hline
$(A_4^{e_8}, 60)$ & \includegraphics[width=5cm]{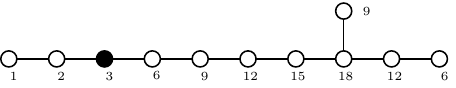} & 92 & $(65; e_8),\ (69; g_2)$ \\ \hline
$(A_4^{e_8}, 61)$ & \includegraphics[width=5cm]{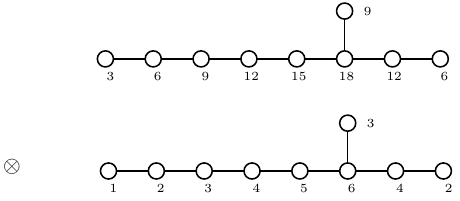} & 118 & $(69; e_8),\ (70; m)$ \\ \hline
$(A_4^{e_8}, 62)$ & \includegraphics[width=5cm]{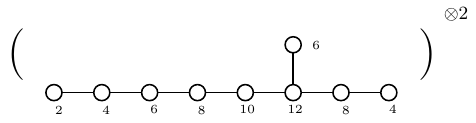} & 118 & $(70; A_1)$ \\ \hline
$(A_4^{e_8}, 63)$ & \includegraphics[width=5cm]{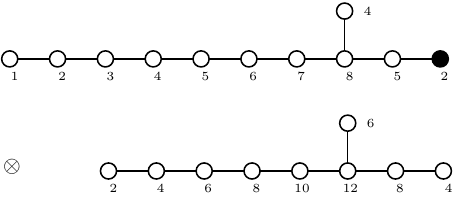} & 105 & $(66; A_1),\ (71; d_{10})$ \\ \hline
$(A_4^{e_8}, 64)$ & \includegraphics[width=5cm]{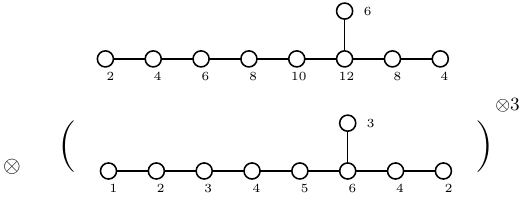} & 146 & $(70; e_8),\ (73; A_1)$ \\ \hline
$(A_4^{e_8}, 65)$ & \includegraphics[width=5cm]{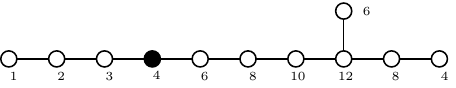} & 63 & $(68; e_7),\ (72; b_3)$ \\ \hline
$(A_4^{e_8}, 66)$ & \includegraphics[width=5cm]{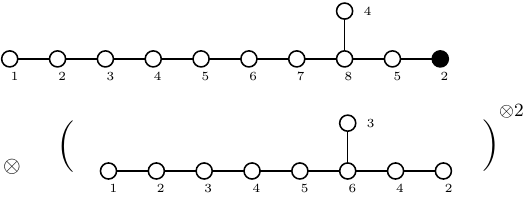} & 104 & $(67; e_8),\ (75; d_{10})$ \\ \hline
$(A_4^{e_8}, 67)$ & \includegraphics[width=5cm]{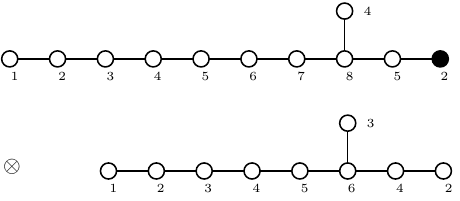} & 159 & $(68; e_8),\ (76; d_{10})$ \\ \hline
$(A_4^{e_8}, 68)$ & \includegraphics[width=5cm]{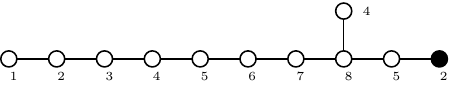} & 46 & $(77; d_{10})$ \\ \hline
$(A_4^{e_8}, 69)$ & \includegraphics[width=5cm]{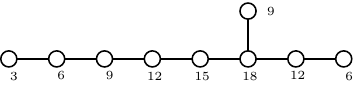} & 89 & $(71; m)$ \\ \hline
$(A_4^{e_8}, 70)$ & \includegraphics[width=5cm]{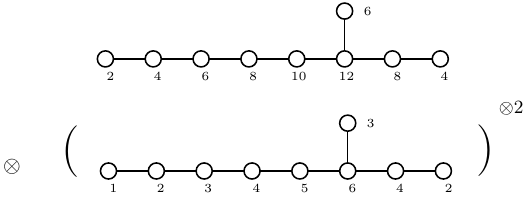} & 117 & $(71; e_8),\ (74; a_1)$ \\ \hline
$(A_4^{e_8}, 71)$ & \includegraphics[width=5cm]{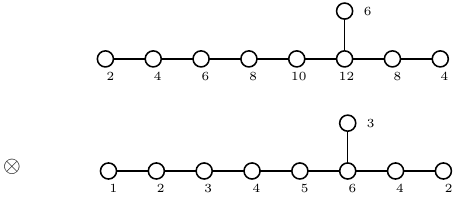} & 88 & $(72; e_8),\ (75, A_1)$ \\ \hline
$(A_4^{e_8}, 72)$ & \includegraphics[width=5cm]{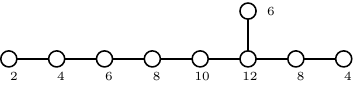} & 59 & $(76; A_1)$ \\ \hline
$(A_4^{e_8}, 73)$ & \includegraphics[width=5cm]{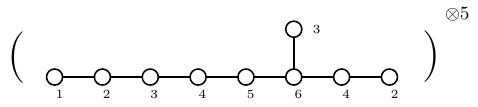} & 145 & $(74; e_8)$ \\ \hline
$(A_4^{e_8}, 74)$ & \includegraphics[width=5cm]{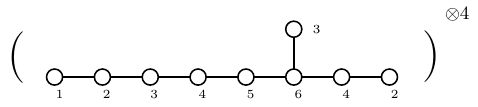} & 116 & $(75; e_8)$ \\ \hline
$(A_4^{e_8}, 75)$ & \includegraphics[width=5cm]{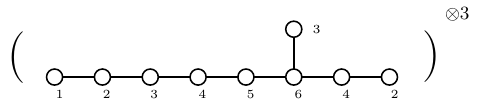} & 87 & $(76; e_8)$ \\ \hline
$(A_4^{e_8}, 76)$ & \includegraphics[width=5cm]{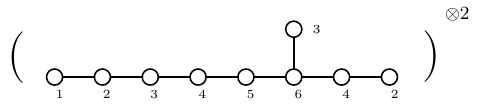} & 58 & $(77; e_8)$ \\ \hline
$(A_4^{e_8}, 77)$ & \includegraphics[width=5cm]{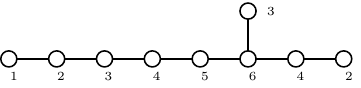} & 29 & $(78; e_8)$ \\ \hline
$(A_4^{e_8}, 78)$ & $\varnothing$ & 0 & IR theory \\ \hline
\caption{Magnetic quivers corresponding to the nodes in the Hasse diagram of the $A_4$-type orbi-instanton theory. The labels in blue correspond to the homomorphisms $\mathbb{Z}_4\rightarrow E_8$.}\label{A4_orbi-instanton_MQs_table}
\end{longtable}

\subsubsection{SU-Type Theories on \texorpdfstring{$-2$}{-2} Curves}\label{SUtypeMQs}
Let us now consider the $A_3^3$ theory as discussed in \S\ref{SUtype}:
\begin{equation}
    [\SU(4)] \ \ \overset{\ksu(4)}{2} \ \ \overset{\ksu(4)}{2} \ \ \overset{\ksu(4)}{2} \ \ [\SU(4)].
\end{equation}
The Hasse diagram is obtained in Figure \ref{A3rank3Hasse}. From the Type IIA brane system
\begin{equation}
    \includegraphics[width=10cm]{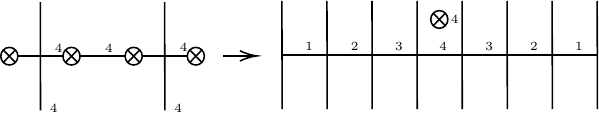},
\end{equation}
we have the magnetic quiver
\begin{equation}
    \includegraphics[width=5cm]{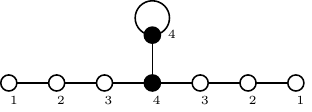}.
\end{equation}
The Hasse diagram obtained from this magnetic quiver can be found in Figure \ref{A3rank3MQHasse}. One can see that the two Hasse diagrams indeed agree\footnote{Notice that the Hasse diagram in Figure \ref{A3rank3MQHasse} has a different convention compared to those in the main context, where it is inverted.}.

\subsection{Orthosymplectic Magnetic Quivers}\label{orthosymplecticMQs}
By introducing the orientifolds to the brane systems, we are allowed to construct magnetic quivers for more theories. Now, these quivers have orthogonal and symplectic nodes \cite{Cabrera:2019dob}, which have also been extensively studied in the literature. However, as opposed to the unitary cases, many tools for the orthosymplectic magnetic quivers are less known and are still under development. Here, we shall apply our current knowledge and see what we can learn from comparing our algorithm for the 6d generalized quivers with the orthosymplectic magnetic quivers.

\subsubsection{Rank 0 D-Type Conformal Matters Revisited}\label{rank0DkospMQs}
Our first example would again be the rank 0 D-type conformal matter theories \eqref{Dkrank0}. Besides the unitary magnetic quivers in \S\ref{rank0DkuMQs}, they also admit orthosymplectic descriptions. For $D_k^0$, we have
\begin{equation}
    \includegraphics[width=13cm]{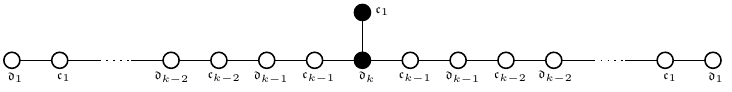}.
\end{equation}
Comparing this with (the decay of) the unitary magnetic quivers, we can verify that the $d_n$ slice is given by the difference of the quivers
\begin{equation}
    \includegraphics[width=7cm]{figures/dnosp1.pdf}.\label{dnosp}
\end{equation}
For $k=4$, the $\mathfrak{d}_4$ node in the magnetic quiver also becomes balanced. This is indeed the $e_8$ slice \cite{Bourget:2020gzi,Bourget:2020xdz} as given in Table \ref{slicesMQs}.

\subsubsection{D-Type Conformal Matters of Higher Ranks}\label{higherrankDospMQs}
For more general conformal matters, the Higgsings would certainly be more complicated. If we have a conformal matter theory of classical type, the magnetic quiver, which would be orthosymplectic, can be obtained from the Type IIA construction by allowing negatively charged branes \cite{Mekareeya:2016yal,Hanany:2022itc}. The algorithm for the generalized quivers should be consistent with the orthosymplectic quiver decays and fissions in the Hasse diagrams under Higgsings. We shall illustrate this with the $(D_4,D_4)$ conformal matter theories here. We also discuss a C-type conformal matter example in Appendix \ref{Sp3conformalmatters}.

\paragraph{Long quivers} As we have seen \S\ref{Dtypes}, the Higgsed theories that are not obtained from endpoint-changing flows all live in the nilpotent hierarchy \cite{Heckman:2016ssk}. In other words, they are labelled by a pair of nilpotent orbits in $\kso(8)$. Analogous to the unitary cases, we shall view them as the orthosymplectic quiver decays. Let us first start with the long quivers. In the Type IIA brane picture, this means that the D8-branes on the left and right sides encoding the nilpotent orbits do not ``cross'' each other. More specifically, the rightmost D8-branes for the left orbit $[m_1^{p_1},\dots]$ should be on the left to the leftmost D8-branes for the right orbit $[n_1^{q_1},\dots]$ (but they are allowed to live in the same NS5 interval):
\begin{equation}
    N\geq m_1+n_1-1,
\end{equation}
where $N=2r+1$ is the number of NS5 intervals for the rank $r$ conformal matter theory\footnote{Therefore, the long quiver condition such that all the nilpotent orbits would satisfy is $N\geq13$, namely rank no less than 6, coming from the (largest) principal orbit $[7,1]$. Of course, for a specific nilpotent orbit, it could be a shorter quiver.}.

The magnetic quivers are then given as follows. There is a $\mathfrak{d}_4$ node in the middle connected to a $\mathfrak{c}_{n=r+1}$ node with an antisymmetric. The two nilpotent orbits give two tails connected to the $\mathfrak{d}_4$ node. Let us now list the brane systems and the tails (including the $\mathfrak{d}_4$ and $\mathfrak{c}_n$ nodes) for the nilpotent orbits:
\begin{itemize}
    \item $[1^8]$:
    \begin{align}
        &[\SO(8)] \ \ 1 \ \ \overset{\kso(8)}{4} \ \ 1 \ \ \overset{\kso(8)}{4} \ \ \dots,\nonumber\\
        &\includegraphics[width=15cm]{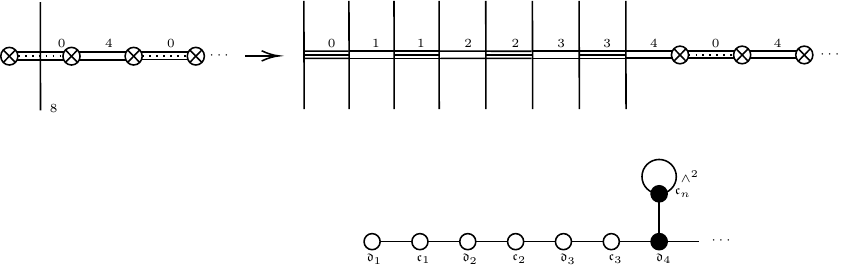};
    \end{align}
    
    \item $[2^2,1^4]$:
    \begin{align}
        &[\SU(2)\times\SU(2)\times\SU(2)] \ \ \overset{\kso(8)}{3} \ \ 1 \ \ \overset{\kso(8)}{4} \ \ 1 \ \ \dots,\nonumber\\
        &\includegraphics[width=15cm]{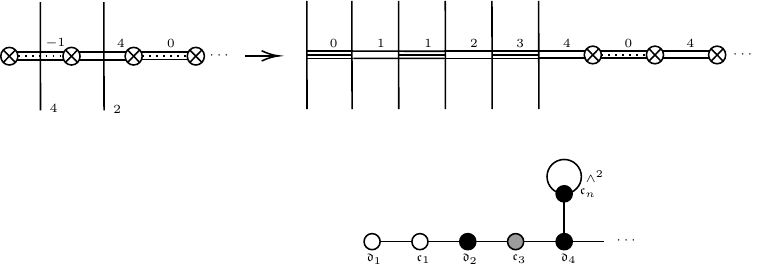};
    \end{align}

    \item $[2^4]^\text{I,II}$:
    \begin{align}
        &[\Sp(2)] \ \ \overset{\kso(7)}{3} \ \ 1 \ \ \overset{\kso(8)}{4} \ \ 1 \ \ \dots,\nonumber\\
        &\includegraphics[width=13cm]{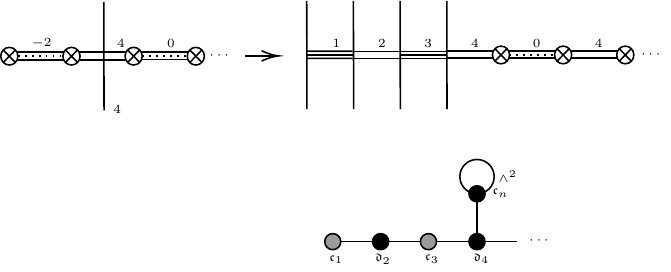};
    \end{align}

    \item $[3,1^5]$:
    \begin{align}
        &[\Sp(2)] \ \ \overset{\kso(7)}{3} \ \ 1 \ \ \overset{\kso(8)}{4} \ \ 1 \ \ \dots,\nonumber\\
        &\includegraphics[width=15cm]{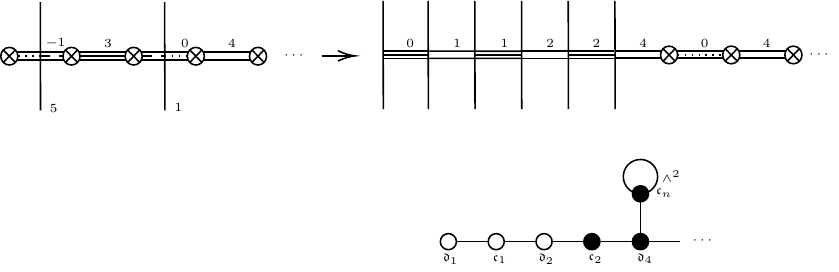};
    \end{align}

    \item $[3,2^2,1]$ (non-special):
    \begin{align}
        &[\SU(2)] \ \ \overset{\kg_2}{3} \ \ 1 \ \ \overset{\kso(8)}{4} \ \ 1 \ \ \dots,\nonumber\\
        &\includegraphics[width=13cm]{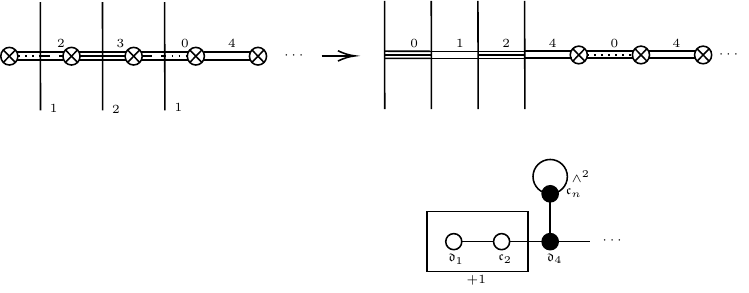};
    \end{align}

    \item $[3^2,1^2]$:
    \begin{align}
        & \overset{\ksu(3)}{3} \ \ 1 \ \ \overset{\kso(8)}{4} \ \ 1 \ \ \dots,\nonumber\\
        &\includegraphics[width=13cm]{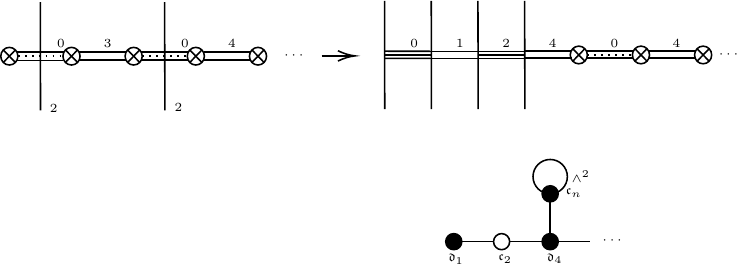};
    \end{align}

    \item $[4^2]^\text{I,II}$:
    \begin{align}
        & \overset{\ksu(2)}{2} \ \ \underset{[\SU(2)]}{\overset{\kso(7)}{3}} \ \ 1 \ \ \overset{\kso(8)}{4} \ \ 1 \ \ \dots,\nonumber\\
        &\includegraphics[width=9cm]{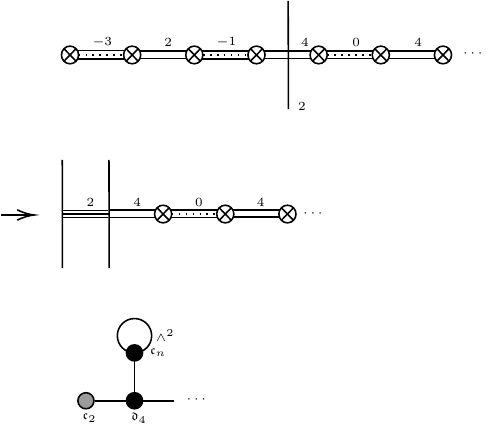};
    \end{align}

    \item $[5,1^3]$:
    \begin{align}
        & \overset{\ksu(2)}{2} \ \ \underset{[\SU(2)]}{\overset{\kso(7)}{3}} \ \ 1 \ \ \overset{\kso(8)}{4} \ \ 1 \ \ \dots,\nonumber\\
        &\includegraphics[width=9cm]{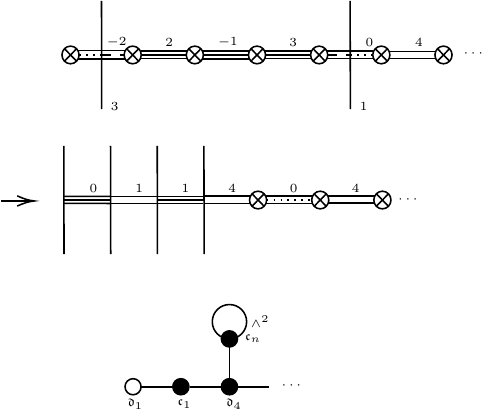};
    \end{align}

    \item $[5,3]$:
    \begin{align}
        & \overset{\ksu(2)}{2} \ \ \overset{\kg_2}{3} \ \ 1 \ \ \overset{\kso(8)}{4} \ \ 1 \ \ \dots,\nonumber\\
        &\includegraphics[width=9cm]{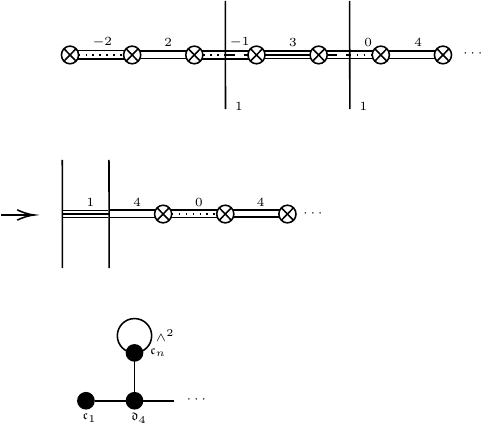};
    \end{align}

    \item $[7,1]$:
    \begin{align}
        & 2 \ \ \overset{\ksu(2)}{2} \ \ \overset{\kg_2}{3} \ \ 1 \ \ \overset{\kso(8)}{4} \ \ 1 \ \ \dots,\nonumber\\
        &\includegraphics[width=11cm]{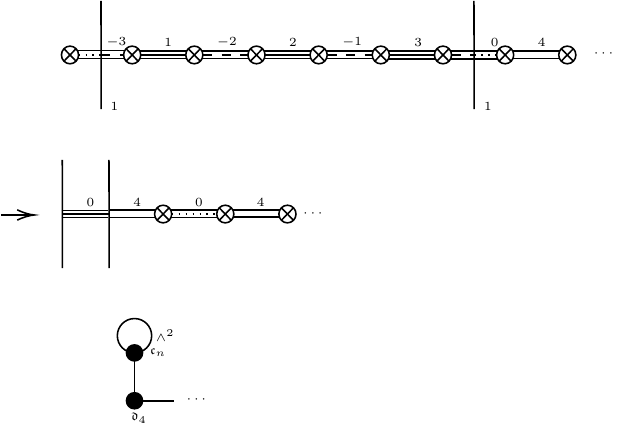}.
    \end{align}
\end{itemize}

Among these tails, there is a nilpotent orbit that is non-special. For any non-special orbit, the magnetic quiver does not give the desired Coulomb branch that is equal to the Higgs branch of the 6d theory as moduli spaces\footnote{For the very even orbits which are special, there are also some subtleties as already mentioned in \S\ref{intro}, and we shall not repeat this here.}. In fact, the resulting quiver is always the same as the one associated with the special orbit which is the LS dual orbit of the non-special one. From the perspective of the Type IIA brane set-ups, this is because the brane transitions in such cases do not involve creations and annihilations of physical D6-branes, and hence the moduli cannot be distinguished from its special LS dual. Such phenomenon has already been observed in \cite{Cabrera:2017njm,Hanany:2022itc} (see also \cite{Chacaltana:2012zy,Balasubramanian:2023iyx} for relevant discussions).

Therefore, in the tail associated with the non-special orbit, we put a box surrounding it with a number attached. This number indicates the difference between the dimension of the actual moduli space and the dimension from the naive quiver. For instance, in the $[3,2^2,1]$ case here, the magnetic quiver obtained from the brane picture (which is the same as the one for $[3^2,1^2]$) has the Coulomb branch of dimension $7+n+\dots$, but the dimension is $8+n+\dots$ for the Higgs branch of the corresponding 6d theory. We will make a further comment on the non-special orbits at the end of this subsubsection.

From the nilpotent Higgsings, we can also tell what the difference of the magnetic quivers looks like for a corresponding slice. Again, we can see that the $d_n$ slices are those given in \eqref{dnosp} and in \S\ref{nilporbslices}. For the flows involving the non-special one, the slice is straightforward from the generalized quivers as they are minimal nilpotent orbit Higgsings. Moreover, there are differences between the parent and child magnetic quivers given by either a single $\mathfrak{c}_1$ node or a single $\mathfrak{d}_1$ node. Since they all belong to the nilpotent hierarchy, we can see that they are actually different slices:
\begin{itemize}
    \item $[3^2,1^2]\rightarrow[4,^2]^\text{I,II}$: The difference is a single $\mathfrak{d}_1$ node. The slice is $2A_3$.
    \item $[3^2,1^2]\rightarrow[5,1^3]$: The difference is a single $\mathfrak{c}_1$ node. The slice is $C_2$.
    \item $[5,3]\rightarrow[7,1]$: The difference is a single $\mathfrak{c}_1$ node. The slice is $D_4$.
\end{itemize}

Now, let us consider the quiver fissions in this example. They correspond to the endpoint-changing Higgsings. It turns out that there are no fissions from the induced orbit/partition splitting trick, and the only possible fissions are
\begin{equation}
    \includegraphics[width=10cm]{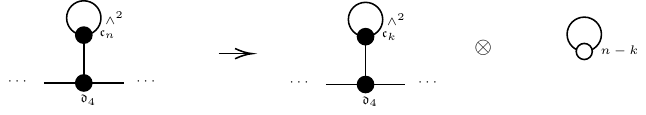}.
\end{equation}
To separate a unitary piece, we need $n-k\geq2$. To maintain the long quiver condition, we need $N'\geq m_1+n_1-1$, where $N'=2r'+1$ and $k=r'+1$. In other words, $2k\geq m_1+n_1$. Therefore, we have $\frac{m_1+n_1}{2}\leq k\leq n-2$. We shall discuss more non-trivial fissions in \S\ref{ospfissions}. Here, let us just mention that although it is clear that the slice for this Higgsing is 1-dimensional as expected, the exact nature of this slice is still not known. For unitary magnetic quivers, such fission gives the non-normal singularity $m$. It is natural to expect that orthosymplectic quiver fission would have something similar, and it is denoted as $\widetilde{m}$ in the Hasse diagram.

From the above discussions, we can see that there are only finitely many Higgsed theories from the quiver decays. On the other hand, as the rank grows larger, the number of possible theories from the quiver fissions increases.

\paragraph{Rank 3} For the short quivers, let us illustrate this with the case of rank 3. The Hasse diagram is given in Figure \ref{D4rank3Hasse}. We shall only consider the magnetic quivers with the pair of nilpotent orbits satisfying $m_1+n_1>8$ here. The other cases in the decay process, as well as those in the fission process, are exactly the same as the above discussions for the long quivers.

The first example would be the nilpotent orbit pair $[5,1^3]\text{-}[5,1^3]$. If one simply takes the tail for the long quiver, which is a $\mathfrak{d}_1$ node connected to a $\mathfrak{c}_1$ node, the resulting quiver does not give the correct Coloumb branch. This can be most easily seen from the dimension. The naive quiver has the Coulomb branch of dimension 12, but the 6d Higgs branch is of dimension 11. To get the right magnetic quiver, we start with the corresponding brane system
\begin{equation}
    \includegraphics[width=9cm]{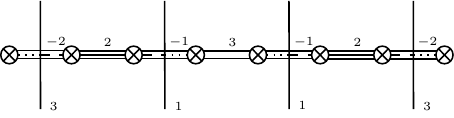}.
\end{equation}
After brane transitions, the magnetic quiver reads
\begin{equation}
    \includegraphics[width=4cm]{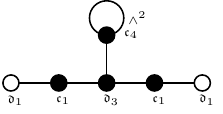}.
\end{equation}
In fact, by looking at the brane system, we might think of this as the same as the child (long quiver) theory with nilpotent orbit pair $[3,1^3]\text{-}[3,1^3]$ of the rank 3 ``$(\SO(6),\SO(6))$ conformal matter theory''.

Likewise, for the pair $[5,1^3]\text{-}[4^2]^\text{I,II}$, we have the brane system
\begin{equation}
    \includegraphics[width=9cm]{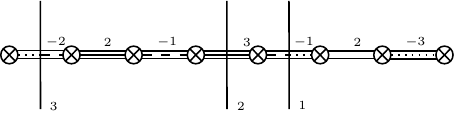}.
\end{equation}
After brane transitions, the magnetic quiver reads
\begin{equation}
    \includegraphics[width=3cm]{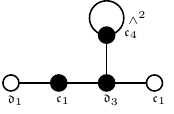}.\label{D4rank3ex2MQ}
\end{equation}
By looking at the brane system, we might think of this as the partition pair $[4,1^3]\text{-}[4,3]$. Notice that, however, they are not nilpotent orbits in $\kso(7)$.

Let us also mention the flow from the magnetic quivers for the pairs $[5,1^3]/\left[4^2\right]^\text{I,II}\text{-}\left[3^2,1^2\right]$,
\begin{equation}
    \includegraphics[width=9cm]{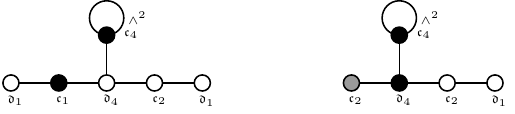},
\end{equation}
to the quiver in \eqref{D4rank3ex2MQ}. We can see that the two differences of the quivers
\begin{equation}
    \includegraphics[width=7cm]{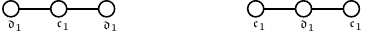}
\end{equation}
both give the $d_3$ slice.

However, there still exist cases where the brane systems cannot give the magnetic quivers with the right Coulomb branches, even if the orbits are special. Surprisingly, we find that for the pair $[4^2]^\text{I,II}\text{-}[4^2]^\text{I,II}$ (ignoring the subtleties of the very even orbits), the magnetic quiver obtained from the brane system does not coincide with the corresponding 6d Higgs branch (although the magnetic quiver derived from the brane system is consistent with the tails for the long quivers since $m_1+n_1=8$). Nevertheless, in this case, we still have a correct description from the magnetic quiver, which is \eqref{D4rank3ex2MQ}, since this is the same as the case with $[5,1^3]\text{-}[4^2]^\text{I,II}$.

If one continues the process of Higgsings, the brane constructions would lose the effect, and there could be cases that are not associated with $\kso(8)$ nilpotent orbits appearing. For instance, the theory for the nilpotent orbit pair $[5,3]\text{-}[5,1^3]/[4^2]^\text{I,II}$, which has label $(D_4^3,37)$, has two corresponding brane systems, but neither of them has the orthosymplectic magnetic quiver that recovers the corresponding 6d Higgs branches as can be easily seen from the dimension check. More explicitly, we have
\begin{equation}
    \includegraphics[width=15cm]{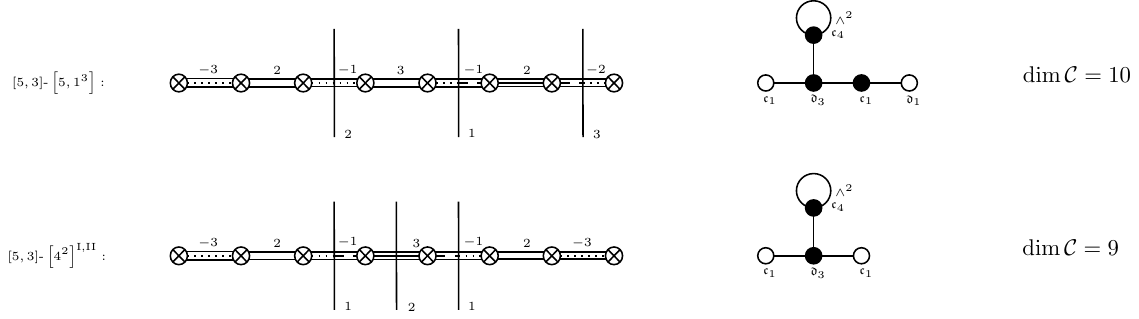}.
\end{equation}
However, the dimension of the 6d Higgs branch is 8. For $(D_4^3,33)$, this is not even associated with any $\kso(8)$ nilpotent orbits (and it is not the only case here). Nevertheless, for all but one of them, there is a unitary magnetic quiver since they are also in the Hasse diagram of the $A_3^3$ theory. Determining the quiver decays and the transverse slices certainly becomes more subtle when one goes between an orthosymplectic quiver and a unitary one. It turns out that in this example, all such unknown slices are 1-dimensional (which may or may not be the same) as labelled by the question marks in Table \ref{D4rank3_table}. In this paper, we shall not pursue the solution to this problem. Let us simply write the flow, for instance, from $(D_4^3,6)$ to $(D_4^3,16)$, schematically as
\begin{equation}
    \includegraphics[width=13cm]{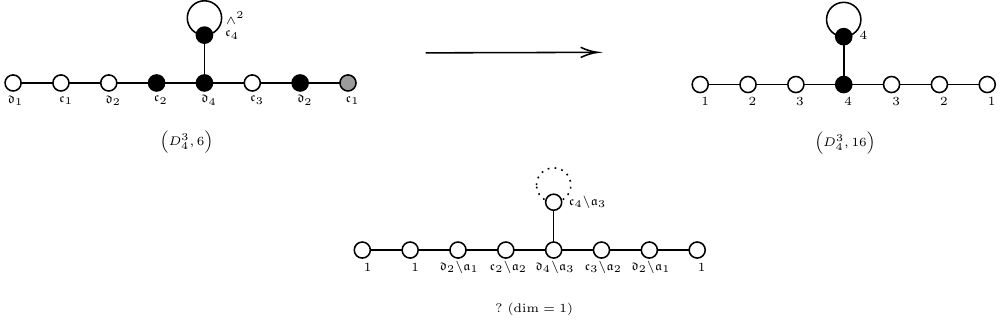}.\label{D4rank3ospuex}
\end{equation}

The only theory whose magnetic quiver is unknown (besides the ones associated with the non-special orbits), either unitary or orthosymplectic, is $(D_4^3,37)$. In fact, this was already encountered in \cite{Hanany:2022itc}, where the Higgsings $(D_4^3,35)\rightarrow(D_4^3,37)\rightarrow(D_4^3,38)$ were discussed. It is worth noticing that in the Hasse diagram, its adjacent parent theories, $(D_4^3,27)$ and $(D_4^3,35)$ have orthosymplectic magnetic quiver descriptions while its adjacent child theory, $(D_4^3,38)$, has a unitary magnetic quiver description. In particular, the slice from Higgsing $(D_4,35)$ is $f_4$ whose orthosymplectic version is still not clear, and the 1-dimensional slice from Higgsing $(D_4^3,27)$ is unknown. The slice transverse to $(D_4^3,38)$ in $(D_4^3,37)$ is $c_3$ whose unitary magnetic quiver is certainly known, but one cannot determine a quiver from the inverse process of decay.

\paragraph{Non-special orbits} To the best of our knowledge, there is still no solution to finding the actual 3d $\mathcal{N}=4$ magnetic quivers when non-special nilpotent orbits are involved. It is natural to wonder if we could guess what the magnetic quivers look like following the strategy of quiver decays. More concretely, the Higgsed theory after decay should be a smaller quiver in the sense that the ranks of the gauge nodes should be smaller than or equal to those in the parent quiver. In terms of the brane system, some of the D6-branes are ``thrown away'' under Higgsing. However, it seems that the non-special cases do not fit into this simple pattern.

This is most obvious by considering the example with the non-special orbit $\left[3,2^2,1^5\right]$ in $\mathfrak{so}(12)$. By comparing the quivers and brane systems of its adjacent nodes in the Hasse diagram\footnote{Notice that the one associated to $\left[3,2^2,1^5\right]$ can also be Higgsed to the one associated to $[\left[3,2^4,1\right]$. However, since $\left[3,2^4,1\right]$ is also non-special, we shall omit it here.}
\begin{equation}
    \includegraphics[width=15cm]{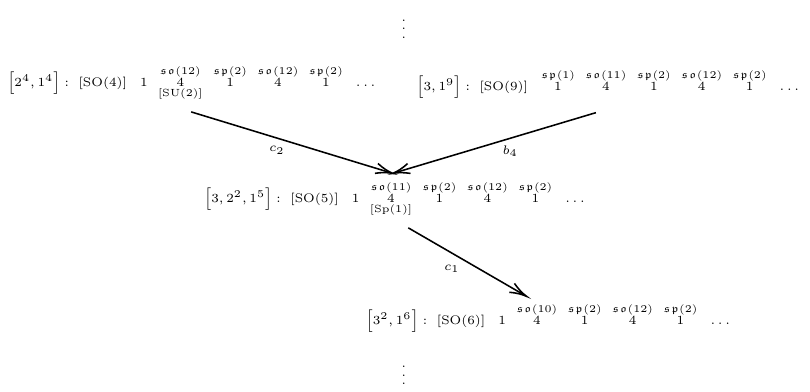},
\end{equation}
one would get
\begin{equation}
    \includegraphics[width=15cm]{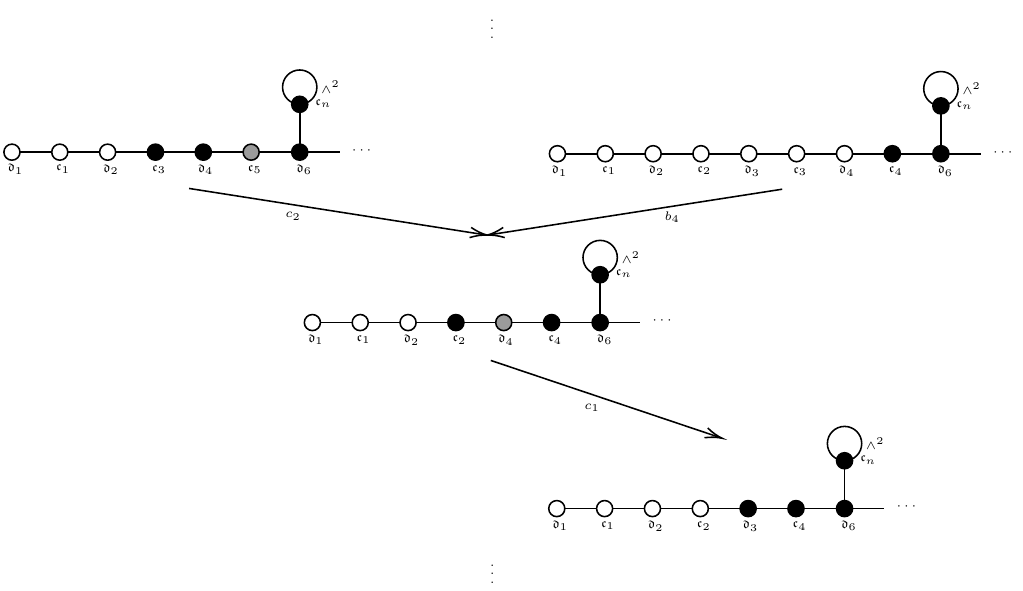}.
\end{equation}
However, if this is the case, then the $c_2$ slice would be given by two disconnected $\mathfrak{c}_1$ nodes. If one continues the similar guessing for the non-special orbit $\left[3,2^4,1\right]$, then the $c_3$ flow from $\left[2^6\right]$ to it would be given by $\mathfrak{d}_1-\mathfrak{c}_1-\mathfrak{d}_1$ (which should actually represent the $d_3$ slice), and the slice to $\left[3,2^4,1\right]$ in $\left[3,2^2,1^5\right]$ would be given by two disconnected $\mathfrak{d}_1$ nodes. Therefore, it remains unclear whether/how one could construct the magnetic quivers associated with the non-special nilpotent orbits, and we leave this to future work.

\subsubsection{Rank 0 \texorpdfstring{$(E_6,E_6)$}{(E6,E6)} Conformal Matter}\label{rank0E6ospMQ}
For E-type conformal matter theories, the magnetic quivers are still not known. Here, we shall construct the magnetic quiver for the rank 0 $(E_6,E_6)$ conformal matter theory
\begin{equation}
    [E_6] \ \ 1 \ \ \overset{\ksu(3)}{3} \ \ 1 \ \ [E_6],\label{E6rank0}
\end{equation}
whose Hasse diagram is the sub-diagram composed of nodes with blue labels in Figure \ref{E7rank0Hasse}.

In fact, the magnetic quiver for its only minimal Higgsing descendant, $(E_7^0,24)=(E_6^0,1)$ in Table \ref{E7rank0_table}, is already known\footnote{Notice that this is also the $(A_4^{e_8},46)$ theory in Table \ref{A4_orbi-instanton_MQs_table}.} \cite{Bourget:2024mgn}:
\begin{equation}
    \includegraphics[width=9cm]{figures/A4orbi-instanton46.pdf}.\label{E6rank0MQ1}
\end{equation}
The dimension of its Coulomb branch, i.e., the dimension of the Higgs branch of $(E_6^0,1)$, is 68. From the computation of the anomaly polynomial, we know that the dimension of the Higgs branch of $(E_6^0,0)$ in \eqref{E6rank0} is 79. They differ by 11 as expected since the slice is $e_6$. It is then natural to wonder if we can guess the magnetic quiver for $(E_6^0,0)$ by adding an $e_6$ piece to the $(E_6^0,1)$ one, using the corresponding affine Dynkin quiver, either untwisted or twisted. However, by checking the resulting Hasse diagrams, we find that none of the possible guesses would give the right result.

Of course, using the brane system is a standard way to construct the magnetic quivers. It is believed that E-type conformal matter theories may not have Type IIA brane constructions. However, including the negatively charged branes renders this possible. For \eqref{E6rank0}, the main difficulty comes from the $E_6$ flavours. Luckily, there is a similar case that appears in the Higgsings of the rank 1 $(D_4,D_4)$ conformal matter theory. In Appendix \ref{D4lowranks}, the one associated with the nilpotent orbit pair $[3^2,1^2]\text{-}[1^8]$ is basically ``half'' of \eqref{E6rank0}. There is a $-1$ curve with an $E_6$ flavour intersecting a $-3$ curve decorated by an $\mathfrak{su}(3)$ gauge algebra. It is not easy to directly add things to the other side of this brane system to get another copy of the $E_6$ flavour, but we can treat this symmetric configuration in \eqref{E6rank0} as a bifurcation and then use the ON-planes. More explicitly, we have
\begin{equation}
    \includegraphics[width=13cm]{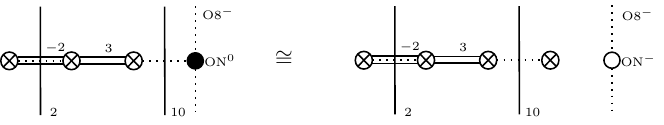}.
\end{equation}
After HW transitions, we get
\begin{equation}
    \includegraphics[width=11cm]{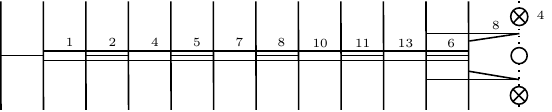}.
\end{equation}
From this, we can read off the magnetic quiver\footnote{While we were about to submit this paper, we noticed that the same magnetic quiver was also obtained in \cite{Bennett:2024llh} using a slightly different method.}:
\begin{equation}
    \includegraphics[width=10cm]{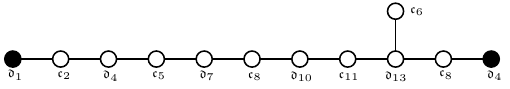}.\label{E6rank0MQ}
\end{equation}
One may check that the dimension of the Coulomb branch is exactly 79. Although one is orthosymplectic while the other is unitary, the magnetic quivers \eqref{E6rank0MQ} and \eqref{E6rank0MQ1} are of the same shape. As listed in Table \ref{E6rank0MQ_table}, most of the Higgsed theories can be obtained by the quiver decay and fission algorithm as all but one magnetic quiver are unitary.
\begin{longtable}{|c|c|c|c|}
\hline
Label &  Magnetic quiver & $d_{\mathbb{H}}$ & (descendant \#; flow type) \\ \hline \hline
$(E_6^0,0)=(E_7^0, 18)$ & \includegraphics[width=5cm]{figures/E6rank0MQ.pdf} & 79 & (1;$e_6$) \\ \hline
$(E_6^0,1)=(E_7^0, 24)$ & \includegraphics[width=5cm]{figures/A4orbi-instanton46.pdf} & 68 & (2;$e_6$), (3;$a_5$) \\ \hline
$(E_6^0,2)=(E_7^0, 31)$ & \includegraphics[width=5cm]{figures/A4orbi-instanton57.pdf} & 57 & (4;$a_{11}$) \\ \hline
$(E_6^0,3)=(E_7^0, 32)$ & \includegraphics[width=5cm]{figures/A4orbi-instanton65.pdf} & 63 & (4;$e_7$), (5;$b_3$) \\ \hline
$(E_6^0,4)=(E_7^0, 38)$ & \includegraphics[width=5cm]{figures/A4orbi-instanton68.pdf} & 46 & (7,$d_{10}$) \\ \hline
$(E_6^0,5)=(E_7^0, 39)$ & \includegraphics[width=5cm]{figures/A4orbi-instanton72.pdf} & 59 & (6,$A_1$) \\ \hline
$(E_6^0,6)=(E_7^0, 40)$ & \includegraphics[width=5cm]{figures/A4orbi-instanton76.pdf} & 58 & (7,$e_8$) \\ \hline
$(E_6^0,7)=(E_7^0, 41)$ & \includegraphics[width=5cm]{figures/e8u.pdf} & 29 & (8,$e_8$) \\ \hline
$(E_6^0,8)=(E_7^0, 42)$ & $\varnothing$ & 0 & IR theory \\ \hline
\caption{The magnetic quivers for the rank 0 $E_6$ conformal matter theory and its Higgsed theories.}\label{E6rank0MQ_table}
\end{longtable}
For the Higgsing from $(E_6^0,0)$ to $(E_6^0,1)$, we start from an orthosymplectic quiver and get a unitary quiver, which is not standard in all the magnetic quiver manipulations. Nevertheless, we know that this is the $e_6$ slice. Therefore, let us schematically write it as\footnote{Alternatively, we could apply the quiver subtractions to this case. One can check that the Hasse diagram can be recovered by subtracting the unitary magnetic quiver slices (where we get the orthosymplectic version of $e_6$ to subtract at the last step). We would like to thank Deshuo Liu for pointing this out.}
\begin{equation}
    \includegraphics[width=13cm]{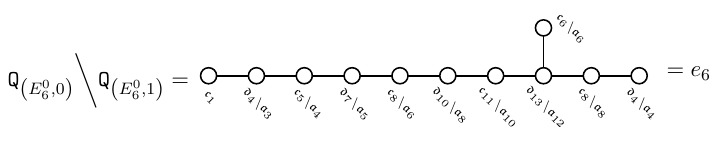}.\label{e6fromE6rank0}
\end{equation}
Recall that the theory has an $E_6\times E_6$ flavour symmetry which is manifest in the generalized quiver in \eqref{E6rank0}. This indicates that $(E_6^0,0)$ can be Higgsed to two $(E_6^0,1)$ theories corresponding to two $e_6$ slices (and hence also two $(E_6^0,2)\sim(E_6^0,7)$ theories) although we are omitting the isomorphic nodes in the Hasse diagrams throughout the main context of the paper. Therefore, the magnetic quiver in \eqref{E6rank0MQ} should implicitly contain two copies of \eqref{e6fromE6rank0} at the same place. This is also clear from the brane system since it is realized by a bifurcation.

For the $E_6$ conformal matters of higher ranks or the $E_7$, $E_8$ conformal matters, the magnetic quivers are still not known. The bifurcation trick here does not seem to apply to these cases. Another possible approach is to consider their compactified 5d theories as in \cite{Hayashi:2021pcj}. However, how to obtain the magnetic quivers from the web diagrams with multi-valent gluings is still not clear.

\subsubsection{The Tri-Leg \texorpdfstring{$D_4$}{D4} Theory}\label{sec:D4hatrank1MQ}
Let us now consider the $\widehat{D}_4^1$ theory:
\begin{align}
    [\SO(8)] \ \ 1 \ \ &\overset{\kso(8)}{4} \ \ 1 \ \ [\SO(8)]\nonumber\\
    \quad\quad\quad\quad\quad\quad&\ \ 1\\ 
    [&\SO(8)]\nonumber,
\end{align}
whose Hasse diagram can be found in \S\ref{D4hatrank1}. To obtain its magnetic quiver, we can again treat two of the legs as a bifurcation. The brane system reads
\begin{equation}
    \includegraphics[width=13cm]{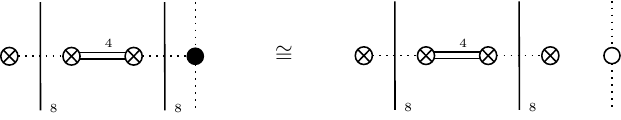}.
\end{equation}
After HW transitions, we get
\begin{equation}
    \includegraphics[width=13cm]{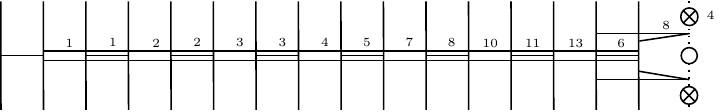}.
\end{equation}
The magnetic quiver is then
\begin{equation}
    \includegraphics[width=10cm]{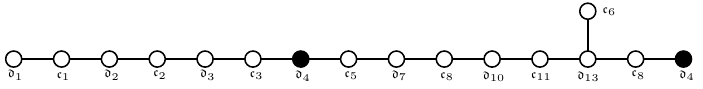}\label{D4hatrank1MQ}
\end{equation}
whose Coulomb branch is indeed of dimension 88.

As the $\widehat{D}_4^1$ theory can be viewed as the $D_4^1$ conformal matter theory with an extra leg added yielding a bifurcation, many of its descendant theories in Table \ref{D4hatrank1_table} would have similar structures. As a result, it is possible to obtain the magnetic quivers in the same manner for some of these theories. For $\left(\widehat{D}_4^1,1\right)$, we have
\begin{align}
    &[\SO(8)] \ \ 1 \ \ \underset{[\SU(2)^3]}{\overset{\kso(8)}{3}} \ \ 1 \ \ [\SO(8)],\\
    &\includegraphics[width=10cm]{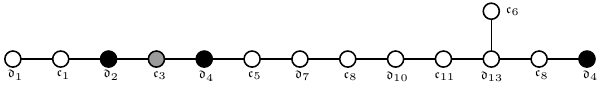}.
\end{align}
For $\left(\widehat{D}_4^1,2\right)$, we have
\begin{align}
    &[\Sp(2)^3] \ \ \overset{\kso(8)}{2} \ \ 1 \ \ [\SO(8)],\\
    &\includegraphics[width=10cm]{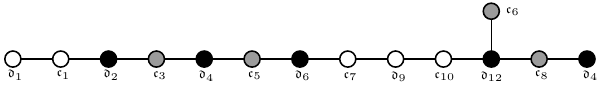}.
\end{align}
For $\left(\widehat{D}_4^1,3\right)$, we have
\begin{align}
    &[\SO(9)] \ \ 1 \ \ \underset{[\Sp(2)]}{\overset{\kso(7)}{2}} \ \ 1 \ \ [\SO(9)],\\
    &\includegraphics[width=10cm]{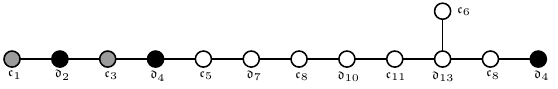}.
\end{align}
The slices of the transitions are clear from the magnetic quivers. For the remaining theories, some also appear in the Higgsings of the rank 0 $(E_6,E_6)$ conformal matter theory and/or the $A_4$ orbi-instanton theory. Therefore, we shall not repeat them here. We have not found the magnetic quiver descriptions for the other theories yet. For instance, although the $\left(\widehat{D}_4^1,6\right)$ theory can have a brane construction in a similar way, it is a non-special case in the following sense. The brane system of this theory can be obtained by introducing the ON-plane to the $(D_4^1,5)$ theory (see Table \ref{D4rank1_table} and Appendix \ref{D4lowranks}):
\begin{equation}
    [\Sp(1)]\ \ \overset{\kg_2}{3} \ \ 1 \ \ [F_4]\ \ \xrightarrow{\text{bifurcation}} \ \ [F_4] \ \ 1 \ \ \underset{[\Sp(1)]}{\overset{\kg_2}{3}} \ \ 1 \ \ [F_4].
\end{equation}
However, the $(D_4^1,5)$ theory is associated to the nilpotent orbit pair $[3,2^2,1]\text{-}[1^8]$, where $[3,2^2,1]$ is non-special. This renders the magnetic quiver obtained this way not being the desired one for $\left(\widehat{D}_4^1,6\right)$. Likewise, we do not have a magnetic quiver for $\left(\widehat{D}_4^1,5\right)$. Although the corresponding $(D_4^1,4)$ theory is associated with the special nilpotent orbit pairs, such construction somehow loses its effect for this specific short quiver.

Let us also make a comment on the quiver fission for $\left(\widehat{D}_4^1,0\right)$. It should flow to a configuration with 3 disconnected $e_8$ pieces, with the 1-dimensional transverse slice denoted as $\widetilde{A}_1$. Although there could be problems in recovering the full Hasse diagram using quiver subtractions, for this flow, we may subtract $e_8$ and then $2e_8$ to obtain the corresponding leaf. It turns out that this leaf has the same magnetic quiver as the quiver for the rank 1 $(D_4,D_4)$ conformal matter theory, which does not admit a $3e_8$ flow (if we insist on subtracting an (orthosymplectic) $e_8$ magnetic quiver, we would then obtain the Kleinian singularity $D_4$). Therefore, the $\widetilde{A}_1$ flow seems to have some subtleties from the perspective of quiver subtractions. Nevertheless, suppose this flow does exist. Then it is tempting to identify $\widetilde{A}_1$ with $D_4$ despite these subtleties.

\subsubsection{Orthosymplectic Quiver Fissions}\label{ospfissions}
As discussed in \S\ref{endptchanging}, there is a type of flow that would change the curve configurations in the F-theory descriptions, and hence the name endpoint-changing flows. In many cases of such flows, the moduli spaces after Higgsings are then the product spaces of several pieces. For unitary magnetic quivers, this corresponds to the quiver fissions. We expect that these flows are also reflected by fissions in the orthosymplectic cases. Here, let us give an example.

The orthosymplectic magnetic quivers for theories under minimal nilpotent orbit Higgsings from the $(D_6,D_6)$ conformal matter theory were summarized in \cite[Appendix B]{Hanany:2022itc} for the long quivers\footnote{Since the magnetic quiver descriptions for the non-special orbits are still not complete, we shall consider the special ones here.}. Consider the theory associated with the nilpotent orbit pair $[5,3,1^4]\text{-}[3^2,1^6]$:
\begin{equation}
    [\SU(2)\times\SU(2)] \ \ \overset{\kso(8)}{3} \ \ \overset{\ksp(1)}{1} \ \ \overset{\kso(11)}{4} \ \ \overset{\ksp(2)}{1} \ \ \overset{\kso(12)}{4} \ \ \dots \ \ \overset{\kso(12)}{4} \ \ \overset{\ksp(2)}{1} \ \ \overset{\kso(10)}{4} \ \ 1 \ \ [\SO(6)],
\end{equation}
whose magnetic quiver reads
\begin{equation}
    \includegraphics[width=10cm]{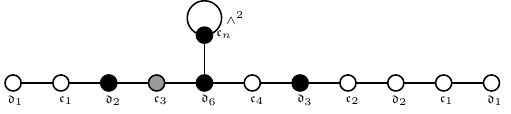}.\label{ospfissionex0}
\end{equation}

For theories of this form, an obvious fission would be separating a single unitary node $(n-k)$ with an adjoint. The other piece would be simply changing the $\mathfrak{c}_n$ node to $\mathfrak{c}_k$. This requires $2k\geq8$ to remain the long quiver condition, and $n-k\geq2$ to have a unitary quiver part\footnote{In other words, at least 4 half NS5-branes would be separated so that they would combine into 2 full NS5-branes that form the brane system for the unitary magnetic quiver.} Therefore, $4\leq k\leq n-2$.

There could also be more non-trivial fissions. For $\mathfrak{g}=\kso(12)$, it has a Levi subalgebra $\mathfrak{l}=\mathfrak{u}(2)\oplus\kso(8)$. Therefore, we expect that this theory can be split into an $\SO(8)$-type theory and an $\SU(2)$-type theory. Indeed, the above nilpotent orbits are some induced orbits from the orbits in $\mathfrak{l}$:
\begin{equation}
    [5,3,1^4]=\text{Ind}_{\mathfrak{l}}^{\mathfrak{g}}(\mathcal{O}_{\mathfrak{l}})=\text{Ind}_{\mathfrak{l}}^{\mathfrak{g}}(\mathcal{O}'_{\mathfrak{l}}),\quad[3^2,1^6]=\text{Ind}_{\mathfrak{l}}^{\mathfrak{g}}(\mathcal{O}''_{\mathfrak{l}}),
\end{equation}
where
\begin{equation}
    \mathcal{O}_{\mathfrak{l}}=\mathcal{O}_{[2]}\oplus\mathcal{O}_{[2^2,1^4]},\quad\mathcal{O}'_{\mathfrak{l}}=\mathcal{O}_{[1^2]}\oplus\mathcal{O}_{[3,1^5]},\quad\mathcal{O}''_{\mathfrak{l}}=\mathcal{O}_{[1^2]}\oplus\mathcal{O}_{[1^8]}.
\end{equation}
In each of these decompositions, the first part is an orbit in $\mathfrak{sl}(2)$ and the second part is an orbit in $\mathfrak{so}(8)$. As a result, the parent theory can be Higgsed to an $\SO(8)$-type theory whose nilpotent orbit pair is $[2^2,1^4]\text{-}[1^8]$ (resp.~$[3,1^5]\text{-}[1^8]$) with an $\SU(2)$-type theory whose nilpotent orbit pair is $[2]\text{-}[1^2]$ (resp.~$[1^2]\text{-}[1^2]$).

To summarize, the magnetic quiver in \eqref{ospfissionex0} can be split into
\begin{equation}
    \includegraphics[width=10cm]{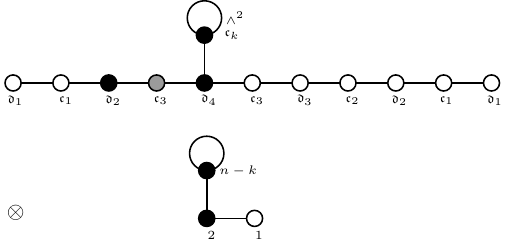}
\end{equation}
for $5\leq k\leq n-3$, or
\begin{equation}
    \includegraphics[width=10cm]{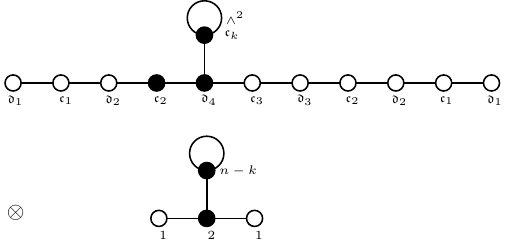}
\end{equation}
for $6\leq k\leq n-2$. The generalized quivers are
\begin{align}
    &[\SU(2)\times\SU(2)\times\SU(2)] \ \ \overset{\kso(8)}{3} \ \ 1 \ \ \overset{\kso(8)}{4} \ \ 1 \ \ \overset{\kso(8)}{4} \ \ \dots \ \ \overset{\kso(8)}{4} \ \ 1 \ \ \overset{\kso(8)}{4} \ \ 1 \ \ [\SO(8)]\nonumber\\
    \sqcup& \ \ 2 \ \ \underset{[N_f=1]}{\overset{\ksu(2)}{2}} \ \ \overset{\ksu(2)}{2} \ \ \overset{\ksu(2)}{2} \ \ \dots \ \ \overset{\ksu(2)}{2} \ \ \overset{\ksu(2)}{2} \ \ [\SU(2)]
\end{align}
and
\begin{align}
    &[\Sp(2)] \ \ \overset{\kso(7)}{3} \ \ 1 \ \ \overset{\kso(8)}{4} \ \ 1 \ \ \overset{\kso(8)}{4} \ \ \dots \ \ \overset{\kso(8)}{4} \ \ 1 \ \ \overset{\kso(8)}{4} \ \ 1 \ \ [\SO(8)]\nonumber\\
    \sqcup& \ \ [\SU(2)] \ \ \overset{\ksu(2)}{2} \ \ \overset{\ksu(2)}{2} \ \ \overset{\ksu(2)}{2} \ \ \dots \ \ \overset{\ksu(2)}{2} \ \ \overset{\ksu(2)}{2} \ \ [\SU(2)]
\end{align}
respectively. One may check that they satisfy the criteria for the endpoint-changing flows. Moreover, the slices are always of dimension 1.

When the theories are associated with nilpotent orbits, the fission process in the Higgsings gives a nice physical interpretation of the induced orbits. Whether the orbit is some induced orbit from certain Levi subalgebras provides a criterion for the possible fissions. Of course, for classical-type theories, we may also take a less cultivated way to split the partitions directly. This may be considered as some sort of ``splitting'' of the brane system. First, the LS duals of the two orbits are
\begin{equation}
    \mathtt{d}\left([5,3,1^4]\right)=[5,3,1^4],\quad\mathtt{d}\left([3^2,1^6]\right)=[7,3,1^2].
\end{equation}
The possible splittings are
\begin{equation}
    [5,3,1^4]\rightarrow\left([5,3],[1^4]\right),\quad[5,3,1^4]\rightarrow\left([5,1^3],[3,1]\right),\quad[7,3,1^2]\rightarrow\left([7,1],[3,1]\right),
\end{equation}
which give rise to an $\kso(8)$ orbit in the first entry of each pair. We shall view the second orbit in each pair as an $\kso(4)$ orbit (rather than an A-type orbit), but the $\kso(8)$ orbit would be sufficient to determine the fission. Now, the LS duals of these $\kso(8)$ orbits are
\begin{equation}
    \mathtt{d}\left([5,3]\right)=[2^2,1^4],\quad\mathtt{d}\left([5,1^3]\right)=[3,1^5],\quad\mathtt{d}\left([7,1]\right)=[1^8].
\end{equation}
As we can see, this gives the orthosymplectic quivers after the fissions. The unitary quiver under each fission can then be obtained from the difference between the parent quiver and the orthosymplectic quiver in the Higgsed theory. Let us also make a comment on the $\kso(4)$ orbit part. Their LS dual orbits are
\begin{equation}
    \mathtt{d}\left([1^4]\right)=[3,1],\quad\mathtt{d}\left([3,1]\right)=[1^4].
\end{equation}
Roughly speaking, they can be thought of as the branes that are ``separated'' far away from the original brane system. Then in the absence of orientifolds, these half branes would recombine into full branes\footnote{In a very imprecise and loose way, we could schematically treat them as follows. For $[3,1]$, we have
\begin{equation}
    \includegraphics[width=7cm]{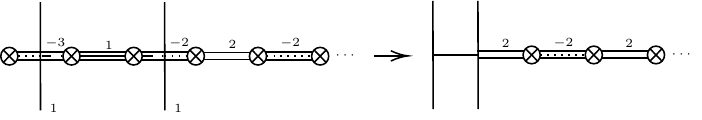}.\nonumber
\end{equation}
Likewise, for $[1^4]$, we have
\begin{equation}
    \includegraphics[width=7cm]{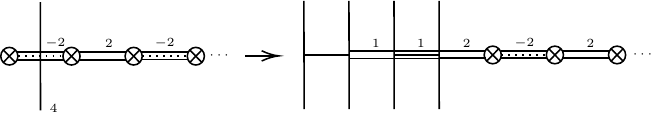}.\nonumber
\end{equation}
In other words, there are some Higgsed theories of the ``$(\kso(4),\kso(4))$ conformal matter theory''. Now, remove all the orientifolds, and combine all the neighbouring half-branes. This gives rise to the $\SU(2)$-type theories with nilpotent orbits $[2]$ and $[1^2]$ respectively.}. This would turn the above partitions into $[2]$ and $[1^2]$ with $\SU(2)$-type theories whose unitary magnetic quivers are shown above.

\section*{Acknowledgement} We are grateful to Jacques Distler, Amihay Hanany, Jonathan J.~Heckman, Craig Lawrie, Deshuo Liu, Dmytro Matvieievskyi, Noppadol Mekareeya, Hiraku Nakajima, Marcus Sperling, Futoshi Yagi, Zhenghao Zhong for fruitful discussions and detailed explanations on various topics. JB is supported by a JSPS fellowship. HYZ is supported by WPI Initiative, MEXT, Japan at Kavli IPMU, the University of Tokyo.

\appendix


\section{Matter Contents and Flavour Symmetries in 6D SCFTs}\label{table6d}
In this appendix, we review the matter content and the flavour symmetry of a 6d SCFT in the F-theory classification, i.e., as a function of the curve self-intersection and the gauge symmetry. The result was obtained in \cite[Table 5.1]{Bertolini:2015bwa}. It is also reviewed in \cite{Heckman:2018jxk} (where Table 3 gives the matter content for each self-intersection and paired gauge symmetry) and in \cite{Frey:2018vpw} (where Table 1 gives the flavour symmetry for each specific matter content). Here, we reproduce this in Table \ref{matterflavour}.
\begin{table}[h]
    \centering
    \begin{tabular}{|c|c|c|} \hline
        $\kg$ & matter for $-n$ curve & global symmetry for $-n$ curve \\ \hline
        $\ksu(2)$ & $(32 - 12n) \frac{1}{2}F$     &   $\kso(32 - 12n)$          \\ \hline
        $\ksu(3)$ & $(18-6n) F$     &   $\ku(18-6n)$          \\ \hline
        $\ksu(4)$ & $(16-4n)F + (2-n) \Lambda^2$     &   $\ku(16-4n) \times \ksp(2-n)$          \\ \hline
        $\ksu(5)$ & $(16-3n)F + (2-n) \Lambda^2$ &   $\ku(16-3n) \times \ksu(2-n)$          \\ \hline
        $\ksu(6)$ & $(16-2n)F + (2-n) \Lambda^2$     &   $\ku(16-2n) \times \ksu(2-n)$         \\ \hline
        $\ksu(6)^* \ \ (n=1)$ & $(16-n)F + \frac{1}{2}(2-n) \Lambda^4$      &   $\ku(16-n) \times \kso(2-n)$          \\ \hline
        $\ksu(k) \ \ k \geq 7$ & $(16-n(8-k))F + (2-n)\Lambda^2$     &   $\ku(16-n(8-k)) \times \ku(2-n)$          \\ \hline
        $\ksp(k) \ \ k \geq 2 \ \ (n=1)$ & $(16+4k)\frac{1}{2}F$     &   $\kso(16+4k)$          \\ \hline
        $\kso(7)$ & $(3-n)V + 2(4-n)S_*$     &   $\ksp(3-n) \times \ksp(8-2n)$          \\ \hline
        $\kso(8)$ & $(4-n)(V + S_+ + S_-)$     &   $\ksp(4-n)^{\otimes 3}$          \\ \hline
        $\kso(9)$ & $(5-n)V + (4-n)S_*$     &   $\ksp(5-n) \times \ksp(4-n)$          \\ \hline
        $\kso(10)$ & $(6-n)V + (4-n)S_*$     &   $\ksp(6-n) \times \ksu(4-n)$          \\ \hline
        $\kso(11)$ & $(7-n)V + (4-n)\frac{1}{2} S_*$     &   $\ksp(7-n) \times \kso(4-n)$          \\ \hline
        $\kso(12)$ & $(8-n)V + (4-n)\frac{1}{2} S_*$     &   $\ksp(8-n) \times \kso(4-n)$          \\ \hline
        $\kso(13)$ & $(9-n)V + (2-\frac{n}{2})\frac{1}{2} S_*$     &   $\ksp(9-n) \times \kso(2-\frac{1}{2}n)$          \\ \hline
        $\kso(k) \ \ k \geq 14 \ \ (n=4)$ & $(k-8)V$     &   $\ksp(k-8)$         \\ \hline
        $\ke_6$ & $(6-n)F$     &   $\ksu(6-n)$          \\ \hline
        $\ke_7$ & $(8-n)\frac{1}{2}F$     &   $\kso(8-n)$          \\ \hline
        $\ke_8 \ \ (n=12)$ & none     &   none          \\ \hline
        $\kf_4$ & $(5-n)F$     &   $\ksp(5-n)$          \\ \hline
        $\kg_2$ & $(10-3n)F$     &   $\ksp(10-3n)$          \\ \hline
    \end{tabular}
    \caption{The matter contents and global symmetries for given curves and gauge symmetries. For each matter gauge symmetry $\kg$, it needs to be paired with the matter content such that the multiplicities in front of each matter content are non-negative - a constraint that we leave implicit. Sometimes, there are extra independent constraints, which we label explicitly in the $\kg$ column. The notations $\ksu(6)$ and $\ksu(6)^*$ simply distinguish the different matter representations. For $\ku$-type flavour symmetries on $\ksu$-type gauge symmetries, some linear combinations would get broken by ABJ anomalies, but such distinction between $\ksu$ and $\ku$ flavour symmetries will not play a big role in our analysis (see \cite{Apruzzi:2020eqi} for more details).
    }\label{matterflavour}
\end{table}

\section{Type IIA Brane Constructions}\label{TypeIIAbranes}
When studying some 6d theories, it would be of great help if the Type IIA brane setups are available. For instance, we can go to the magnetic phase under HW transitions and apply our knowledge of the 3d $\mathcal{N}=4$ Coulomb branch/moduli space of dressed monopole operators. Let us list some basic facts and conventions used in this paper here.

We have different types of branes in the construction. Their occupations of the spacetime directions are summarized in Table \ref{braneoccupations}.
\begin{table}[h]
\centering
\begin{tabular}{c|ccccccccccc}
Branes & $x^0$ & $x^1$ & $x^2$ & $x^3$ & $x^4$ & $x^5$ & $x^6$ & $x^7$ & $x^8$ & $x^9$ \\ \hline
NS5, ON & $\times$ & $\times$ & $\times$ & $\times$ & $\times$ & $\times$ &  &  &  &  \\
D8, O8 & $\times$ & $\times$ & $\times$ & $\times$ & $\times$ & $\times$ &  & $\times$ & $\times$ & $\times$ \\
D6, O6 & $\times$ & $\times$ & $\times$ & $\times$ & $\times$ & $\times$ & $\times$ &  &  &  \\ \hline
F1 & $\times$ &  &  & $\times$ &  &  &  &  &  &  \\
D4 & $\times$ &  &  & $\times$ & $\times$ & $\times$ &  & $\times$ &  & 
\end{tabular}
\caption{The spacetime directions along which the branes extend. The magnetic degrees of freedom arise from the suspensions of the virtual D4-branes in the D6-D8-NS5 systems. Likewise, the virtual F1 strings lead to the electric theories.}\label{braneoccupations}
\end{table}
When drawing the brane systems, the horizontal direction is for $x^6$ while the vertical direction collects $x^{7,8,9}$. In other words, the vertical solid (resp.~dashed) lines will be used to denote the D8-branes (resp.~O8-planes), and the horizontal solid lines will be used to denote D6-branes. The NS5-branes (ON$^0$-planes, resp.~ON$^-$-planes) are denoted as crossed circles (empty circles, resp.~solid circles). The numbers next to the branes indicate their numbers (including negatively charged branes) although we shall often omit the number 1 for a single (full or half) D8-brane (in the magnetic phase) or a single (full or half) NS5-brane. For instance,
\begin{equation}
    \includegraphics[width=2.5cm]{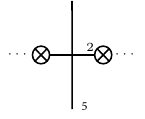}
\end{equation}
shows two D6-branes stretched between the NS5-branes, and there is one D8-brane in the NS5 interval. The D8 branes serve as the flavour branes while the D6-branes give rise to unitary gauge groups. In the above example, the interval gives rise to an $\SU(2)$ gauge group with 5 flavours. The matters transformed (non-trivially) under two adjacent gauge nodes come from the F1-strings stretched between two neighbouring NS5 intervals.

When there are orientifolds, the branes are half-branes. For O6-planes, we follow the convention in \cite{Gaiotto:2008ak} and denote them by\footnote{In the infinite coupling phase, the number labelled next to the NS5-branes would also denote twice the number of half NS5-branes, and we would draw the branes and their images like the D6-branes here.}
\begin{equation}
    \includegraphics[width=13cm]{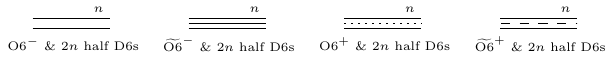}.
\end{equation}
For the D8-branes, the numbers labelled next to them would denote the numbers of the half D8-branes. In an NS5 interval, the above configurations give $\SO(2n)$, $\SO(2n+1)$, $\Sp(n)$, $\Sp'(n)$ gauge groups from the left to the right. If there are $k$ half D8-branes in the interval, there would be an $\Sp(k/2)$ (resp.~$\SO(k)$) flavour symmetry for the negative (resp.~positive) O6-planes.

To set-up a brane configuration for a supersymmetric gauge theory, the cosmological constant needs to satisfy certain constraints \cite{Hanany:1997gh}. Starting from the left boundary and moving to the right, every time one passes through a (full) D$(p+2)$-brane, the cosmological constant $m$ is increased by one unit. Given an NS5-brane, the numbers of D$p$-branes ending on its two sides are determined by the cosmological constant $m$ at the position of the NS5-brane:
\begin{equation}
    m=L_{\text{D}_p}-R_{\text{D}_p},
\end{equation}
where $L_{\text{D}_p}$ (resp.~$R_{\text{D}_p}$) denotes the number of (full) D$p$-branes ending on its left (resp.~right). When there are O$p$-planes, they would also contribute according to their charges \cite{Hanany:1999sj,Feng:2000eq}:
\begin{equation}
    \text{charge}\left(\text{O}p^{\pm}\right)=\pm2^{p-5},\quad\text{charge}\left(\widetilde{\text{O}p}^-\right)=\frac{1}{2}-2^{p-5},\quad\text{charge}\left(\widetilde{\text{O}p}^+\right)=2^{p-5}.
\end{equation}
An O$p^{\pm}$-plane (resp.~$\widetilde{\text{O}p}^{\pm}$-plane) becomes an O$p^{\mp}$-plane (resp.~$\widetilde{\text{O}p}^{\mp}$-plane) when passing through a half NS5-brane. An O$p^{\pm}$-plane becomes an $\widetilde{\text{O}p}^{\pm}$ when passing through a half D$(p+2)$-brane and vice versa.

To move to the magnetic phase, we need to perform HW transitions \cite{Hanany:1996ie}. The brane creations and annihilations should follow the conservation of the linking numbers $l$:
\begin{equation}
    l_{\text{NS}5}=\frac{1}{2}\left(R_{\text{D}(p+2)}-L_{\text{D}(p+2)}\right)+\left(L_{\text{D}p}-R_{\text{D}p}\right),\quad l_{\text{D}(p+2)}=\frac{1}{2}\left(R_{\text{NS}5}-L_{\text{NS}5}\right)+\left(L_{\text{D}p}-R_{\text{D}p}\right),
\end{equation}
where $L_{\text{D}(p+2)}$ (resp.~$R_{\text{D}(p+2)}$) denotes the number of (full) D$(p+2)$-branes left (resp.~right) to the NS5-brane, and likewise for $L_{\text{NS}5}$ and $R_{\text{NS}5}$.

In this paper, we mainly focus on the infinite couplings where the D6-branes would suspend between D8-branes in the magnetic phase\footnote{The complete rule for reading off the magnetic quiver can be found in \cite[\S2.5]{Cabrera:2019dob}.}. To read off the magnetic quiver, we have the following rules \cite{Cabrera:2019dob}:
\begin{equation}
    \includegraphics[width=13cm]{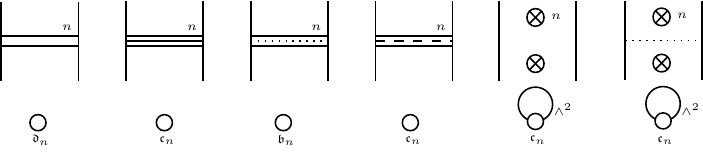}.
\end{equation}
When there are no orientifolds, $n$ D6-branes in a D8 interval would lead to a $\U(n)$ gauge group, and a stack of $n$ NS5-branes would give rise to a $\U(n)$ gauge group with an adjoint.

\subsection{The \texorpdfstring{$(D_4,D_4)$}{(D4,D4)} Conformal Matters of Low Ranks}\label{D4lowranks}
For the conformal matter theories, we would also need the negatively charged branes in the Type IIA constructions \cite{Mekareeya:2016yal,Hanany:2022itc}. In the nilpotent hierarchies, the brane systems can be obtained from the corresponding partitions in a straightforward manner for the classical algebras. Given a partition $[m_1^{p_1},m_2^{p_2},\dots]$, we simply put $p_i$ (half) D8-branes in the $m_i^\text{th}$ NS5 interval. The numbers of D6-branes are determined by the cosmological constant as above. Sometimes, the theory would have curves with different self-intersection numbers decorated by various gauge algebras. Let us just collect some examples here that appear in the nilpotent hierarchies of the $(\SO(8),\SO(8))$ conformal matter theories with one or two $-4$ curves:
\begin{itemize}
    \item rank 1; $[3,1^5]\text{-}[1^8]$:
    \begin{equation}
        \includegraphics[width=8cm]{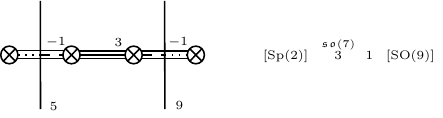},
    \end{equation}
    \item rank 1; $[2^4]^\text{I,II}\text{-}[1^8]$:
    \begin{equation}
        \includegraphics[width=8cm]{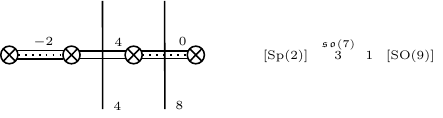},
    \end{equation}
    \item rank 1; $[3,1^5]\text{-}[2^2,1^4]$:
    \begin{equation}
        \includegraphics[width=7cm]{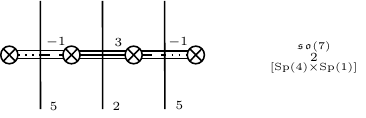},
    \end{equation}
    \item rank 1; $[2^4]^\text{I,II}\text{-}[2^2,1^4]$:
    \begin{equation}
        \includegraphics[width=7cm]{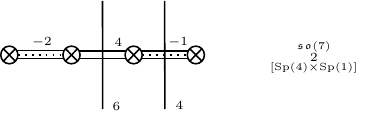},
    \end{equation}
    \item rank 1; $[3,2^2,1]\text{-}[1^8]$:
    \begin{equation}
        \includegraphics[width=8cm]{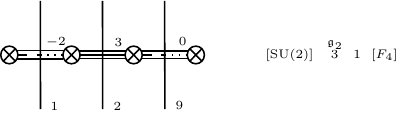},
    \end{equation}
    \item rank 1; $[3,1^5]\text{-}[3,1^5]$:
    \begin{equation}
        \includegraphics[width=6.5cm]{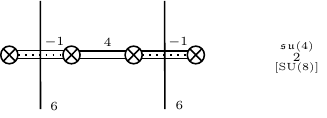},
    \end{equation}
    \item rank 1; $[2^4]^\text{I,II}\text{-}[3,1^5]$:
    \begin{equation}
        \includegraphics[width=6.5cm]{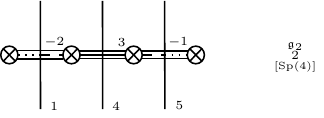},
    \end{equation}
    \item rank 1; $[2^4]^\text{I,II}\text{-}[2^4]^\text{I,II}$:
    \begin{equation}
        \includegraphics[width=6.5cm]{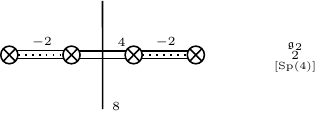},
    \end{equation}
    \item rank 1; $[3^2,1^2]\text{-}[1^8]$:
    \begin{equation}
        \includegraphics[width=7cm]{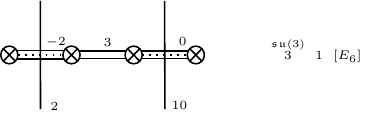},
    \end{equation}
    \item rank 1; $[3,2^2,1]\text{-}[3,1^5]$:
    \begin{equation}
        \includegraphics[width=6.5cm]{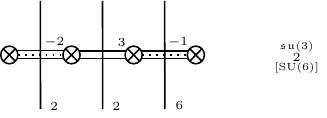},
    \end{equation}
    \item rank 1; $[3,2^2,1]\text{-}[2^4]^\text{I,II}$:
    \begin{equation}
        \includegraphics[width=6.5cm]{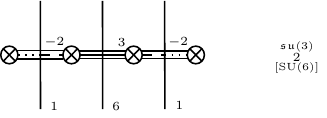},
    \end{equation}
    \item rank 1; $[3,2^2,1]\text{-}[3,2^2,1]$:
    \begin{equation}
        \includegraphics[width=6.5cm]{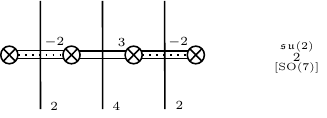},
    \end{equation}
    \item rank 1; $[3^2,1^2]\text{-}[3,1^5]$:
    \begin{equation}
        \includegraphics[width=6.5cm]{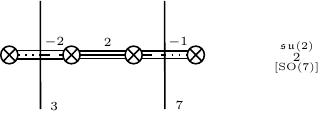},
    \end{equation}
    \item rank 1; $[3^2,1^2]\text{-}[3,2^2,1]$:
    \begin{equation}
        \includegraphics[width=6.5cm]{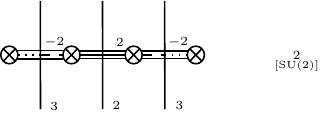},
    \end{equation}
    \item rank 2; $[5,3]\text{-}[3^2,1^2]$:
    \begin{equation}
        \includegraphics[width=9cm]{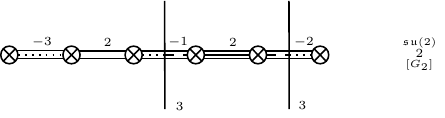},
    \end{equation}
    \item rank 2; $[3^2,1^2]\text{-}[3^2,1^2]$:
    \begin{equation}
        \includegraphics[width=10.5cm]{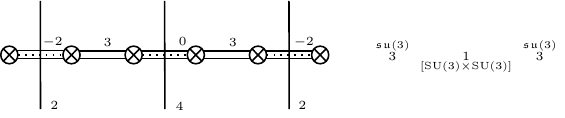},
    \end{equation}
    \item rank 2; $[3,2^2,1]\text{-}[3,2^2,1]$:
    \begin{equation}
        \includegraphics[width=10.5cm]{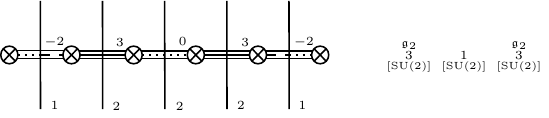}.
    \end{equation}
\end{itemize}

One can then move to the magnetic phases and read off the magnetic quivers. For ``very short'' quivers, the magnetic quivers obtained this way may not have the Coulomb branches that are equal to the 6d Higgs branches. This is illustrated in \S\ref{higherrankDospMQs} and \S\ref{sec:D4hatrank1MQ}, and we shall not repeat this for the examples here. The Hasse diagrams of the rank 1 and rank 2 cases can be found in \S\ref{Dtypes}.

\section{The \texorpdfstring{$(C_3,C_3)$}{(C3,C3)} Conformal Matters}\label{Sp3conformalmatters}
In this Appendix, let us illustrate the magnetic quivers with a C-type conformal matter example. We shall focus on the long quivers in the case of $(C_3,C_3)$ conformal matter theories. For quiver decays, they are associated to the nilpotent orbits in $\ksp(3)$:
\begin{itemize}
    \item $[1^6]$:
    \begin{align}
        &[\Sp(3)] \ \ \overset{\kso(14)}{4} \ \ \overset{\ksp(3)}{1} \ \ \overset{\kso(14)}{4} \ \ \overset{\ksp(3)}{1} \ \ \dots,\nonumber\\
        &\includegraphics[width=15cm]{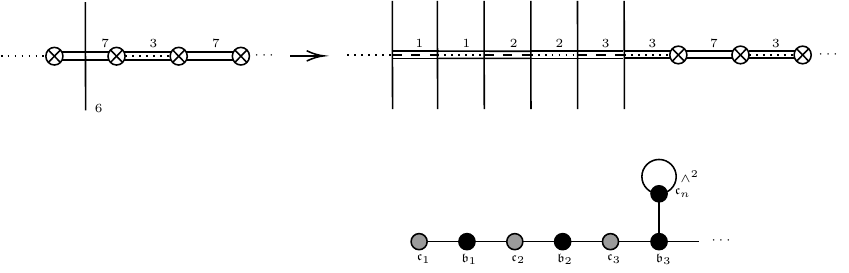};
    \end{align}
    \item $[2,1^4]$ (non-special):
    \begin{align}
        &[\Sp(2)] \ \ \overset{\kso(13)}{4} \ \ \underset{[N_f=1/2]}{\overset{\ksp(3)}{1}} \ \ \overset{\kso(14)}{4} \ \ \overset{\ksp(3)}{1} \ \ \dots,\nonumber\\
        &\includegraphics[width=15cm]{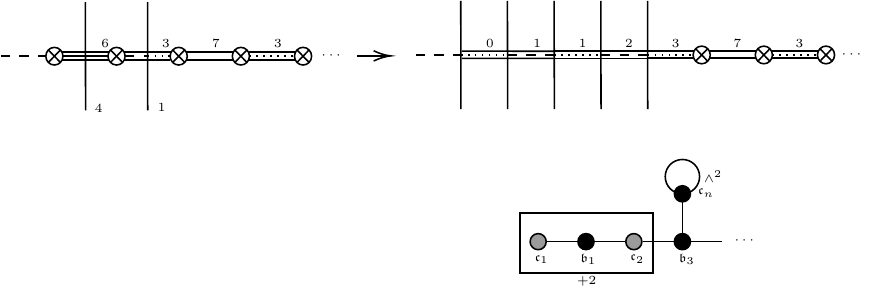},
    \end{align}
    with the slice being $c_3$ for the flow from $[1^6]$ to $[2,1^4]$;
    \item $[2^2,1^2]$:
    \begin{align}
        &[\Sp(1)] \ \ \overset{\kso(12)}{4} \ \ \underset{[N_f=1]}{\overset{\ksp(3)}{1}} \ \ \overset{\kso(14)}{4} \ \ \overset{\ksp(3)}{1} \ \ \dots,\nonumber\\
        &\includegraphics[width=15cm]{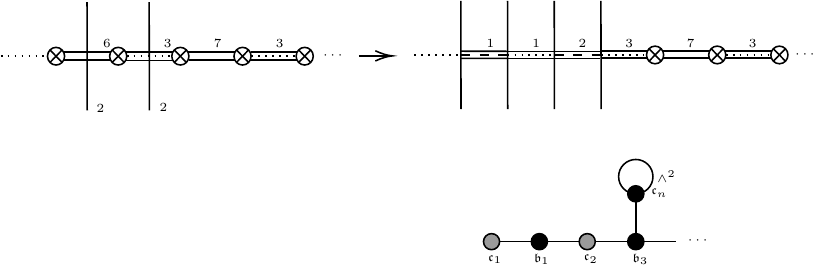},
    \end{align}
    with the slice being $c_2$ for the flow from $[2,1^4]$ to $[2^2,1^2]$;
    \item $[2^3]$:
    \begin{align}
        &\overset{\kso(11)}{4} \ \ \underset{[\SO(3)]}{\overset{\ksp(3)}{1}} \ \ \overset{\kso(14)}{4} \ \ \overset{\ksp(3)}{1} \ \ \dots,\nonumber\\
        &\includegraphics[width=15cm]{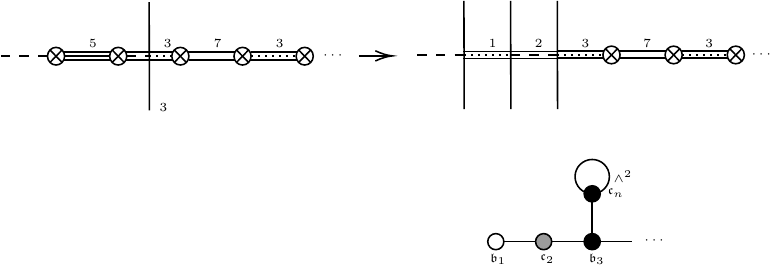},
    \end{align}
    with the slice being $c_1$ for the flow from $[2^2,1^2]$ to $[2^3]$;
    \item $[3^2]$:
    \begin{align}
        &\overset{\kso(10)}{4} \ \ \overset{\ksp(2)}{1} \ \ \underset{[\Sp(1)]}{\overset{\kso(14)}{4}} \ \ \overset{\ksp(3)}{1} \ \ \overset{\kso(14)}{4} \ \ \dots,\nonumber\\
        &\includegraphics[width=15cm]{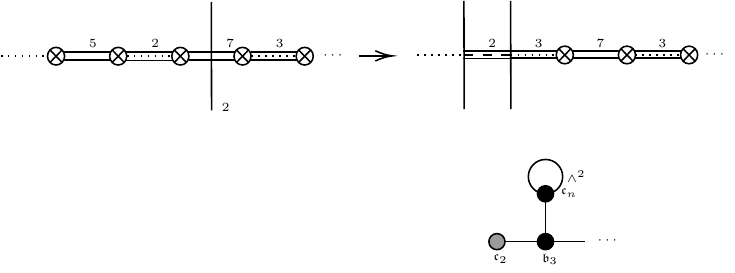},
    \end{align}
    with the slice being $a_1$ for the flow from $[2^3]$ to $[3^2]$ (in this case, the difference of the magnetic quivers is given by a single $\mathfrak{b}_1$ node);
    \item $[4,1^2]$ (non-special):
    \begin{align}
        &[\Sp(1)] \ \ \overset{\kso(11)}{4} \ \ \overset{\ksp(2)}{1} \ \ \overset{\kso(13)}{4} \ \ \underset{[N_f=1/2]}{\overset{\ksp(3)}{1}} \ \ \overset{\kso(14)}{4} \ \ \overset{\ksp(3)}{1} \ \ \dots,\nonumber\\
        &\includegraphics[width=11cm]{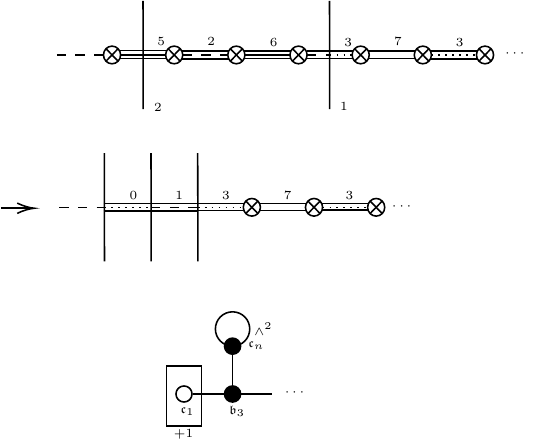},
    \end{align}
    with the slice being $A_1$ for the flow from $[2^3]$ to $[4,1^2]$;
    \item $[4,2]$:
    \begin{align}
        &\overset{\kso(10)}{4} \ \ \underset{[N_f=1/2]}{\overset{\ksp(2)}{1}} \ \ \overset{\kso(13)}{4} \ \ \underset{[N_f=1/2]}{\overset{\ksp(3)}{1}} \ \ \overset{\kso(14)}{4} \ \ \overset{\ksp(3)}{1} \ \ \dots,\nonumber\\
        &\includegraphics[width=11cm]{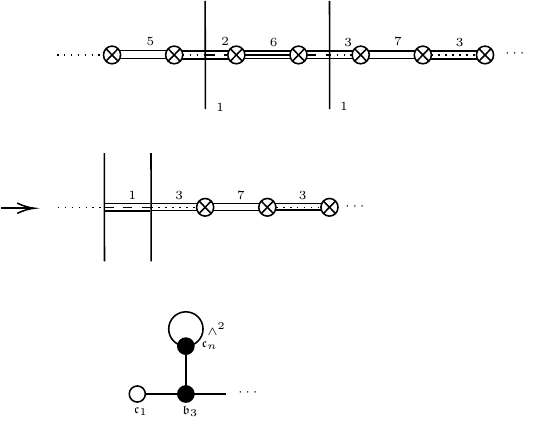},
    \end{align}
    with the slices being $c_1$ for the flows from both $[3^2]$ and $[4,1^2]$ to $[4,2]$;
    \item $[6]$:
    \begin{align}
        &\overset{\kso(9)}{4} \ \ \overset{\ksp(1)}{1} \ \ \overset{\kso(11)}{4} \ \ \overset{\ksp(2)}{1} \ \ \overset{\kso(13)}{4} \ \ \underset{[N_f=1/2]}{\overset{\ksp(3)}{1}} \ \ \overset{\kso(14)}{4} \ \ \overset{\ksp(3)}{1} \ \ \dots,\nonumber\\
        &\includegraphics[width=13cm]{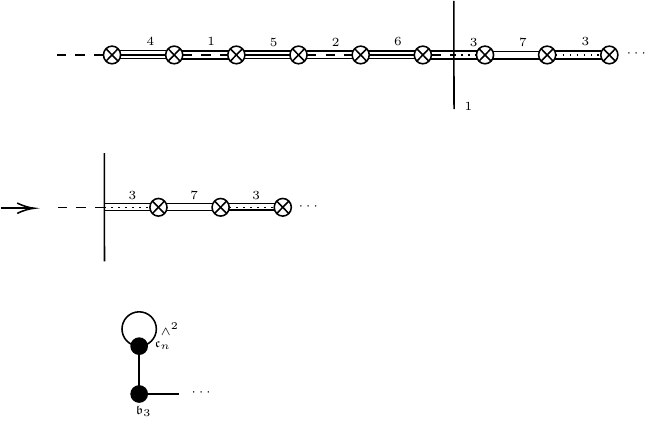},
    \end{align}
    with the slice being $C_3$ for the flow from $[4,2]$ to $[6]$ (although the difference is given by a single $\mathfrak{c}_1$ node).
\end{itemize}

For quiver fissions, there are no fissions from the induced orbits. The only cases would be
\begin{equation}
    \includegraphics[width=13cm]{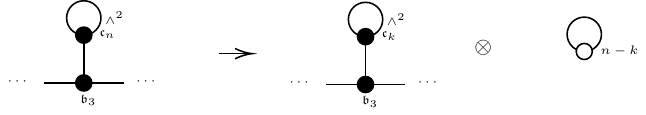}
\end{equation}
with $\frac{m_1+n_1}{2}\leq k\leq n-2$ for the nilpotent orbit pair $[m_1^{p_1},\dots]\text{-}[n_1^{q_1},\dots]$.

\section{Decays and Fissions for Unitary Magnetic Quivers}\label{quiverdecayfission}
Given a quiver, write the ranks of the gauge nodes as a vector $\bm{K}\in\mathbb{Z}^n$. We say that $\bm{K}'\leq\bm{K}$ if $\bm{K}-\bm{K}'$ has non-negative entries. Take the set $\mathcal{V}_0$ of the vectors $\bm{K}'\neq\bm{0}$, satisfying the following conditions:
\begin{itemize}
    \item they are smaller than the vector $\bm{K}$ of the UV theory, that is, $\bm{K}'\leq\bm{K}$;
    \item the corresponding quiver with the vector $\bm{K}'$ is good or ugly, that is, no nodes are underbalanced;
    \item the corresponding quiver does not contain any sub-quiver of U(1) with one or more adjoints;
    \item the corresponding quiver does not contain any sub-quiver that gives the moduli space of instantons, that is, there is no U(1) node connected to a sub-quiver by one simply-laced node, which is fully balanced after deleting this U(1) node.
\end{itemize}
Consider the set $\mathcal{L}_m$ ($m\geq0$) of multisets
\begin{equation}
    \mathcal{L}_m=\{\{\!\!\{\bm{K}_1',\dots,\bm{K}'_m\}\!\!\}\ |\ \bm{K}'_{1,\dots,m}\in\mathcal{V}_0,~\bm{K}_1'+\dots+\bm{K}_m'\leq\bm{K}\},
\end{equation}
and write $\mathcal{L}$ as the union of all $\mathcal{L}_m$. There is a partial order on $\mathcal{L}$ given as follows. Given $l_1,l_2\in\mathcal{L}$, there is a reflexive, antisymmetric but not transitive relation $\leadsto$ such that $l_1\leadsto l_2$ if
\begin{equation}
    |\text{length}(l_1)-\text{length}(l_2)|\leq1,\quad\text{length}(l_1\cap l_2)=\text{length}(l_1)-1,\quad\sum_{\bm{K}'\in l_1}\bm{K}'\geq\sum_{\bm{K}'\in l_2}\bm{K}',
\end{equation}
where $\text{length}(l)=m$ if $l\in\mathcal{L}_m$. Then $l_1\succcurlyeq l_2$ if there exists a chain $l_1\leadsto\dots\leadsto l_2$. The poset $(\mathcal{L},\succcurlyeq)$ coincides with the structure of the Hasse diagram from Higgsing the UV theory. In other words, $\mathcal{L}$ collects the symplectic leaves in the Coulomb branch of the magnetic quiver.

Now, if $l_1\succcurlyeq l_2$ such that $l_1\neq l_2$ and there does not exist any $l_3\neq l_1,l_2$ with $l_1\succcurlyeq l_3\succcurlyeq l_2$, then there are three scenarios depending on $\Delta l=\text{length}(l_1)-\text{length}(l_2)$:
\begin{itemize}
    \item Terminal decay: We have $l_2\subset l_1$ and $\Delta l=1$. In other words, being terminal is in the sense that a sub-quiver in $l_1$ disappears in $l_2$ but this step is not necessarily the last Higgsing in the full Hasse diagram. The transverse slice is the union of $\mu$ copies of the Coulomb branch of the vanishing quiver, where $\mu$ is the multiplicity of the vector for this quiver in $l_1$ (we would omit the multiplicities when drawing the Hasse diagrams in this paper).
    \item Non-terminal decay: We have $\Delta l=0$. There is a unique element $\bm{K}'_1\in l_1\backslash l_2$ and a unique element $\bm{K}'_2\in l_2\backslash l_1$. The transverse slice is given by $\mu$ copies of the Coulomb branch of the quiver determined by $\bm{K}'_1-\bm{K}'_2$ with certain rebalancing (where $\mu$ is the multiplicity of $\bm{K}'_1$ in $l_1$). The rebalancing is done by adding a U(1) node for each connected component $\bm{K}'_{2,\alpha}$ in $\bm{K}'_2$, and there is a $\gcd(\bm{K}'_{2,\alpha})$-laced edge pointing towards the U(1) node.
    \item Fission: We have $\Delta l=-1$. There is exactly one element $\bm{K}'_1\in l_1\backslash l_2$ and two elements $\bm{K}'_2,\bm{K}''_2\in l_2\backslash l_1$. If the quiver corresponding to $\bm{K}'_1$ (with multiplicity $\mu$ in $l_1$) contains 0 or 1 loop, the transverse slice is known. It is $\mu\cdot A_1$ (resp.~$\mu\cdot m$) if $\gcd(\bm{K}'_2)=\gcd(\bm{K}''_2)$ (resp.~$\gcd(\bm{K}'_2)\neq\gcd(\bm{K}''_2)$).
\end{itemize}

\section{The \texorpdfstring{$A_3^3$}{A3**3} Theory and Magnetic Quivers}\label{A3rank3sympsing}
\begin{figure}[H]
    \centering
    \includegraphics[width=0.9\linewidth]{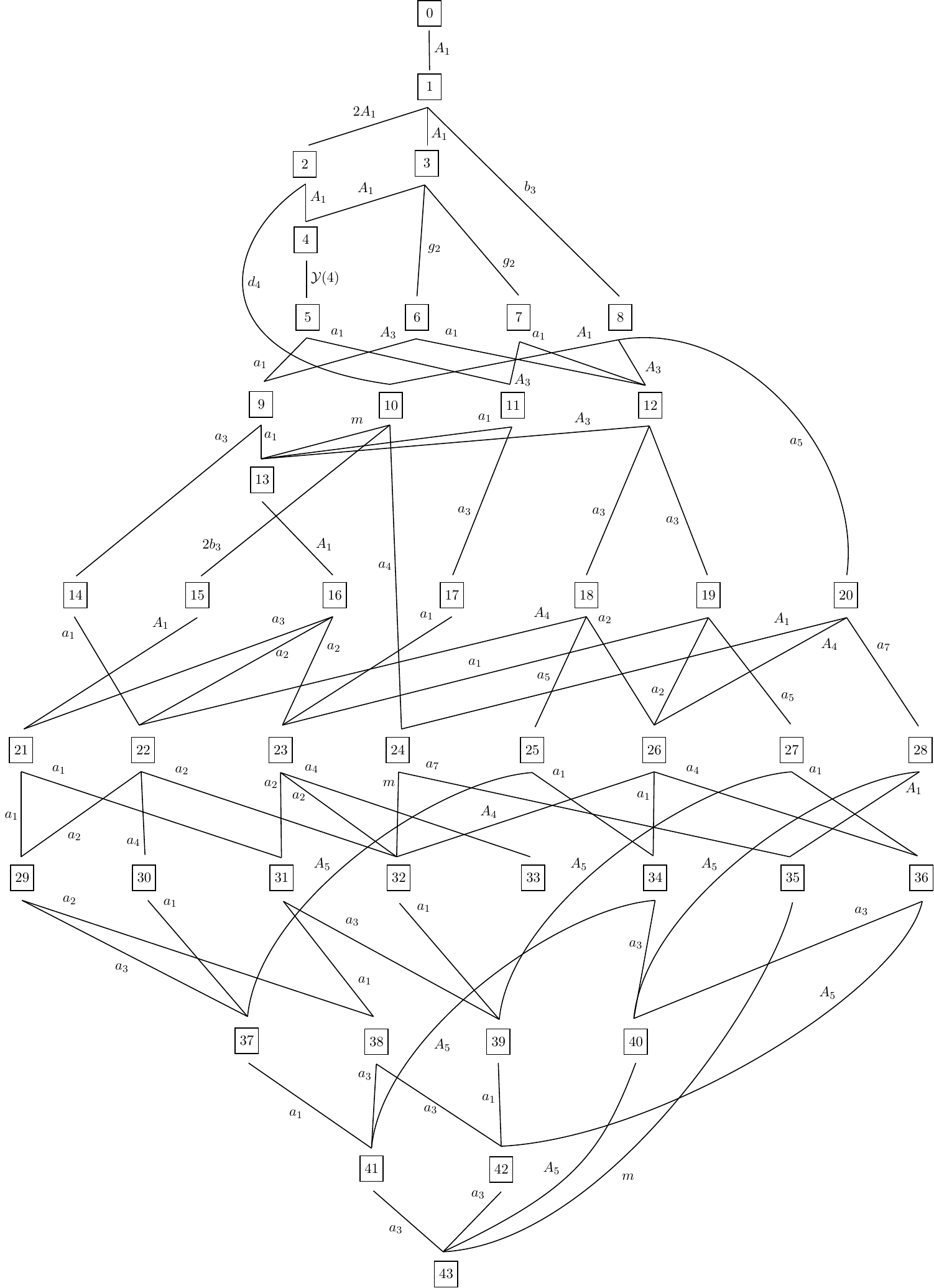}
    \caption{The Hasse diagram for the magnetic quiver \eqref{A3rank3MQ}. In the main context, the parent theory is at the top of the Hasse diagram where the Higgsings follow the arrows. Here, the Hasse diagram follows the stratification of the symplectic singularity to emphasize the geometric feature.}\label{A3rank3MQHasse}
\end{figure}

Recall that the $A_3^3$ theory has the magnetic quiver
\begin{equation}
    \includegraphics[width=5cm]{figures/A3rank3MQ.pdf}.\label{A3rank3MQ}
\end{equation}
To indicate the structure of the symplectic singularity, we use a different convention for the Hasse diagram in Figure \ref{A3rank3MQHasse}, where the parent magnetic quiver is at the bottom.

Notice that in this Hasse diagram, we still omit the multiplicities of the edges. More specifically, there should be double arrows between node 1 and node 2, as well as between node 10 and node 15. This should be clear from the magnetic quivers in Table \ref{A3rank3MQ}. However, unlike the ones in the main context, we separate the different nodes that have the same magnetic quiver. For instance, node 41 and node 42 both correspond to $(A_3^3,1)$ in Figure \ref{A3rank3Hasse}.

\begin{longtable}{|c|c|c|c|c|}
\hline
Label &  Magnetic quiver & $d_{\mathbb{H}}$ & (descendant \#; flow type) & $A_3^3$ label \\ \hline \hline
43 & \includegraphics[width=4.5cm]{figures/A3rank3MQ.pdf} & 19 & (42; $a_3$), (41; $a_3$), (40; $A_5$), (35; $m$) & $(A_3^3,0)$ \\ \hline
42 & \includegraphics[width=3.9cm]{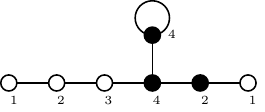} & 16 & (39; $a_1$), (38; $a_3$), (36, $A_5$) & $(A_3^3,1)$ \\ \hline
41 & \includegraphics[width=3.9cm]{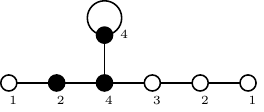} & 16 & (38; $a_3$), (37; $a_1$), (34; $A_5$) & $(A_3^3,1)$ \\ \hline
40 & \includegraphics[width=4.5cm]{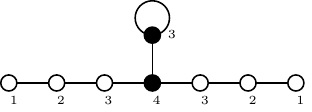} & 18 & (36; $a_3$), (34; $a_3$), (28; $A_5$) & $(A_3^3,2)$ \\ \hline
39 & \includegraphics[width=3.4cm]{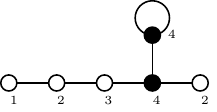} & 15 & (32; $a_1$), (33; $a_3$), (27; $A_5$) & $(A_3^3,3)$ \\ \hline
38 & \includegraphics[width=3.4cm]{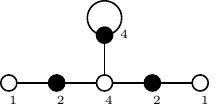} & 13 & (31; $a_1$), (29; $a_2$) & $(A_3^3,4)$ \\ \hline
37 & \includegraphics[width=3.4cm]{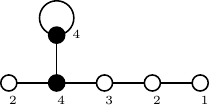} & 15 & (30; $a_1$), (29; $a_3$), (25; $A_5$) & $(A_3^3,3)$ \\ \hline
36 & \includegraphics[width=4cm]{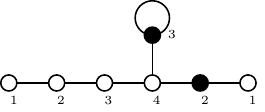} & 15 & (27; $a_1$), (26; $a_4$) & $(A_3^3,5)$ \\ \hline
35 & \includegraphics[width=5cm]{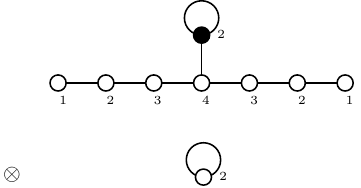} & 17 & (28; $A_1$), (24; $a_7$) & $(A_3^3,6)$ \\ \hline
34 & \includegraphics[width=3.9cm]{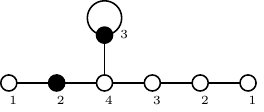} & 15 & (26; $a_1$), (25; $a_1$) & $(A_3^3,5)$ \\ \hline
33 & \includegraphics[width=3.4cm]{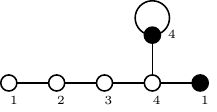} & 14 & (23; $a_4$) & $(A_3^3,7)$ \\ \hline
32 & \includegraphics[width=3.4cm]{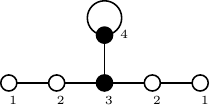} & 12 & (26; $A_4$), (24; $m$), (23; $a_2$), (22; $a_2$) & $(A_3^3,8)$ \\ \hline
31 & \includegraphics[width=2.5cm]{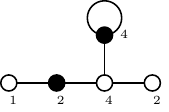} & 12 & (23; $a_2$), (21; $a_1$) & $(A_3^3,9)$ \\ \hline
30 & \includegraphics[width=3.4cm]{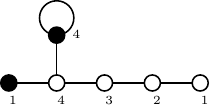} & 14 & (22; $a_4$) & $(A_3^3,7)$ \\ \hline
29 & \includegraphics[width=2.5cm]{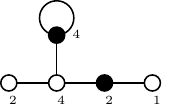} & 12 & (22; $a_2$), (21, $a_1$) & $(A_3^3,9)$ \\ \hline
28 & \includegraphics[width=4.5cm]{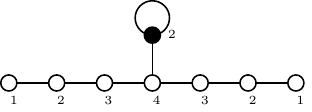} & 17 & (20; $a_7$) & $(A_3^3,10)$ \\ \hline
27 & \includegraphics[width=3.4cm]{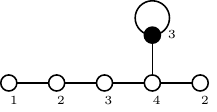} & 14 & (19; $a_5$) & $(A_3^3,11)$ \\ \hline
26 & \includegraphics[width=3.4cm]{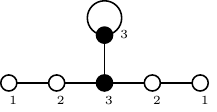} & 11 & (20; $A_4$), (19; $a_2$), (18; $a_2$) & $(A_3^3,12)$ \\ \hline
25 & \includegraphics[width=3.4cm]{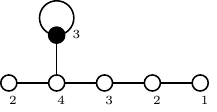} & 14 & (18; $a_5$) & $(A_3^3,11)$ \\ \hline
24 & \includegraphics[width=4.5cm]{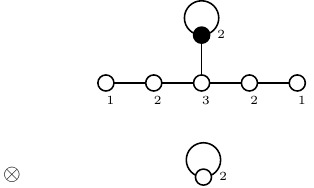} & 11 & (20; $A_1$), (10; $a_4$) & $(A_3^3,13)$ \\ \hline
23 & \includegraphics[width=2.4cm]{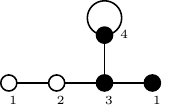} & 10 & (19; $a_1$), (17; $a_1$), (16; $a_2$) & $(A_3^3,14)$ \\ \hline
22 & \includegraphics[width=2.4cm]{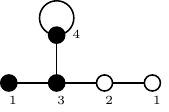} & 10 & (18; $A_4$), (16; $a_2$), (14; $a_1$) & $(A_3^3,14)$ \\ \hline
21 & \includegraphics[width=1.9cm]{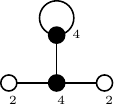} & 11 & (16; $a_3$), (15; $A_1$) & $(A_3^3,15)$ \\ \hline
20 & \includegraphics[width=3.4cm]{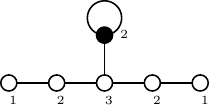} & 10 & (8; $a_5$) & $(A_3^3,16)$ \\ \hline
19 & \includegraphics[width=2.5cm]{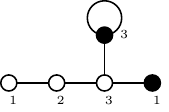} & 9 & (12; $a_3$) & $(A_3^3,17)$ \\ \hline
18 & \includegraphics[width=2.5cm]{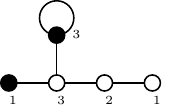} & 9 & (12; $a_3$) & $(A_3^3,17)$ \\ \hline
17 & \includegraphics[width=2cm]{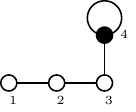} & 9 & (11; $a_3$) & $(A_3^3,18)$ \\ \hline
16 & \includegraphics[width=2cm]{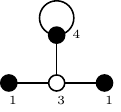} & 8 & (13; $A_1$) & $(A_3^3,19)$ \\ \hline
15 & \includegraphics[width=2.9cm]{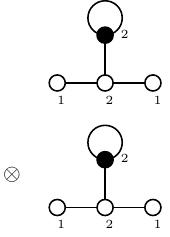} & 10 & (10; $2b_3$) & $(A_3^3,20)$ \\ \hline
14 & \includegraphics[width=2cm]{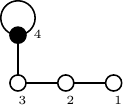} & 9 & (9; $a_3$) & $(A_3^3,18)$ \\ \hline
13 & \includegraphics[width=2cm]{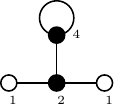} & 7 & (12; $A_3$), (11; $a_1$), (10; $m$), (9, $a_1$) & $(A_3^3,21)$ \\ \hline
12 & \includegraphics[width=2cm]{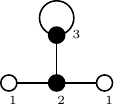} & 6 & (8; $A_3$), (7; $a_1$), (6; $a_1$) & $(A_3^3,22)$ \\ \hline
11 & \includegraphics[width=1.5cm]{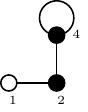} & 6 & (7; $A_3$), (5; $a_1$) & $(A_3^3,23)$ \\ \hline
10 & \includegraphics[width=3.9cm]{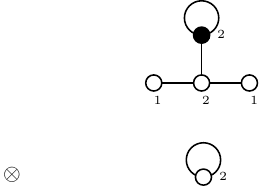} & 6 & (8; $A_1$), (2; $d_4$) & $(A_3^3,24)$ \\ \hline
9 & \includegraphics[width=1.5cm]{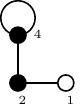} & 6 & (6; $A_3$), (5; $a_1$) & $(A_3^3,23)$ \\ \hline
8 & \includegraphics[width=2cm]{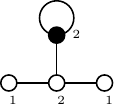} & 5 & (1; $b_3$) & $(A_3^3,25)$ \\ \hline
7 & \includegraphics[width=1.5cm]{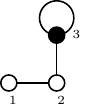} & 5 & (3; $g_2$) & $(A_3^3,26)$ \\ \hline
6 & \includegraphics[width=1.5cm]{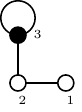} & 5 & (3; $g_2$) & $(A_3^3,26)$ \\ \hline
5 & \includegraphics[width=1cm]{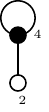} & 5 & (4; $\mathcal{Y}(4)$) & $(A_3^3,27)$ \\ \hline
4 & \includegraphics[width=1cm]{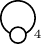} & 3 & (3; $A_1$), (2; $A_1$) & $(A_3^3,28)$ \\ \hline
3 & \includegraphics[width=1cm]{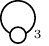} & 2 & (1; $A_1$) & $(A_3^3,29)$ \\ \hline
2 & \includegraphics[width=2cm]{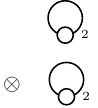} & 2 & (1; $2A_1$) & $(A_3^3,30)$ \\ \hline
1 & \includegraphics[width=1cm]{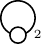} & 1 & (0; $A_1$) & $(A_3^3,31)$ \\ \hline
0 & $\varnothing$ & 0 & IR theory & $(A_3^3,32)$ \\ \hline
\caption{The Hasse diagram for the magnetic quiver \eqref{A3rank3MQ}.}\label{A3rank3MQs_table}
\end{longtable}

\addcontentsline{toc}{section}{References}
\bibliographystyle{utphys}
\bibliography{ref}

\end{document}